\documentclass[preprint,3p,12pt]{elsarticle}
\usepackage{lineno,hyperref}
\usepackage{amsmath,bm,geometry}
\usepackage{amsfonts}
\usepackage{color}
\usepackage{nomencl,etoolbox,ragged2e,siunitx}
\usepackage{mathtools}
\allowdisplaybreaks[0]
\modulolinenumbers[5]

\renewcommand\nomgroup[1]{%
  \item[\bfseries
  \ifstrequal{#1}{N}{Notation}{%
  \ifstrequal{#1}{C}{Constants}{%
  \ifstrequal{#1}{V}{Variables}{%
  \ifstrequal{#1}{S}{Sup/Sub-scripts}{{}}}}}]
  }

\makenomenclature

\nomenclature[N]{$k_B$}{Boltzmann constant}
\nomenclature[N]{$k_B$}{Boltzmann constant}
\nomenclature[N]{$\ell_s$}{mean free path of solid phase}
\nomenclature[N]{$\ell_g$}{mean free path of gas phase}
\nomenclature[N]{$Kn_s$}{Knudsen number of solid phase}
\nomenclature[N]{$Kn_g$}{Knudsen number of gas phase}
\nomenclature[N]{$\epsilon_s$}{volume fraction of solid phase}
\nomenclature[N]{$\epsilon_g$}{volume fraction of gas phase}
\nomenclature[N]{$t$}{Time}
\nomenclature[N]{$T_{ref}$}{characteristic time scale}
\nomenclature[N]{$\vec{x}$}{spacial variable}
\nomenclature[N]{$\vec{v}$}{velocity}
\nomenclature[N]{$\vec{U}_s$}{macroscopic velocity of solid phase}
\nomenclature[N]{$\vec{U}_g$}{macroscopic velocity of gas phase}
\nomenclature[N]{$\vec{g}$}{gravitational acceleration}
\nomenclature[N]{$\rho_s$}{material density of solid phase}
\nomenclature[N]{$\rho_g$}{density of gas phase}
\nomenclature[N]{$p_g$}{pressure of gas phase}
\nomenclature[N]{$\vec{F}_{hydro}$}{hydrodynamic force}
\nomenclature[N]{$m_s$}{mass of one solid particle}
\nomenclature[N]{$\mathcal{Q}$}{particle-particle collision term}
\nomenclature[N]{$T_s^M$}{solid phase material temperature}
\nomenclature[N]{$r_{T_m}$}{proportion of the kinetic energy loss transferred into the material thermal energy in inelastic collision}
\nomenclature[N]{$r$}{restitution coefficient in inelastic collision}
\nomenclature[N]{$f_s$}{velocity distribution function of solid partial phase}
\nomenclature[N]{$g_s$}{Maxwellian distribution of solid phase}
\nomenclature[N]{$g_g$}{Maxwellian distribution of gas phase}
\nomenclature[N]{$T_s$}{granular temperature of solid phase}
\nomenclature[N]{$\tau_s$}{relaxation time of particle phase}
\nomenclature[N]{$C_s$}{heat capacity of solid phase}
\nomenclature[N]{$T_g$}{temperature of gas phase}
\nomenclature[N]{$\tau_T$}{response time of interphase heat conduction}
\nomenclature[N]{$\lambda$}{Reciprocal of normalized temperature}
\nomenclature[N]{$\vec{\phi}$}{conservative moments}
\nomenclature[N]{$d_s$}{diameter of solid particle}
\nomenclature[N]{$C_d$}{drag coefficient}
\nomenclature[N]{$Re_s$}{particle Reynolds number}
\nomenclature[N]{$\nu_g$}{kinematic viscosity of gas phase}
\nomenclature[N]{$\tau_{st}$}{response time of solid particle}
\nomenclature[N]{$E_s$}{energy per unit mass of solid phase}
\nomenclature[N]{$E_g$}{energy per unit mass of gas phase}
\nomenclature[N]{$C_v$}{the specific heat capacity of gas phase at constant volume}
\nomenclature[N]{$\mathbb{I}$}{identity matrix}
\nomenclature[N]{$\mu_g$}{dynamic viscosity of gas phase}
\nomenclature[N]{$\kappa_g$}{thermal conductivity of gas phase}
\nomenclature[N]{$\sigma(\vec{U}_g)$}{strain rate tensor of gas phase}
\nomenclature[N]{$f_{s,ij}$}{cell averaged value of velocity distribution function of solid phase}
\nomenclature[N]{$\vec{W}_{s,ij}$}{cell averaged value of macroscopic variables of solid phase}
\nomenclature[N]{$T_{s,ij}$}{Cell averaged value of material temperature of solid phase}
\nomenclature[N]{$\vec{W}_{g,ij}$}{cell averaged value of macroscopic variables of gas phase}
\nomenclature[N]{$\vec{\omega}_1$}{particle velocity-independent part of the acceleration}
\nomenclature[N]{$\vec{\omega}_2$}{particle velocity-dependent part of the acceleration}
\nomenclature[N]{$\Omega_i$}{volume of $i$-th cell in physical space}
\nomenclature[N]{$\Omega_j$}{volume of $j$-th cell in velocity space}
\nomenclature[N]{$d_s$}{solid particle diameter}
\nomenclature[N]{$\vec{D}$}{interspecies fractional force}
\nomenclature[N]{$p_s$}{granular pressure}
\nomenclature[N]{$\vec{\psi}$}{conservative moments}
\nomenclature[N]{$St_s$}{particle Stokes number}
\nomenclature[N]{$St_c$}{critical Stokes number}
\nomenclature[N]{$\nabla$}{gradient}
\nomenclature[N]{$\partial \Omega$}{boundary of control volume}
\nomenclature[N]{$f_{sT}$}{velocity distribution of solid material temperature}
\nomenclature[S]{$s$}{solid phase}
\nomenclature[S]{$g$}{gas phase}
\nomenclature[S]{$M$}{material (temperature)}
\nomenclature[S]{$i$}{averaged value on $i$-th cell in physical space}
\nomenclature[S]{$j$}{averaged value on $j$-th cell in velocity space}
\nomenclature[S]{$n$}{value at $n$-th time step}
\nomenclature[S]{$n+1$}{value at $(n+1)$-th time step}
\nomenclature[S]{$*$}{intermediary value for an operator splitting method}
\begin{document}

\title{A Unified Gas-kinetic Scheme for Continuum and Rarefied Flows VI: Dilute Disperse Gas-Particle Multiphase System}

\author[HKUST]{Chang Liu}
\ead{cliuaa@ust.hk}

\author[NWPU]{Zhao Wang}
\ead{wzhao@mail.nwpu.edu.cn}

\author[HKUST,HKUST2]{Kun Xu\corref{cor1}}
\ead{makxu@ust.hk}

\address[HKUST]{Mathematics Department, Hong Kong University of Science and Technology, Clear Water Bay, Kowloon, Hong Kong, China}
\cortext[cor1]{Corresponding author}

\address[NWPU]{School of Power and Energy, Northwestern Polytechnical University, 127 Youyi Xilu, Xi'an, Shaanxi, 710072,China}

\address[HKUST2]{HKUST Shenzhen Research institute, Shenzhen 518057, China}

\begin{abstract}
In this paper, a unified gas kinetic scheme (UGKS) for multiphase dilute gas-particle system is proposed.
The UGKS multiphase (UGKS-M) is a finite volume method, which captures flow physics in the regimes
from collisionless multispecies transport to the two-fluid hydrodynamic Navier-Stokes (NS) solution  with the variation of Knudsen number,
and from granular flow regime to dusty gas dynamics with the variation of Stokes number.
The reason for preserving the multiscale nature in UGKS-M  is mainly coming from the direct modeling of the flow physics
in the scales of discrete cell size and time step, where the ratio of the time step over the particle collision time determines flow behavior
in different regimes.
For the particle phase, the integral solution of the kinetic equation is used in the construction of numerical flux,
which takes into account the particle transport, collision, and acceleration.
The gas phase, which is assumed to be in the continuum flow regime,
evolves numerically by the gas kinetic scheme (GKS), which is a subset of the UGKS for the Navier-Stokes solutions.
The interaction between the gas and particle phase is calculated based on a velocity space mapping method,
which solves accurately the kinetic acceleration process.
The stability of UGKS-M is determined by the CFL condition only.
With the inclusion of the material temperature evolution equation of solid particles,
once the total energy loss in inelastic collision transfers into particle material thermal energy,
the UGKS-M conserves the total mass, momentum, and energy for the whole multiphase system.
In the numerical tests, the UGKS-M shows good multiscale property in capturing the particle trajectory crossing (PTC), particle wall reflecting phenomena, and vortex-induced segregation of inertial particles under different Stokes numbers.
The scheme is also applied to simulate shock induced fluidization problem and the
simulation results agree well with experimental  measurement.
\end{abstract}


\maketitle

\section{Introduction}
Gas-particle multiphase flow represents an important class of multiphase flow, which is widely applied in many fields of engineering applications, such as the medicine, chemical process industry, aerospace engineering, and environment science \cite{snider1998sediment,benyahia2007study,dobran1993numerical}.
{
The solid particle phase behaves as a granular flow. In the dilute limit, the rapid granular flow is described by the kinetic theory and has been an active research area in the past several decades \cite{campbell1990rapid,goldhirsch2003rapid,brilliantov2010kinetic}.
The mechanics of a rapid granular flow or granular gas is analogous to that of a classical molecular gas,
but the grain size of the granular flow is much larger than the gas molecular gas,
and the grain collisions are typically inelastic.
In spite of the differences, the Chapman-Enskog kinetic theory of dense gas \cite{chapman1990mathematical} can still be applied to the granular flow,
according to which the Eulerian granular models have been developed for the past decades, a pioneering work is done by Jenkins and Savage \cite{jenkins1983theory}.
Similar to the gas flow, the flow regime of granular gas is characterized by the granular Knudsen number ($Kn_s$), which represents the importance of the particle collisions relative to the particle transport \cite{fox2008}.
When the granular Knudsen number is small and particle collisions are dominant, the dynamics of the granular flow follows the Eulerian-granular hydrodynamic models.
Following the similar asymptotic analysis as Jenkins and Savage, several Eulerian-granular models have been developed by Lun et al. \cite{lun1984kinetic}, Syamlal et al. \cite{syamlal1985hydrodynamics}, Ding and Gidaspow \cite{ding1990bubbling}, which can be viewed as generalizations of the granular Navier-Stokes equations with modified constitutive relations.
The Eulerian-granular models effectively predict the dynamics of the granular gas in the continuum regime with small particle Knudsen, however, when the granular Knudsen number is large, and for wall bounded granular gases, the non-equilibrium effects will appear and the Eulerian hydrodynamic models are no longer a good approximation \cite{fox2008}. In the rarefied regime, the kinetic equation and multiscale models are required in order to capture the physically consistent non-equilibrium phenomena.}

The multiphase flow physics is greatly enriched when accounting the interstitial gas field. The interphase interaction is due to the hydrodynamic force and heat conduction. The hydrodynamic force includes the buoyancy force and drag force that is determined by the particle Reynolds number ($Re_s$) \cite{schiller1935,dobran1993numerical}. Besides the Knudsen number and particle Reynolds number, another important parameter is the normalized particle inertial response time, or the particle Stokes number ($St_s$) \cite{fox2008}. If the Stokes number is sufficiently small, the particles are easily driven by the gas field. For the particle-laden turbulent flow, the particle will not go through the impingement plane, and the phenomenon of preferential concentration of particles emerges \cite{squires1991preferential}. If the Stokes number is large, the gas field can barely affect the particle motion. The phenomenon of particle trajectory crossing (PTC) occurs, and the velocity distribution function will be far from the local equilibrium. In such regime, the nonphysical $\delta$-shock may be predicted by the hydrodynamic models, where the assumption of quasi-equilibrium distribution function breaks.
{
One important gas-particle multiphase flow is the dusty-gas flow, which has wide applications in the industry and environmental science \cite{saito1995numerical,dobran1993numerical,dartevelle2004numerical,saito2001hazard}. In the dusty-gas model, the random motion of dusty phase is weak and the granular temperature is negligible. The gas phase follows the Euler equations, and the dust phase is modeled by a pressureless Euler equations.
The interaction between gas phase and dust phase is due to the drag force and interspecies heat conduction. Since the pressureless Euler equations may develop $\delta$-shocks at isolated points or along the surfaces of co-dimension one \cite{chertock2017hybrid}, to numerically solve the dusty-gas equations is challenging. Several robust and accurate numerical schemes have been developed for the dusty-gas model, including the finite volume schemes developed by T. Saito \cite{saito2002numerical}, T. Saito et al. \cite{saito2003numerical}, M. Pelanti, and R. J. Leveque \cite{pelanti2006high}, and the finite-volume-particle hybrid scheme developed by A. Chertock et al. \cite{chertock2017hybrid}.}
Besides the dusty-gas flow model, there has been continuous interest and efforts on the development of numerical schemes for different flow regimes of gas-particle system, such as the direct numerical simulation (DNS) \cite{uhlmann2005,hoomans1996}, direct simulation Monte Carlo (DSMC) \cite{bird1976}, multiphase particle in cell (MP-PIC) \cite{patankar2001,snider2001,Rourke1996,Rourke2010,Rourke2012}, method of moments (MOM) \cite{desjardins2008,mcgraw1997,fox2008,fox2009}, and hydrodynamic two-fluid solvers \cite{saurel1999}. Due to the rich flow physics and complex flow regimes, the development of efficient multiscale numerical methods is still an active research direction with extremely high demanding in both scientific research and engineering applications.

In this paper, we propose an effective multiscale scheme for monodisperse dilute gas-particle multiphase system.
{
The UGKS-M is applicable for a wide range of Knudsen and Stokes number,
being able to capture the nonequilibrium flow effect such PTC and particle wall refection \cite{marchisio2013computational}.
In the continuum regime, the UGKS-M can recover the hydrodynamic models effectively.
Our scheme is constructed based on the kinetic equation for the solid phase, and the Navier-Stokes equations for the gas phase.
The kinetic equation for solid phase is more fundamental than the Eulerian hydrodynamic equations, especially in the rarefied regime
when the particle Knudsen is large and the local velocity distribution function deviates from the equilibrium.
The nonequilibrium flow physics can be naturally captured by the kinetic equation for solid particles.
The construction of the multiscale method is based on the direct modeling methodology of UGKS.
The integral-solution based UGKS flux is able to provide a physically consistent multiscale dynamics with the variation of the time step to the
local particle collision time, and naturally bridges the kinetic flux and hydrodynamic flux.
Therefore in the continuum regime current scheme can effectively recover the hydrodynamic solution including the viscous effect and heat conduction. Similarly to the original UGKS for neutral gas \cite{ugks2010},
the cell size and time step are not limited by the mean free path and local collision time.}

In the past years, based on the Boltzmann and kinetic model equations,
the gas-kinetic scheme (GKS) for the Euler and Navier-Stokes equations \cite{xu2001},
and the UGKS for the flow simulation in the whole flow regimes have been systematically developed \cite{ugks2010,huang2012,huang2013,xu2017,kun2014book}.
Originally proposed for the neutral gas flow simulation, the UGKS has been successfully developed for many other multiscale transport problems
in the subsequent studies.
For neutral gas transport, the UGKS can capture gas dynamic physics from the highly non-equilibrium regime to continuum regime \cite{wang2017,xiao2017,liu2016,wang2014ugks};
for radiative transport, the UGKS can present solutions from optically thin regime to optically thick regime \cite{sun2015,sun2015asymptotic,sun2017,sun2018};
for plasma transport, the UGKS can capture solution from collisionless Vlasov regime to highly collisional magnetized hydrodynamic regime \cite{liu2017}.
The UGKS has distinguishable advantages compared with many other numerical methods.
Compared with DSMC, UGKS is able to provide more accurate solution without suffering statistical noise.
For the continuum flow and micro flow, the UGKS is much more efficient than DSMC.
For the hypersonic flow simulations, the efficiency of the equation-based UGKS can be much improved through the use of implicit and multigrid techniques \cite{zhu2016,zhu2017}, and becomes comparable to DSMC.
In comparison with asymptotic preserving (AP) schemes, which mostly recover the Euler solutions in the continuum regime,
the UGKS is able to present accurate solutions in the whole flow regimes from the Navier-Stokes to the free molecular transport \cite{chen2015}.
To recover the Navier-Stokes solutions is a basic requirement for a multiscale method to present accurate and reliable solutions in the near-continuum  and transition regimes \cite{xu2017}.
The methodology of direct modeling in UGKS uses cell size and time step to do the modeling,
constructs discrete multiscale governing equations, and gets the numerical evolution solutions.
In other words, the UGKS provides a continuum spectrum of governing equations from the Navier-Stokes to the Boltzmann equation
in a discretized space \cite{liu2016}.
The GKS becomes a subset of UGKS for the continuum flow simulations, especially for the Navier-Stokes solutions.
In this work, we combine the GKS and UGKS to construct a multiscale scheme for gas-particle multiphase flow.
The dynamics of the gas phase is modeled by GKS, and the evolution of particle phase is modeled by UGKS, while they are fully coupled in the
UGKS-Multiphase (UGKS-M) scheme.

The outline of this paper is the following.
{
In Section 2, we introduce the governing equations used for the construction of UGKS-M, namely the kinetic equation and material temperature evolution equation for solid phase and the Navier-Stokes equations for gas phase. The asymptotic behavior of the solid kinetic equation in the continuum regime is also presented.}
The UGKS-M is introduced in Section 3, including the UGKS for particle phase and GKS for gas phase.
In Section 4, numerical experiments in a wide range of flow regimes are studied to validate the multiscale method.
Section 5 is the conclusion.

\section{Kinetic model for dilute disperse gas-particle flow}
{
The unified gas kinetic scheme for multiphase flow is construct based on the kinetic equation for the solid phase and the Navier-Stokes equations for the gas phase.
The time dependent integral solution of kinetic equation is used in the construction of the multiscale flux. Therefore, UGKS-M not only consists with the kinetic equation in the rarefied regime, not also preserves the hydrodynamic models in the continuum regime.
According to Chapman-Enskog theory \cite{chapman1990mathematical}, the hydrodynamic models is the asymptotic limit of the kinetic equation in the continuum limit.
In this section, we first introduce the kinetic equation for particle phase, and then study its asymptotic property in the continuum regime.
The gas phase in continuum flow regime is solved by the gas kinetic scheme. GKS is a kinetic-based Navier-Stokes solver, and therefore we will also briefly introduce the gas phase kinetic equation.}

\subsection{Kinetic equations for particle phase}
Here we consider a dilute monodisperse gas particle multiphase flow.
The solid particle has a single diameter $d_s$, which is much smaller than the solid particle mean free path $\ell_{s}$, i.e.,
the traveling distance in subsequent collision between solid particles,
and the solid phase can be described by a velocity distribution function of its apparent density $f_s(\vec{x},t,\vec{v})$.
The dynamics of solid particle phase is composed of a Williams-Boltzmann type kinetic equation \cite{marchisio2013computational},
coupled with the equation for solid temperature $T_s^M$. The equation system can be written as
{
\begin{align}
  &\frac{\partial f_s}{\partial t}+\nabla_{\vec{x}}\cdot(\vec{v} f_s)+
  \nabla_{\vec{v}} \cdot \left[\left(\vec{g}-\frac{1}{\rho_s} \nabla_{\vec{x}} p_g\right) f_s\right]+
  \nabla_{\vec{v}} \cdot \left(\frac{\vec{D}}{m_s} f\right)=\mathcal{Q},\label{kinetic-s}\\
  &\frac{\partial C_s\rho_s \epsilon_s T_s^M}{\partial t}+\nabla_{\vec{x}}\cdot(C_s \rho_s \epsilon_s T_s^M \vec{U}_s)
  =r_{T_m}(r^2-1)\frac{3\epsilon_s\rho_sk_BT_s}{4\tau_sm_s}+C_s\epsilon_s\rho_s\frac{T_g-T_s^M}{\tau_T}.\label{temp-s}
\end{align}
In Eq.\eqref{kinetic-s}, $\vec{v}$ is the particle velocity vector;
$\vec{g}$ is the gravitational force;
$\rho_s$ is the material density of the solid particle;
$p_g$ the gas phase pressure;
$m_s$ is the mass of one solid particle;
the subscript $s$ denotes the solid phase variables;
and the subscript $g$ denotes the gas phase variables.
The interaction between the gas phase and solid phase is modeled by a hydrodynamic force
\begin{equation}
\vec{F}_{hydro}=-\frac{m_s}{\rho_s} \nabla_{\vec{x}} p_g+\vec{D},
\end{equation}
where $-\frac{m_s}{\rho_s} \nabla_{\vec{x}} p_g$ is a buoyancy force;
$\vec{D}$ is the drag force from gas phase on a particle;
and term $\mathcal{Q}$ accounts the particle-particle collisions.}
Eq.\eqref{temp-s} includes the temperature transport, the heating from inelastic collision and heat conduction between solid and gas phase.
In Eq. \eqref{temp-s}, $k_B$ is the Boltzmann constant; $C_s$ is the specific heat capacity of solid phase; $\epsilon_s$ is the solid volume fraction; $\vec{U}_s$ is the macroscopic velocity of solid phase;
$T_g$ is the temperature of gas phase;
and $r$ is the restitution coefficient of the inelastic collision.
The heat conduction between phases is modeled by a relaxation time $\tau_T$.
{
$r_{T_m}$ stands for the proportion of the kinetic energy loss in inelastic collision which is transferred into the material thermal energy.
In current computational system, if $r_{T_m}=1$, the total mass, momentum, and energy conserve;
if $r_{T_m}<1$, the density and momentum conserve,
the energy loss due to the shape change and phase change, will not be included in current system.}
The collision term $\mathcal{Q}$ is modeled by a BGK-type relaxation model
\begin{equation}\label{Q}
  \mathcal{Q}=\frac{g_s-f_s}{\tau_s},
\end{equation}
where $\tau_s$ is the relaxation time scale. The unconfined particle Knudsen number can be defined as the ratio between the particle relaxation time and the characteristic time scale $T_{ref}$,
\begin{equation}\label{Kn}
  Kn_s=\frac{\tau_s}{T_{ref}},
\end{equation}
where the subscript ${ref}$ denotes the reference parameters.
The equilibrium Maxwellian distribution function $g_s$ can be written as
\begin{equation}\label{gs}
  g_s(\vec{x},t,\vec{v})=\epsilon_s\rho_s\left(\frac{\lambda_s}{\pi}\right)^{\frac32}
  \exp\left(-\lambda_s(\vec{v}_s-\vec{U}_s)^2\right),
\end{equation}
where $\lambda_s$ relates to the granular temperature $T_s$ of solid phase by
\begin{equation}
  \lambda_s=\frac{m_s}{2k_B T_s}.
\end{equation}
The macroscopic moments of $f_s$ and the Maxwellian distribution $g_s$ satisfy
\begin{equation}
\int \vec{\psi} g_{s} d\Xi=\int \vec{\psi}' f_s  d\Xi,
\end{equation}
where $\vec{\psi}=\left(1,\vec{v},\frac12\vec{v}^2\right)^T$,
$\vec{\psi}'=\left(1,\vec{v},\frac12\vec{v}^2+\frac{r^2-1}{2}(\vec{v}-\vec{U}_s)^2\right)^T$, and $d\Xi = d^3 \vec{v}$.
The restitution coefficient $r$ values from zero to one. $r=0$ represents the fully inelastic collision and $r=1$ represents the elastic collision.
{
For dilute gas-particle flows, the drag force is approximated by \cite{schiller1935}
\begin{equation}
\vec{D}(\vec{U}_g,\vec{v})=\frac{3 m_s \rho_g}{4 d_s \rho_s} C_d \left|\vec{U}_g-\vec{v}\right| (\vec{U}_g-\vec{v}),
\end{equation}
where $\vec{U}_g$ is the gas phase velocity, $\rho_g$ is the gas phase density, and $C_d$ is the particle drag coefficient given by \cite{dobran1993numerical}
\begin{equation}
C_d=\left\{
\begin{aligned}
&\frac{24}{Re_s}(1+0.15Re_s^{0.687}) \quad \emph{\text{if}}\quad Re_{s}<1000 ,\\
&0.44\quad \quad\quad\quad\quad\quad\quad\quad\  \emph{\text{if}} \quad Re_{s}>1000 ,
\end{aligned}
\right.
\end{equation}
where $Re_s=\left|\vec{U}_g-\vec{v}\right|d_s/\nu_g$ is the particle Reynolds number with $\nu_g$ the kinematic viscosity of gas phase.
In this work, an efficient multiscale numerical scheme UGKS-M is constructed.
For large interspecies velocity difference and large drag coefficient, the drag force term becomes stiff and enforces a small explicit time step.
In order to remove the stiffness constraint on time step, an implicit method is used to predict the velocity acceleration.
Currently the implicit treatment requires a simpler form of drag force
\begin{equation}\label{simplified-drag}
\vec{D}(\vec{U}_g,\vec{v})=\frac{m_s}{\tau_{st}}(\vec{U}_g-\vec{v}),
\end{equation}
with a $\vec{v}$-independent particle inertial response time $\tau_{st}$.
For low particle Reynolds number, $\tau_{st}\approx \rho_sd_s^2/(18\mu_g)$, and for large particle Reynolds number
\begin{equation}
\tau_{st}\approx\left\{
\begin{aligned}
&\frac{d_s^2\rho_s}{18\mu_g+2.7(\rho_g d_s|\vec{U}_g-\vec{U}_s|)^{0.687}\mu^{0.313}} , \quad\quad\quad \emph{\text{if}}\quad Re_{s}<1000\\
&\frac{d_s\rho_s}{0.33\rho_g|\vec{U}_g-\vec{U}_s|} , \quad \quad\quad\quad\quad\quad\quad\quad\quad\quad\quad\ \,  \emph{\text{if}} \quad Re_{s}>1000
\end{aligned}
\right.
\end{equation}
where $\mu_g$ is the dynamic viscosity of gas phase.
For the computational tests in this paper, the simplified drag force formulation Eq.\eqref{simplified-drag} is used.}
The dimensionless particle inertial response time is the particle Stokes number,
defined by $St_s=\tau_{st}/T_{ref}$.
{
The above kinetic equation \eqref{kinetic-s} and solid material temperature equation \eqref{temp-s} are the equations we use to construct the UGKS-M. In the following, we give the asymptotic limit of the solid particle kinetic equation following the Chapman-Enskog asymptotic analysis \cite{chapman1990mathematical}.}

When the elastic collision dominates, at the solid particle collision with the restitution coefficient $r=1$,
{
the hydrodynamic equations for the solid phase in the Euler regime as $\tau_s \to 0$ are \cite{chapman1990mathematical}
\begin{equation}\label{Euler-s}\footnotesize
  \begin{aligned}
    &\frac{\partial \epsilon_s \rho_s}{\partial t}+\nabla_{\vec{x}}\cdot(\epsilon_s\rho_s \vec{U}_s)=0,\\
    &\frac{\partial \epsilon_s \rho_s \vec{U}_s}{\partial t}+\nabla_x\cdot(\epsilon_s \rho_s\vec{U}_s\vec{U}_s+p_s\mathbb{I})=
    \frac{\epsilon_s \rho_s (\vec{U}_g-\vec{U}_s)}{\tau_{st}}
    -\epsilon_s \nabla_{\vec{x}}p_g+\epsilon_s\rho_s\vec{g},\\
    &\frac{\partial \epsilon_s \rho_s E_s}{\partial t}+\nabla_{\vec{x}}\cdot((\epsilon_s \rho_s E_s+p_s)\vec{U}_s)=
    \frac{\epsilon_s \rho_s\vec{U}_s}{\tau_{st}}(\vec{U}_g-\vec{U}_s)-\epsilon_s\nabla_{\vec{x}}p_g\cdot \vec{U}_s+\epsilon_s \rho_s\vec{U}_s\vec{g}-\frac{3p_s}{\tau_{st}},
  \end{aligned}
\end{equation}
where $\rho_s$ is the material density of solid particle;
$\epsilon_s\rho_s$ is the apparent density;
$\vec{U}_s$ is the macroscopic velocity;
$T_s$ is the granular temperature;
$p_s=\epsilon_s \rho_s R_s T_s$ is the granular pressure with $R_s=k_B/m_s$;
and $\epsilon_s \rho_s e_s= \frac32p_s$ is the granular internal energy.
The total granular energy density is $\epsilon_s \rho_s E_s=\frac12\epsilon_s\rho_s\vec{u}^2+\epsilon_s \rho_se_s$.
In energy equation, term $-3p_s/\tau_{st}$ stands for the energy loss due to interspecies friction.}
For inelastic collision with $r<1$, the solid phase is governed by the pressureless Euler equation in the continuum regime  \cite{chapman1990mathematical}
\begin{equation}\label{pressureless-Euler-s}\footnotesize
  \begin{aligned}
    &\frac{\partial \epsilon_s \rho_s}{\partial t}+\nabla_{\vec{x}}\cdot(\epsilon_s\rho_s \vec{U}_s)=0,\\
    &\frac{\partial \epsilon_s \rho_s \vec{U}_s}{\partial t}+\nabla_x\cdot(\epsilon_s \rho_s\vec{U}_s\vec{U}_s)=
    \frac{\epsilon_s \rho_s (\vec{U}_g-\vec{U}_s)}{\tau_{st}}
    -\epsilon_s \nabla_{\vec{x}}p_g+\epsilon_s\rho_s\vec{g}.
  \end{aligned}
\end{equation}
The pressureless Euler equations together with the material temperature evolution equation Eq.\eqref{temp-s} are also known as the dusty flow equations.
In the numerical tests, we will also compare the UGKS-M solution with Navier-Stokes and pressureless Euler equations to demonstrate the multiscale property of the UGKS-M.

\subsection{Governing equations for gas phase}
{
The gas phase in the continuum regime is governed by the Navier-Stokes equations
\begin{equation}\label{ns-g}\scriptsize
\begin{aligned}
  \frac{\partial \epsilon_g\rho_g}{\partial t}+\nabla_{\vec{x}}
  \cdot\left(\rho_g \vec{U}_g\right) &=0,\\
  \frac{\partial \epsilon_g\rho_g \vec{U}_g}{\partial t}+\nabla_{\vec{x}}\cdot
  \left(\rho_g \vec{U}_g\vec{U}_g+ p_g\mathbb{I}-\mu_g \sigma(\vec{U}_g)\right)
  &=-\frac{\epsilon_s \rho_s (\vec{U}_g-\vec{U}_s)}{\tau_{st}}
  +\epsilon_g\rho_g \vec{g}+\epsilon_s\nabla_{\vec{x}}p_g,\\
  \frac{\partial \epsilon_g \rho_g E_g}{\partial t}+
  \nabla_x\cdot \left( \vec{U}_g \left(\rho_g E_g+ p_g\right)-\mu_g \sigma(\vec{U}_g) \vec{U}+\kappa_g\nabla_x T_g\right)&=-\frac{\epsilon_s \rho_s\vec{U}_s}{\tau_{st}}(\vec{U}_g-\vec{U}_s)+\epsilon_s\nabla_{\vec{x}}p_g\cdot \vec{U}_s
  +\frac{3p_s}{\tau_{st}}+\epsilon_g\rho_g\vec{U}_g\cdot \vec{g}-C_s\epsilon_s\rho_s\frac{T_g-T_s^M}{\tau_T},\\
\end{aligned}
\end{equation}}
where $\rho_g$ is the density of gas phase;
$\epsilon_g \rho_g$ is the apparent density of gas phase;
$\vec{U}_g$ is the macroscopic velocity;
$T_g$ is the temperature;
$p_g=\rho_gR_gT_g$ is gas pressure;
and $\epsilon_g \rho_g E_g=\epsilon_g \rho_g \vec{U}^2_g/2+\epsilon_g p_g/(\gamma-1)$.
Tensor $\sigma(\vec{U}_g)$ denotes the strain rate tensor given by
\begin{equation}\label{strain}
  \sigma(\vec{U}_g)=\left(\nabla_{\vec{x}} \vec{U}_g+(\nabla_{\vec{x}} \vec{U}_g)^T\right)-\frac23 \text{div}_{\vec{x}} \vec{U}_g \mathbb{I}.
\end{equation}
The viscosity $\mu_g$ and the thermal conductivity $\kappa_g$ are expressed as
\begin{equation}
\mu_g=\tau_g p_g,\quad
\kappa_g=\frac{5}{2} R_g\tau_g p_g.
\end{equation}
In the energy equation, $3p_s/\tau_{st}$ is the energy increase from the interspecies friction;
and $-C_s\epsilon_s\rho_s(T_g-T_s^M)/\tau_T$ is the energy change due to interspecies heat conduction.
{
The above Navier-Stokes equations are solved by GKS \cite{xu2001}, which is a kinetic equation based NS solver.} The kinetic equation and the interspecies heat conduction equation can be written as
{\footnotesize
\begin{align}
  &\frac{\partial \epsilon_g f_g}{\partial t}+\nabla_{\vec{x}}\cdot(\vec{v}f_g )+\nabla_{\vec{v}}\cdot\left\{\left[\frac{1}{\tau_{st}}(\vec{U}_g-\vec{v})-\frac{1}{\rho_s}\nabla p_g\right]f_s\right\}+\nabla_{\vec{v}}\cdot(\vec{g} \epsilon_g f_g)=\frac{\epsilon_g g_g-\epsilon_g f_g}{\tau_g},\label{kinetic-gas}\\
  &\frac{d T_g}{d t}=-\frac{\epsilon_s\rho_sC_s}{\epsilon_g\rho_gC_v}
  \frac{T_g-T^M_s}{\tau_T}, \label{temp-g}
\end{align}}%
where $\epsilon_g=1-\epsilon_s$ is the gas volume fraction;
$g_g$ is the local equilibrium distribution function of gas phase;
$\tau_g$ is the relaxation time of gas phase;
$C_v$ is the specific heat capacity of gas phase at constant volume;
and $g_g$ is the Maxwellian distribution satisfying
$$\int \vec{\psi}_g (g_g-f_g) d\vec{v}d\xi=0,$$
with
$$\vec{\psi}_g=(1,u,v,w,\frac12(\vec{v}^2+\xi^2))^T.$$
The Navier-Stokes equations Eq.\eqref{ns-g} can be recovered from the kinetic equation Eq.\eqref{kinetic-gas} in the continuum regime following the Chapman-Enskog asymptotic analysis \cite{chapman1990mathematical}.

\section{Unified gas-kinetic scheme for gas-particle multiphase flow}
\subsection{Unified gas-kinetic scheme for solid particle phase}
\subsubsection{General framework}
The UGKS for solid particle phase is built on a finite volume framework. The phase space is divided into a set of numerical control volumes $\mathbf{X}\otimes\mathbf{V}=\sum_i\Omega_{x_i}\otimes\sum_j\Omega_{v_j}=\sum_{ij}\Omega_{ij}$.
The scheme evolves the cell averaged distribution
$$f_{s,ij}=\frac{1}{\Omega_{ij}}\int_{\Omega_{ij}} f_s(\vec{x},t,\vec{v}) d\vec{x} d\vec{v},$$
and cell averaged macroscopic variables
\begin{align}\nonumber
\vec{W}_{s,i}&=\frac{1}{\Omega_i}\int \vec{W_s}(\vec{x})d\vec{x},\\
T^M_{s,i}&=\frac{1}{\Omega_i}\int T^M_{s,i}(\vec{x})d\vec{x},
\end{align}
where the macroscopic variables are $\vec{W_s}=(\epsilon_s\rho_s,\epsilon_s\rho_s\vec{U}_s,\epsilon_s\rho_sE_s)^T.$
The particle phase kinetic equation Eq.\eqref{kinetic-s} is split as
\begin{align}
  &\mathcal{L}_{s1}:\frac{\partial f_s}{\partial t}+\nabla_{\vec{x}}\cdot(\vec{v}f_s)+
  \nabla_{\vec{v}}\cdot(\vec{\omega}_1 f_s)=\frac{g_s-f_s}{\tau_s},\label{kinetic-s1}\\
  &\mathcal{L}_{s2}:\frac{\partial f_s}{\partial t}+\nabla_{\vec{v}}\cdot (\vec{\omega}_2 f_s)=0,\label{kinetic-s2}
\end{align}
where $\vec{\omega}_1=\left(\vec{g}-\nabla p_g/\rho_s\right)$ is the particle velocity-independent part of the acceleration, and $\vec{\omega}_2=(\vec{U}_g-\vec{v})/\tau_{st}$ is the particle velocity-dependent part of the acceleration term.
For $\mathcal{L}_{s1}$, the Eulerian finite volume scheme is adopted, while for $\mathcal{L}_{s2}$ a velocity space mapping method is utilized to evolve the distribution function.
The numerical evolution equations for the velocity distribution function are
\begin{align}
    &\mathcal{L}_{sf1}:\quad f^{*}_{s,ij}=f^{n}_{s,ij}+
    \frac{1}{|\Omega_i|}\int_{t^n}^{t^{n+1}}\oint_{\partial\Omega_i}
    f_{s,\partial \Omega_i} (t,\vec{v}_j) \vec{v}_j\cdot d\vec{s} dt\nonumber\\
    &\ \quad\quad\quad\quad\quad\quad\quad\quad+\frac{1}{|\Omega_j|}\int_{t^n}^{t^{n+1}}\oint_{\partial \Omega_j}
    f_{s,\partial \Omega_j} (\vec{x}_i,t) \vec{\omega}_1\cdot d\vec{s} dt,\label{ugks-f1}\\
   &\mathcal{L}_{sf2}:\quad f_{s,ij}^{**}=\frac{1}{|\Omega_{j}|}\int_{P_{\omega_2}(\Omega_{j})} f^{*}_{s,i}(\vec{v})d\vec{v},\label{ugks-f2}\\
    &\mathcal{L}_{sf3}:\quad f^{n+1}_{s,ij}=\left(f^{**}_{s,ij}+\frac{\Delta t}{\tau_{s,ij}}g^{n+1}_{s,ij}\right)/\left(1+\frac{\Delta t}{\tau_{s,ij}}\right).\label{ugks-f3}
\end{align}
The evolution of the velocity distribution function is coupled with the evolution of the macroscopic variables $\vec{W}_{s,i}, T^M_{s,i}$.
{\footnotesize
\begin{align}
&\mathcal{L}_{sw1}:\quad\vec{W}_{s,i}^{*}=\vec{W}_{s,i}^n+
\frac{1}{|\Omega_i|}\oint_{\partial \Omega_i}\int_{t^n}^{t^{n+1}}
\int\vec{\psi} f_{s,\partial \omega_i} (\vec{v},t) \vec{v}\cdot\vec{e}_1 d\Xi dt ds+
\Delta t \vec{S}_{s,i}^n,\label{ugks-w1}\\
&\quad \quad\quad\quad C_s \rho_s \epsilon_{s,i}^{n+1}T_{s,i}^*=C_s \rho_s \epsilon_{s,i}^{n+1}T_{s,i}^n+
\frac{1}{|\Omega_i|}\oint_{\partial \Omega_i}\int_{t^n}^{t^{n+1}}
\int f_{sT,\partial \omega_i} (\vec{v},t) \vec{v}\cdot\vec{e}_1 d\Xi dt ds\label{ugks-T}\\
&\mathcal{L}_{sw2}:\quad\vec{U}^{n+1}_{s,i}=\gamma_{Us1}\vec{U}_{s,i}^*+\gamma_{Us2}\vec{U}^*_{g,i},\label{ugks-w2}\\
&\mathcal{L}_{sw3}:\quad\vec{W}^{**}_{s,i}=\int \vec{\psi}f^{**}_{s,ij}d\Xi,\label{ugks-w3}\\
&\mathcal{L}_{sw4}:\quad e^{n+1}_{s,i}=e^{**}_{s,i}/\left(1+\frac{\Delta t}{2\tau_{s,i}}(1-r^2)\right),\label{ugks-w4}\\
&\mathcal{L}_{sw5}:\quad T^{M,n+1}_{s,i}=\gamma_{T1}\left(T^{M,*}_{s,i}+\frac{\Delta t(1-r^2)e^{**}_{s,i}}{[2\tau_{s,i}+\Delta t(1-r^2)]C_{s,i}}\right)
-\gamma_{T2}\left(T^*_{g,i}+\frac{{\epsilon_s}^{n+1}_i\rho_{s}(E^{**}_{s,i}-E^{*}_{s,i})}{{\epsilon_g}^{n+1}_i\rho^{n+1}_{g,i}C_{v,i}}\right),\label{ugks-w5}
\end{align}
}%
where
\begin{align*}
\vec{S}^n_i=(0,\theta_i \rho_{s} \vec{g}-\theta_{i}\nabla_{\vec{x}}p_{g,i}^n,\theta_i \rho_{s} \vec{U}_{s,i}\cdot\vec{g}-\theta_i\vec{U}_{s,i}\cdot\nabla_{\vec{x}}p_{g,i}^n),
\end{align*}
and the coefficients
\begin{align*}
\gamma_{Us1}&=\frac{\epsilon_{s,i}^{n+1}\rho_s+\epsilon_{g,i}^{n+1}\rho_{g,i}^{n+1}
\exp\left[-\left(1+\frac{\epsilon_{s,i}^{n+1}\rho_s}{\epsilon_{g,i}^{n+1}\rho_{g,i}^{n+1}}\right)\frac{\Delta t}{\tau_{st,i}}\right]}
{\epsilon_{g,i}^{n+1}\rho_{g,i}^{n+1}+\epsilon_{s,i}^{n+1}\rho_s},\\
\gamma_{Us2}&=\frac{\epsilon_{g,i}^{n+1}\rho_{g,i}^{n+1}-\epsilon_{g,i}^{n+1}\rho_{g,i}^{n+1}
\exp\left[-\left(1+\frac{\epsilon_{s,i}^{n+1}\rho_s}{\epsilon_{g,i}^{n+1}\rho_{g,i}^{n+1}}\right)\frac{\Delta t}{\tau_{st,i}}\right]}
{\epsilon_{g,i}^{n+1}\rho_{g,i}^{n+1}+\epsilon_{s,i}^{n+1}\rho_s},\\
\gamma_{Ts1}&=\frac{{\epsilon_s}^{n+1}_i\rho_{s}C_s+{\epsilon_g}^{n+1}_i\rho^{n+1}_{g,i}C_v
\exp\left[-\left(1+\frac{{\epsilon_s}^{n+1}_i \rho_{s}C_s}{{\epsilon_g}^{n+1}_i \rho^{n+1}_{g,i}C_v}\right)\frac{\Delta t}{\tau_{T,i}}\right]}
{{\epsilon_g}^{n+1}_i \rho^{n+1}_{g,i}C_v+{\epsilon_s}^{n+1}_i \rho_{s}C_s},\\
\gamma_{Ts2}&=\frac{-{\epsilon_g}^{n+1}_i\rho^{n+1}_{g,i} C_v+{\epsilon_g}^{n+1}_i\rho^{n+1}_{g,i} C_v
\exp\left[-\left(1+\frac{{\epsilon_s}^{n+1}_i \rho_{s} C_s}
{{\epsilon_g}^{n+1}_i \rho^{n+1}_{g,i} C_v}\right)\frac{\Delta t}{\tau_{T,i}}\right]}
{{\epsilon_g}^{n+1}_i \rho^{n+1}_{g,i} C_v+{\epsilon_s}^{n+1}_i \rho_{s}}.
\end{align*}

\subsubsection{Multiscale flux}
\begin{figure}
\centering
\includegraphics[width=0.8\textwidth]{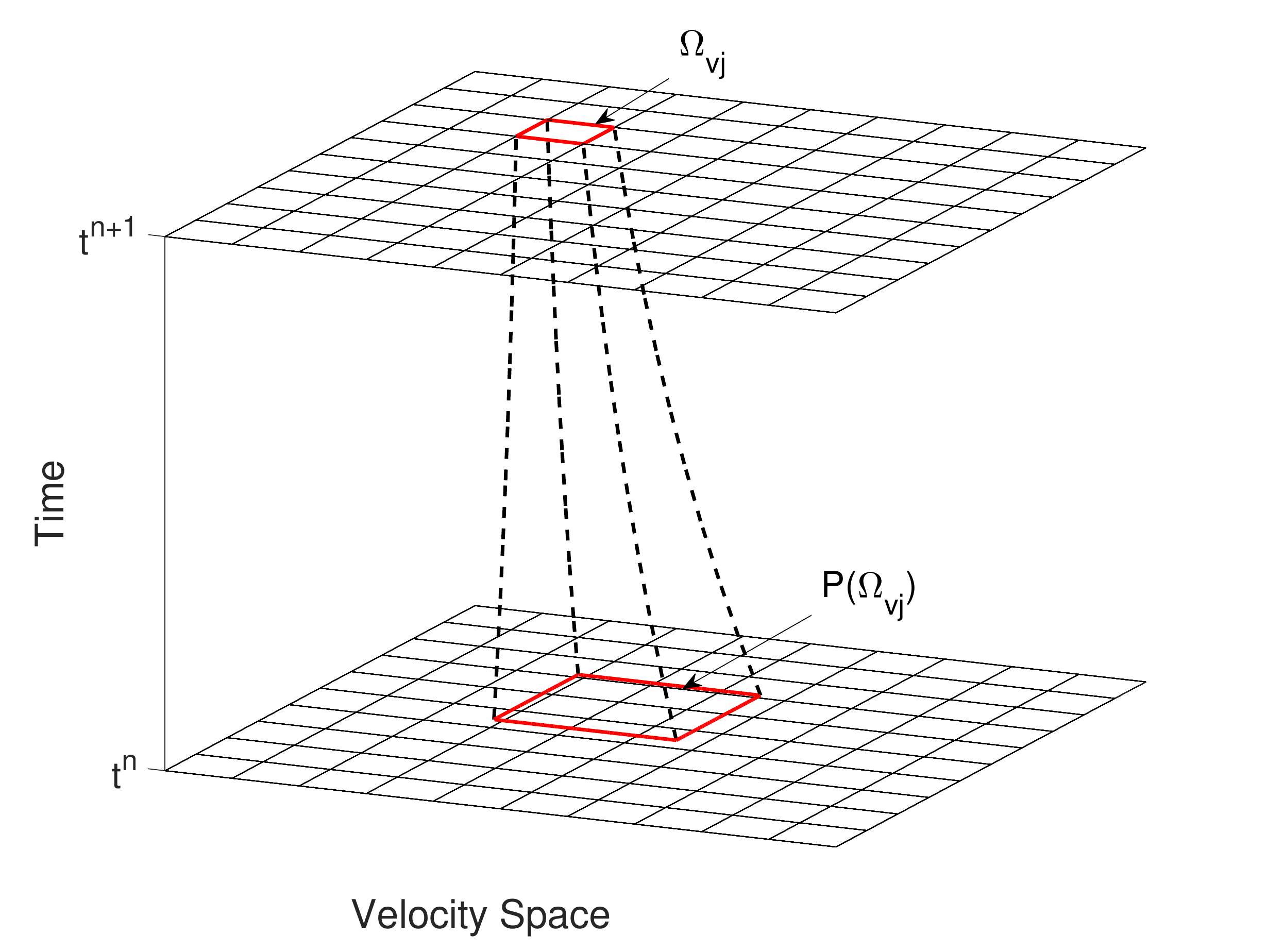}
\caption{Solid lines show the velocity space $\Omega_{v_j}$, and the dashed lines show the mapped velocity space $P_{\omega_2}(\Omega_{v_j})$ from $t^{n+1}$ onto $t^n$.}
\label{vspacemapping}
\end{figure}

The multiscale numerical fluxes in Eq.\eqref{ugks-f1} and Eq.\eqref{ugks-w1} are calculated from the integral solution of Eq.\eqref{kinetic-s1},
which is critical for the multiscale property of UGKS.
Let $t^n=0$, the normal direction of the physical cell interface $\vec{x}_0$ is $\vec{e}_1$, and the local basis of the physical cell interface $\vec{x}_0$ is $(\vec{e}_1,\vec{e}_2,\vec{e}_3)$. The integral solution $f_s(\vec{x}_0,t,\vec{v}_j)$ can be written as
\begin{equation}\label{integral-s}
   f_s(\vec{x}_0,t,\vec{v}_j)=\frac{1}{\tau_s}\int_{0}^t g_s(\vec{x}^\prime,t^\prime,\vec{v}^\prime)\mathrm{e}^{-(t-t^\prime)/\tau}dt^\prime+
   \mathrm{e}^{-t/\tau_s}f_{s,0}(\vec{x}_0-\vec{u}t,\vec{v}_j-\vec{\omega}_1t),
\end{equation}
where $x'=\vec{x}_0-\vec{v}_j(t-t')$ and $\vec{v}'=\vec{v}_j-\vec{\omega}_1(t-t')$ are the characteristics, and $f_0$ is the initial distribution function at time $t^n$.
The initial distribution function is reconstructed as
\begin{equation}\label{ugks-f0}
\begin{aligned}
  f_{s,0}(\vec{x},\vec{v})&=\left(f^l_{s,0}\vec{x}_0+\Delta \vec{x}\cdot\nabla_{\vec{x}} f^l_s+
  \Delta \vec{v}\cdot \nabla_{\vec{v}} f^l_s\right)\left(1-H[\Delta \vec{x}\cdot\vec{e}_1]\right)\\
  &-\left(f^r_{s,0}\vec{x}_0+\Delta \vec{x}\cdot\nabla_{\vec{x}} f^r_s+\Delta \vec{v}\cdot \nabla_{\vec{v}} f^r_s\right)\left(H[\Delta \vec{x}\cdot\vec{e}_1]\right),
\end{aligned}
\end{equation}
where $\Delta \vec{x}=\vec{x}-\vec{x}_0$, $\Delta \vec{v}=\vec{v}-\vec{v}_j$, and $H[x]$ is the Heaviside function
\begin{equation}
  H[x]=\left\{
  \begin{aligned}
    1 \quad   &x>0,\\
    0 \quad  &x\leq0.
  \end{aligned}
  \right.
\end{equation}
Slope limiter, such as the van-Leer limiter \cite{leveque2002finite}, is used to reconstruct the slopes of $\nabla_{\vec{x}} f$ and
$\nabla_{\vec{v}} f$ in each control volume of phase space.
The equilibrium Maxwellian distribution function is expanded in phase space and time as
\begin{equation}\label{ugks-g}
\begin{aligned}
  g_s(\vec{x},t,\vec{v})&=g_s(\vec{x}_0,0,\vec{v}_j)\bigg\{1+[a_{sx}^l(1-H[\Delta x\cdot\vec{e}_1])+
  a_{sx}^rH[\Delta x\cdot\vec{e}_r](x-x_0)\\
  &+a_{sy}(y-y_0)+a_{sz}(z-z_0)
  -2\lambda(\vec{U}_s-\vec{v}_j)(\vec{v}-\vec{v}_j)+a_{st}t\bigg\},
\end{aligned}
\end{equation}
where $x=\vec{x}\cdot\vec{e}_1$, $y=\vec{x}\cdot\vec{e}_2$, $z=\vec{x}\cdot\vec{e}_3$,
The initial Maxwellian distribution $g_s(\vec{x}_0,0,\vec{v}_j)$ at cell interface is fully determined by the macroscopic quantities at $(\vec{x}_0,0)$, and the required macroscopic variables are obtained by taking moments of the reconstructed distribution function $f_{s,0}(\vec{x}_0,\vec{v})$,
\begin{equation}\label{evaluate-g}
\vec{W}_0(\vec{x}_0)=\int_{\vec{v}\cdot\vec{e}_1>0} \vec{\psi}' f_{s,0}^l  d\Xi+
\int_{\vec{v}\cdot\vec{e}_1<0} \vec{\psi}' f_{s,0}^r  d\Xi,
\end{equation}
where $\vec{\psi}'=\left(1,\vec{v},\frac12\vec{v}^2+\frac{r^2-1}{2}(\vec{v}-\vec{U}_s)^2\right)$.
The derivative coefficients $a_{sx}^{l,r}$, $a_{sy}$, $a_{sz}$, $a_{st}$ are related to spatial and time derivatives of $g_s(\vec{x},t,\vec{v})$.
All derivative coefficients are functions of particle velocities in the form of $a=a_i\psi_i$, where $\vec{\psi}=(1,\vec{v},\frac12\vec{v}^2)$ is the collisional invariants, for example
\begin{align}
  \left(\frac{\partial g_s}{\partial x}\right)_{\vec{x}_0,0}^{l,r}&=\left(a_{sx,1}^{l,r}+a_{sx,2}^{l,r}u+a_{sx,3}^{l,r}v
  +a_{sx,4}^{l,r}w+\frac12a_{sx,5}^{l,r}\vec{v}^2\right)g_s(\vec{x}_0,0,\vec{v}),\\
  \left(\frac{\partial g_s}{\partial t}\right)_{\vec{x}_0,0}&=\left(a_{st,1}+a_{st,2}u+a_{st,3}v
  +a_{st,4}w+\frac12a_{st,5}\vec{v}^2\right)g_s(\vec{x}_0,0,\vec{v}),
\end{align}
with $u=\vec{v}\cdot{e}_1$, $v=\vec{v}\cdot{e}_2$, $w=\vec{v}\cdot{e}_3$.
The spatial derivative functions of Maxwellian distribution function are calculated from the linear system,
\begin{equation}
<a_{sx}^{l,r}>=\left(\frac{\partial \vec{W}^g}{\partial x}\right)^{l,r},\\
<a_{sy}>=\left(\frac{\partial \vec{W}^g}{\partial y}\right),\\
<a_{sz}>=\left(\frac{\partial \vec{W}^g}{\partial z}\right),
\end{equation}
where $<...>$ is the notation for the moments of Maxwellian distribution defined by
\begin{equation}
<...>=\int (...) g\psi d\Xi.
\end{equation}
Taking $a_{sx}^{l,r}$ as an example,
\begin{align}\label{evaluate-a}
  &a_{sx,5}^{l,r}=\frac{\rho_s}{3p^2_s}\left[2\left(\frac{\partial \rho_s E_s}{\partial x}\right)^{l,r}
  +\left(\vec{U}^2-\frac{3p_s}{\rho_s}\right)\left(\frac{\partial \rho_s}{\partial x}\right)^{l,r}
  -\sum_{i=1}^32U_{s,i}\left(\frac{\partial \rho_s U_{s,i}}{\partial x}\right)^{l,r}\right],\\
  &a_{sx,i+1}^{l,r}=\frac{1}{p}\left[\left(\frac{\partial \rho_{s}U_{s,i}}{\partial x}\right)^{l,r}
  -U_{s,i}\left(\frac{\partial \rho_s}{\partial x}\right)^{l,r}\right]-U_{s,i}a_{sx,5}^{l,r} \quad (i=1,2,3),\\
  &a_{sx,1}^{l,r}=\frac{1}{\rho_s}\left(\frac{\partial \rho}{\partial x}\right)^{l,r}
  -U_{s,i}a_{sx,i+1}^{l,r}-\frac12\left(\vec{U}^2+\frac{3p_s}{\rho_s}\right)a_{sx,5}^{l,r}.
\end{align}
The time derivatives of Maxwellian satisfy the compatibility condition
\begin{equation}\footnotesize
<(a_{sx}^l (1-H[\vec{v}\cdot\vec{e}_1])+a_{sx}^r H[\vec{v}\cdot\vec{e}_1])u+a_{sy}v+a_{sz}w-2\lambda(\vec{v}-\vec{U}_s)\cdot\vec{\omega}_1+a_{st}>
=(0,0,0,0,\frac{r^2-1}{2\tau_s}{\epsilon_s}^{n}\rho_se^n_s),
\end{equation}
from which the moments $<a_{st}>$ can be obtained, and hence $a_{st}$ can be calculated in a similar manner to the spatial derivative functions \cite{xu2001,ugks2010}.
Substituting Eq.\eqref{ugks-f0} and Eq.\eqref{ugks-g} into Eq.\eqref{integral-s}, the integral solution is
\begin{equation}\label{integral-s1}
\begin{aligned}
  f_s(\vec{x}_0,t,\vec{v}_j)&=\gamma_1 g_{s,0}(\vec{x}_0,\vec{v}_j)\\
  &+\gamma_2((a_{sx}^l H[\vec{v}\cdot\vec{e}_1]
  +a_{sx}^r (1-H[\vec{v}\cdot\vec{e}_1])) u+a_{sy}v+a_{sz}w)g_{s,0}(\vec{x}_0,\vec{v}_j)\\
  &-\gamma_22\lambda(\vec{v}-\vec{U}_s)\omega_{s1} g_{s,0}(\vec{x},\vec{v}_j)
  +\gamma_3 a_{st}g_{s,0}(\vec{x}_0,\vec{v}_j)\\
  &+\gamma_4(f_{s0}^l(\vec{x}_0,\vec{v}_j)H[\vec{v}\cdot\vec{e}_1]
  +f_{s0}^r(\vec{x}_0,\vec{v}_j)(1-H[\vec{v}\cdot\vec{e}_1]))\\
  &+\gamma_5(\vec{v}\cdot\nabla_{\vec{x}}f_{s0}^l+\vec{\omega}_1\cdot\nabla_{\vec{v}}f_{s0}^l)H[\vec{v}\cdot\vec{e}_1]\\
  &+\gamma_5(\vec{v}\cdot\nabla_{\vec{x}}f_{s0}^r+\vec{\omega}_1\cdot\nabla_{\vec{v}}f_{s0}^r)(1-H[\vec{v}\cdot\vec{e}_1]),
\end{aligned}
\end{equation}
where
\begin{equation}
\begin{aligned}
  &\gamma_1=(1-\exp(-t/\tau_s)),  \quad \gamma_2=(t+\tau_s)\exp(-t/\tau_s)-\tau_s,\\
  &\gamma_3=(t+\tau_s(\exp(-t/\tau_s)-1)),\quad
  \gamma_4=\exp(-t/\tau_s),  \quad \gamma_5=-t\exp(-t/\tau_s).
\end{aligned}
\end{equation}
Similarly, the integral solution at velocity cell interface in the velocity space is constructed as
\begin{equation}\label{integral-s2}
\begin{aligned}
  f_s(\vec{x}_i,t,\vec{v}_0)&=\gamma_1 g_{s,0}(\vec{x}_i,\vec{v}_0)+
  \gamma_2((\tilde{a}_{sx}u+\tilde{a}_{sy}v+\tilde{a}_{sz}w)g_{s,0}(\vec{x}_i,\vec{v}_0)\\
  &-\gamma_22\lambda(\vec{v}_0-\vec{U}_s)\vec{\omega}_{s1} g_{s,0}(\vec{x}_i,\vec{v}_0)
  +\gamma_3 \tilde{a}_{st}g_{s,0}(\vec{x}_i,\vec{v}_0)\\
  &+\gamma_4(f_{s0}^l(\vec{x}_i,\vec{v}_0)H[\vec{\omega}_1\cdot\vec{e}_{v1}]
  +f_{s0}^r(\vec{x}_i,\vec{v}_0)(1-H[\vec{\omega}_1\cdot\vec{e}_{v1}]))\\
  &+\gamma_5(\vec{v}\cdot\nabla_{\vec{x}}f_{s0}+\omega_1\cdot\nabla_{\vec{v}}f_{s0}^l)H[\vec{\omega}_1\cdot\vec{e}_{v1}]\\
  &+\gamma_5(\vec{v}\cdot\nabla_{\vec{x}}f_{s0}+\omega_1\cdot\nabla_{\vec{v}}f_{s0}^r)(1-H[\vec{\omega}_1\cdot\vec{e}_{v1}]).
\end{aligned}
\end{equation}
Analogous to velocity distribution function of solid apparent density, the integral solution of the velocity distribution of solid material temperature $f_{sT}(\vec{x}_0,t,\vec{v})$ can be obtained. The distribution of solid material temperature is related to the distribution of apparent solid density by
\begin{equation}
f_{sT}(\vec{x},t,\vec{v})=C_s T_M(\vec{x},t)f_{s}(\vec{x},t,\vec{v}).
\end{equation}
The numerical flux terms in the evolution equations Eq.\eqref{ugks-w1}, Eq.\eqref{ugks-T}, and Eq.\eqref{ugks-f1}, can be calculated from the cell interface integral solutions. The flux of distribution function and conservative variables from $t^n$ to $t^{n+1}$ are
\begin{align}
  &\int_{t^n}^{t^{n+1}}\int  \oint_{\partial \Omega_i} \vec{\psi}f_{s,\partial \Omega_i}(t,\vec{v}_j) \vec{v}_j\cdot d\vec{s} d\Xi dt=
  \sum_{i=1}^{N} s_i \int_{t^n}^{t^{n+1}} \int\vec{\psi}\vec{v}\cdot \vec{n}_i f_s(\vec{x}_0,t,\vec{v}_j)dv dt,\\
  &\int_{t^n}^{t^{n+1}}\int  \oint_{\partial \Omega_i} f_{sT,\partial \Omega_i}(t,\vec{v}_j) \vec{v}_j\cdot d\vec{s} d\Xi dt=
  \sum_{i=1}^{N} s_i \int_{t^n}^{t^{n+1}} \int\vec{v}\cdot \vec{n}_i f_{s,T}(\vec{x}_0,t,\vec{v}_j)dv dt,\\
  &\int_{t^n}^{t^{n+1}}\oint_{\partial \Omega_i} f_{s,\partial \Omega_i}(t,\vec{v}_j) \vec{v}_j\cdot d\vec{s}dt=
  \sum_{i=1}^{N} s_i \int_{t^n}^{t^{n+1}} \vec{v}_j\cdot \vec{n}_i f_s(\vec{x}_0,t,\vec{v}_j)dt,\\
  &\int_{t^n}^{t^{n+1}}\oint_{\partial \Omega_j} f_{s,\partial \Omega_j}(\vec{x}_i,t) \vec{\omega}_1\cdot d\vec{s}dt=
  \sum_{j=1}^{M} s_j \int_{t^n}^{t^{n+1}} \vec{\omega}_1 \cdot \vec{n}_j f_s(\vec{x}_i,t,\vec{v}_0)dt,
\end{align}
where the velocity integration in the macroscopic flux is calculated by numerical quadrature such as Newton-cotes formula or Gauss-Hermite quadrature.
{\subsubsection{Interspecies momentum exchange and velocity mapping method}}
The macroscopic momentum exchange between gas and solid phase is predicted by the ODE system
\begin{equation}\label{momentum-ode}
  \left\{
  \begin{aligned}
    &\frac{d \epsilon_{s,i}^{n+1}\rho_s\vec{U}_{s,i}}{dt}=\epsilon_{s,i}^{n+1}\rho_s\frac{\vec{U}_{g,i}-\vec{U}_{s,i}}{\tau_{st}},\\
    &\frac{d \epsilon_{g,i}^{n+1}\rho_{g,i}^{n+1}\vec{U}_{g,i}}{dt}=-\epsilon_{s,i}^{n+1}\rho_s\frac{\vec{U}_{g,i}-\vec{U}_{s,i}}{\tau_{st}},
  \end{aligned}
  \right.
\end{equation}
from which the macroscopic velocities of gas and solid phase can be obtained at $t^{n+1}$ as given in $\mathcal{L}_{sw2}$.
The evolution of velocity distribution of solid phase follows Eq.\eqref{ugks-f2}, which solves $\mathcal{L}_{s2}$ by the following velocity space mapping method.
{
Following the characteristics of Eq.\eqref{kinetic-s2}
\begin{equation}
\frac{d \vec{v}}{dt}=\frac{\vec{U}^{n+1}_g-\vec{v}}{\tau_{st}},
\end{equation}
the velocity space at $t^{n+1}$ can be mapped onto the velocity space at $t^n$ by
\begin{equation}\label{mapping}
P_{\omega_2}(\vec{v}_j)=\vec{U}_g^{n+1}+e^{\Delta t/\tau_{st}}(\vec{v}_j-\vec{U}^{n+1}_g),
\end{equation}
as shown in Fig. \ref{vspacemapping}.
The velocity distribution function $f^{n+1}_{s,ij}$ can be updated by Eq.\eqref{ugks-f2}.
For a structured rectangular velocity space, the evolution of velocity distribution follows
\begin{equation}
  f_{s,ikl}^{**}=\frac{1}{(u_{k+\frac12}-u_{k-\frac12})(v_{l+\frac12-v_{l-\frac12}})}
  \int_{u'_{k-\frac12}}^{u'_{k+\frac12}}\int_{v'_{l-\frac12}}^{v'_{l+\frac12}} f^{*r}_{s,i}(u,v) du dv,
\end{equation}
where
\begin{equation}
\left\{
 \begin{aligned}
  &u'_{k\pm\frac12}=U_g^{n+1}+e^{\Delta t/\tau_{st}}(u_{k\pm\frac12}-U_g^{n+1}),\\
  &v'_{l\pm\frac12}=U_g^{n+1}+e^{\Delta t/\tau_{st}}(u_{l\pm\frac12}-U_g^{n+1}),
 \end{aligned}
 \right.
\end{equation}
and $f^{*r}_{s,i}(u,v)$ is the reconstructed velocity distribution function in velocity space.
From the updated velocity distribution function, the total energy of solid phase can be updated by Eq.\eqref{ugks-w3}.}

\subsubsection{Particle collision and interspecies heat conduction}
The kinetic energy loss $\mathcal{L}_{sw4}$ due to the inelastic collision is calculated from
\begin{equation}
\frac{e^{n+1}_{s}-e^{*}_s}{\Delta t}=\frac{\alpha^2-1}{2\tau_{st}} e^{n+1}_s,
\end{equation}
which is given in Eq.\eqref{ugks-w4}.
From the updated macroscopic variables $\vec{W}^{n+1}_{s,i}$, the corresponding equilibrium Maxwellian distribution $g_{s,ij}^{n+1}$ can be constructed, and the velocity distribution can be updated to $t^{n+1}$ by Eq.\eqref{ugks-f3}.
The temperature conduction between solid and gas phase is modeled by Eq.\eqref{temp-s} and Eq.\eqref{temp-g}.
By solving the heat conduction ODE system
\begin{equation}
  \left\{
  \begin{aligned}
    &\frac{d \epsilon_{s,i}^{n+1}\rho_sC_s T_{s,i}^M}{dt}=\epsilon_{s,i}^{n+1}\rho_sC_s\frac{T_{g,i}-T_{s,i}^M}{\tau_{T}},\\
    &\frac{d \epsilon_{g,i}^{n+1}\rho_{g,i}^{n+1}C_v T_{g,i}}{dt}=-\epsilon_{s,i}^{n+1}\rho_sC_s\frac{T_{g,i}-T_{s,i}^M}{\tau_{T}},
  \end{aligned}
  \right.
\end{equation}
with initial condition
$$T^{M}_{s,i}(t^n)=T^{M,*}_{s,i}+\frac{\Delta t(1-r^2)e^{**}_{s,i}}{[2\tau_{s,i}+\Delta t(1-r^2)]C_{s,i}},
\quad T_{g,i}(t^n)=T^*_{g,i}+\frac{{\epsilon_s}^{n+1}_i\rho_{s}(E^{**}_{s,i}-E^{*}_{s,i})}{{\epsilon_g}^{n+1}_i\rho^{n+1}_{g,i}C_{v,i}},$$
the material temperature of solid phase is evolved based on Eq.\eqref{ugks-w5}.
In summary Eq.\eqref{ugks-f1}-\eqref{ugks-w5} compose the UGKS scheme for solid particle phase.

\subsection{Gas kinetic scheme for gas phase}
\subsubsection{General framework}
The numerical scheme for the gas phase is also built on a finite volume framework, using the same physical space division as the solid phase, namely $\mathbf{X}=\sum_i\Omega_{x_i}$.
Since the gas phase is in continuum regime governed by the Navier-Stokes equations \eqref{ns-g}, the GKS is utilized to evolve the cell averaged macroscopic variables,
\begin{align}\nonumber
\vec{W}_{g,i}&=\frac{1}{\Omega_i}\int \vec{W_g}(\vec{x})d\vec{x},
\end{align}
where $\vec{W}_g=(\epsilon_g \rho_g,\epsilon_g \rho_g\vec{U}_g,\epsilon_g \rho_gE_g)^T.$ The gas phase kinetic equation is split into
\begin{align}
&\mathcal{L}_{g1}:\frac{\partial \epsilon_g f_g}{\partial t}+\nabla_{\vec{x}}\cdot(\vec{v}f_g)+\nabla_{\vec{v}}\cdot(\vec{g} \epsilon_g f_g)=\frac{\epsilon_g g_g-\epsilon_g f_g}{\tau_g},\label{kinetic-g1}\\
&\mathcal{L}_{g1}:\frac{\partial \epsilon_g f_g}{\partial t}
  +\nabla_{\vec{v}}\cdot\left\{f_s\left[\frac{1}{\tau_{st}}(\vec{U}_g-\vec{v})-\frac{1}{\rho_s}\nabla p_g\right]\right\}=0.\label{kinetic-g2}
\end{align}
The numerical evolution equations for gas phase macroscopic variables are
{\footnotesize
\begin{align}
&\mathcal{L}_{gw1}:\quad\vec{W}^{n+1}_{g,i}= \vec{W}^{n}_{g,i}+\frac{1}{|\Omega_i|}\int_{t^n}^{t^{n+1}}\int
\oint_{\partial \Omega_i}\vec{\psi} f_{g,\partial \omega_i}(\vec{v},t) \vec{u}\cdot d\vec{s} d\Xi dt
+\Delta t\vec{S}_{g,i}^n-\Delta t\vec{S}_{s,i}^n,\label{gks-w1}\\
&\mathcal{L}_{gw2}:\quad\vec{U}^{n+1}_{g,i}=\gamma_{ug1}\vec{U}^n_{s,i}+\gamma_{ug2}\vec{U}^n_{g,i},\label{gks-w2}\\
&\mathcal{L}_{gw3}:\quad\vec{T}^{n+1}_{g,i}=\gamma_{Tg3}\left(\vec{T}^{M,*}_{s,i}+\frac{\Delta t(1-r^2)e^{**}_{s,i}}{\tau_s+\Delta t(1-r^2)}\rho_{s}C_{s,i}\right)+\gamma_{Tg4}\left(\vec{T}^*_{g,i}+\frac{\epsilon_s^{n+1}\rho_{s}(E^{**}_{s,i}-E^{*}_{s,i})}{{\epsilon_g}^{n+1}_i\rho^{n+1}_{g,i}C_v}\right) , \label{gks-w3}
\end{align}
}%
where $\vec{S}_{g,i}^n=(0,\epsilon_{gi}^{n}\rho^n_{g,i} \vec{g},\epsilon_{gi}^{n}\rho^n_{g,i}\vec{U}_{g,i}\cdot\vec{g})$,
and the coefficients
$$\gamma_{ug1}=\frac{\theta_i^{n+1}\rho_{s}-\theta_i^{n+1}\rho_{s}\exp\left[-\left(1+\frac{\theta_i^{n+1} \rho_{s}}{\epsilon_g^{n+1} \rho^{n+1}_{g,i}}\right)\frac{\Delta t}{\tau_{st,i}}\right]}{\epsilon_g^{n+1} \rho^{n+1}_{g,i}+\theta_i^{n+1} \rho_{s}},$$
$$\gamma_{ug2}=\frac{\theta_i^{n+1}\rho_{s}+\epsilon_g^{n+1}\rho^{n+1}_{g,i}\exp\left[-\left(1+\frac{\theta_i^{n+1} \rho_{s}}{\epsilon_g^{n+1} \rho^{n+1}_{g,i}}\right)\frac{\Delta t}{\tau_{st,i}}\right]}{\epsilon_g^{n+1} \rho^{n+1}_{g,i}+\theta_i^{n+1} \rho_{s}},$$
$$\gamma_{Tg3}=\frac{\theta_i^{n+1}\rho_{s}C_{s,i}-\theta_i^{n+1}\rho_{s}C_s\exp\left[-\left(1+\frac{\theta_i^{n+1} \rho_{s}C_s}{\epsilon_g^{n+1} \rho^{n+1}_{g,i}C_v}\right)\frac{\Delta t}{\tau_{T,i}}\right]}{\epsilon_g^{n+1} \rho^{n+1}_{g,i}C_v+\theta_i^{n+1} \rho_{s}C_s},$$
$$\gamma_{Tg4}=\frac{\theta_i^{n+1}\rho_{s}C_{s,i}+\epsilon_g^{n+1}\rho^{n+1}_{g,i}C_v\exp\left[-\left(1+\frac{\theta_i^{n+1} \rho_{s}C_s}{\epsilon_g^{n+1} \rho^{n+1}_{g,i}C_v}\right)\frac{\Delta t}{\tau_{T,i}}\right]}{\epsilon_g^{n+1} \rho^{n+1}_{g,i}C_v+\theta_i^{n+1} \rho_{s}C_s}.$$

\subsubsection{Numerical flux}
\begin{figure}
\centering
\includegraphics[width=0.8\textwidth]{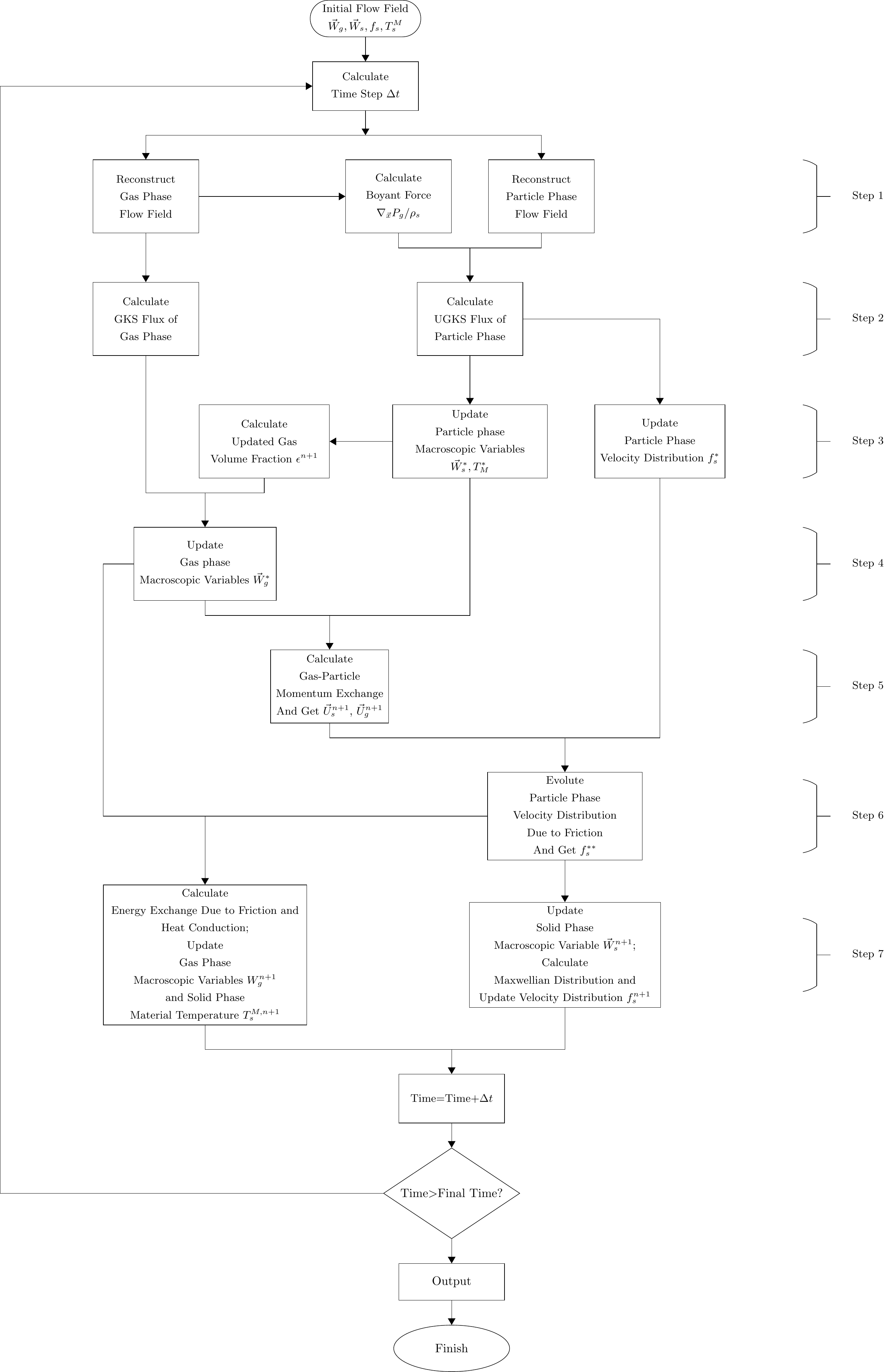}
\caption{Flow chart for UGKS-M.}
\label{chart}
\end{figure}

The numerical flux is calculated from the integral solution of Eq.\eqref{kinetic-g1}.
Assuming that the cell interface is located at $\vec{x_0}$ with normal direction $\vec{e}_1$ and local basis $(\vec{e}_1,\vec{e}_2,\vec{e}_3)$ and $t^n=0$, the integral solution is
\begin{equation}\label{integral-g}
   f_g(\vec{x}_0,t,\vec{v})=\frac{1}{\tau}\int_{0}^t g_g(\vec{x}^\prime,t^\prime,\vec{v}^\prime)\mathrm{e}^{-(t-t^\prime)/\tau}dt^\prime+
   \mathrm{e}^{-t/\tau}f_{g0}(\vec{x}_0-\vec{v}t,\vec{v}-\vec{g}t),
\end{equation}
where $\vec{x}'=\vec{x}_0-\vec{v}(t-t')$ and $\vec{v}'=\vec{v}_0-\vec{g}(t-t')$ are characteristics.
The initial distribution is expanded as
\begin{equation}\label{gks-f0}
\begin{aligned}
  f_{0,g}(\vec{x},\vec{v})&=g^l_{0,g}[1-\tau_g(\vec{a}_{gx}^l\cdot \vec{v}+a_{gt}-2\lambda(\vec{v}-\vec{U}_g))+\vec{a}^l_{gx}\cdot \Delta \vec{x}](1-H[\Delta \vec{x}\cdot \vec{e}_1])\\
  &+g^r_{0,g}[1-\tau_g(\vec{a}_{gx}^r\cdot \vec{v}+a_{gt}-2\lambda(\vec{v}-\vec{U}_g))+\vec{a}^r_{gx}\cdot \Delta \vec{x}]H[\Delta \vec{x}\cdot \vec{e}_1],
\end{aligned}
\end{equation}
where $\vec{a}_{gx}^{l,r}=(a_{gx}^{l,r},a_{gy}^{l,r},a_{gz}^{l,r})$ are the derivative coefficient functions.
And the equilibrium distribution function is expanded as
\begin{equation}\label{gks-g}
\begin{aligned}
  g_g(\vec{x},t,\vec{v})&=g_g(\vec{x}_0,0,\vec{v}_j)\bigg\{1+[a_{gx}^l(1-H[\Delta x\cdot\vec{e}_1])+
  a_{gx}^rH[\Delta x\cdot\vec{e}_r](x-x_0)\\
  &+a_{gy}(y-y_0)+a_{gz}(z-z_0)
  -2\lambda(\vec{U}_s-\vec{v}_j)(\vec{v}-\vec{v}_j)+a_{gt}t\bigg\}.
\end{aligned}
\end{equation}
The Maxwellian distribution function as well as the derivative coefficient functions can be evaluated in a similar way as the solid phase Eq.\eqref{evaluate-g}-\eqref{evaluate-a}.
Substituting Eq.\eqref{gks-f0} and Eq.\eqref{gks-g} into Eq.\eqref{integral-g}, the integral solution is expressed as
\begin{equation}
\begin{aligned}
  f_g(\vec{x}_0,t,\vec{v})&=\gamma_1 g_{g0}(\vec{x}_0,\vec{v})\\
  &+\gamma_2((a_{gx}^l H[\vec{v}\cdot\vec{e}_1]
  +a_{gx}^r (1-H[\vec{v}\cdot\vec{e}_1])) u+a_{gy}v+a_{gz}w)g_{g0}(\vec{x}_0,\vec{v})\\
  &-\gamma_22\lambda(\vec{v}-\vec{U}_s)\cdot \vec{g} g_{g0}(\vec{x}_0,\vec{v})
  +\gamma_3 a_{gt}g_{g0}(\vec{x}_0,\vec{v})\\
  &+\gamma_4(g_{g0}^l(\vec{x}_0,\vec{v})H[\vec{v}\cdot\vec{e}_1]
  +g_{g0}^r(\vec{x}_0,\vec{v})(1-H[\vec{v}\cdot\vec{e}_1]))\\
  &+\gamma_5g_{g0}^r(\vec{a}^r_{g}\cdot\vec{v}-2\lambda^r(\vec{v}-\vec{U}_g)\cdot\vec{g})H[\vec{v}\cdot\vec{e}_1]\\
  &+\gamma_5g_{g0}^l(\vec{a}^l_{g}\cdot\vec{v}-2\lambda^l(\vec{v}-\vec{U}_g)\cdot\vec{g})(1-H[\vec{v}\cdot\vec{e}_1])\\
  &+\gamma_6(a^r_{gt}g_{g0}^rH[\vec{v}\cdot\vec{e}_1]+a^l_{gt}g_{g0}^l(1-H[\vec{v}\cdot\vec{e}_1])),
\end{aligned}
\end{equation}
where
\begin{equation}
\begin{aligned}
  &\gamma_1=(1-\exp(-t/\tau_g)),  \quad \gamma_2=(t+\tau_g)\exp(-t/\tau_g)-\tau_s,\\
  & \gamma_3=(t+\tau_s(\exp(-t/\tau_g)-1)), \quad \gamma_4=\exp(-t/\tau_g),  \\
  & \gamma_5=-(t+\tau_g)\exp(-t/\tau_g),\quad \gamma_6=-\tau_g\exp(-t/\tau_g).
\end{aligned}
\end{equation}
The numerical flux terms in the evolution equation Eq.\eqref{gks-w1} can be calculated from the cell interface integral solutions. The flux of conservative variables from $t^n$ to $t^{n+1}$ is
\begin{align}
  \int_{t^n}^{t^{n+1}}\int  \oint_{\partial \Omega_i} \vec{\psi}f_{g,\partial \Omega_i}(t,\vec{v}_j) \vec{v}_j\cdot d\vec{s} d\Xi dt=
  \sum_{i=1}^{N} s_i \int_{t^n}^{t^{n+1}} \int\vec{\psi}\vec{v}\cdot \vec{n}_i f_g(\vec{x}_0,t,\vec{v})dv dt.
\end{align}
In summary, Eq.\eqref{gks-w1}-\eqref{gks-w3} compose of the GKS for gas phase, and the flow chart for UGKS-M is shown in Fig. \ref{chart}.

\subsection{Limiting solutions of UGKS-M}
The UGKS-M for multiphase flow simulations preserves the flow regime for a wide range of particle Knudsen number $Kn_s$ and particle Stokes number/normalized particle response time $\tau_{st}$.
In the rarefied regime with $Kn_s\gg1$, the integral solutions Eq.\eqref{integral-s1} and Eq.\eqref{integral-s2} become
\begin{equation}\label{integral-s3}
\begin{aligned}
  f_s(\vec{x}_0,t,\vec{v}_j)&=(f_{s0}^l(\vec{x}_0,\vec{v}_j)H[\vec{v}\cdot\vec{e}_1]
  +f_{s0}^r(\vec{x}_0,\vec{v}_j)(1-H[\vec{v}\cdot\vec{e}_1]))\\
  &+t(\vec{v}\cdot\nabla_{\vec{x}}f_{s0}^l+\vec{\omega}_1\cdot\nabla_{\vec{v}}f_{s0}^l)H[\vec{v}\cdot\vec{e}_1]\\
  &+t(\vec{v}\cdot\nabla_{\vec{x}}f_{s0}^r+\vec{\omega}_1\cdot\nabla_{\vec{v}}f_{s0}^r)(1-H[\vec{v}\cdot\vec{e}_1]),
\end{aligned}
\end{equation}
and
\begin{equation}\label{integral-s4}
\begin{aligned}
  f_s(\vec{x}_i,t,\vec{v}_0)&=f_{s0}^r(\vec{x}_i,\vec{v}_0)(1-H[\vec{\omega}_1\cdot\vec{e}_{v1}]))\\
  &+t(\vec{v}\cdot\nabla_{\vec{x}}f_{s0}+\vec{\omega}_1\cdot\nabla_{\vec{v}}f_{s0}^l)H[\vec{\omega}_1\cdot\vec{e}_{v1}]\\
  &+t(\vec{v}\cdot\nabla_{\vec{x}}f_{s0}+\vec{\omega}_1\cdot\nabla_{\vec{v}}f_{s0}^r)(1-H[\vec{\omega}_1\cdot\vec{e}_{v1}]).
\end{aligned}
\end{equation}
The solid particle collision equation Eq.\eqref{ugks-f3} degenerates to
\begin{equation}\label{ugks-f4}
f_{s,ij}^{n+1}=f^*_{s,ij}.
\end{equation}
The numerical governing equations for solid phase in rarefied regime consist of Eq.\eqref{ugks-f1},\eqref{ugks-f2},\eqref{ugks-f4}, \eqref{ugks-w1}-\eqref{ugks-w5} with the numerical flux calculated from  Eq.\eqref{integral-s3}-\eqref{integral-s4}, which converge to a consistent numerical scheme for the collisionless Boltzmann equation
\begin{align}
  \frac{\partial f_s}{\partial t}+\nabla_{\vec{x}}\cdot(\vec{v} f_s)+
  \nabla_{\vec{v}} \cdot \left(\vec{g}-\frac{1}{\rho_s} \nabla_{\vec{x}} p_g f_s\right)+
  \nabla_{\vec{v}} \cdot \left(\frac{\vec{D}}{m_s} f\right)=0.\label{Vlasov}
\end{align}
According to the analysis in \cite{liu2016}, the integral solution in the continuous regime with $Kn_s\ll1$ becomes
\begin{equation}\label{integral-s5}
\begin{aligned}
  f_s(\vec{x}_0,t,\vec{v}_j)=&g_{s0}(\vec{x}_0,\vec{v}_j)\\
  &-\tau_s((a_{sx}^l H[\vec{v}\cdot\vec{e}_1]
  +a_{sx}^r (1-H[\vec{v}\cdot\vec{e}_1])) u+a_{sy}v+a_{sz}w)g_{s0}(\vec{x}_0,\vec{v}_j)\\
  &+\tau_s\lambda(\vec{v}-\vec{U}_s)\omega_{s1} g_{s0}(\vec{x}_0,\vec{v}_j)
  +t a_{st}g_{s0}(\vec{x}_0,\vec{v}_j)\\
  =&f_{s}^{NS}(\vec{x}_0,0,\vec{v}_j)+t a_{st}g_{s0}(\vec{x}_0,\vec{v}_j),
\end{aligned}
\end{equation}
and
\begin{equation}\label{integral-s6}
\begin{aligned}
  f_s(\vec{x}_i,t,\vec{v}_0)=& g_{s0}(\vec{x}_i,\vec{v}_0)
  -\tau_s((\tilde{a}_{sx}u+\tilde{a}_{sy}v+\tilde{a}_{sz}w)g_{s0}(\vec{x}_i,\vec{v}_0)\\
  &+\tau_s\lambda(\vec{v}_0-\vec{U}_s)\omega_{s1} g_{s0}(\vec{x}_i,\vec{v}_0)
  +t \tilde{a}_{st}g_{s0}(\vec{x}_i,\vec{v}_0)\\
  =&f_{s}^{NS}(\vec{x}_i,0,\vec{v}_0)+t a_{st}g_{s0}(\vec{x}_i,\vec{v}_0).
\end{aligned}
\end{equation}
Therefore, UGKS-M provides a consistent NS flux in the continuum regime.
In the Euler regime with $Kn_s\to0$, Eq.\eqref{ugks-f3} converges to
\begin{equation}\label{ugks-f6}
  f^{n+1}_{s,ij}=g^{n+1}_{ij},
\end{equation}
which shows that the UGKS-M recovers the Euler equations in the Euler limiting regime.

In the granular flow regime with $\tau_{st}\to\infty$, Eq.\eqref{ugks-w2} and Eq.\eqref{ugks-f2} degenerate to
\begin{equation}
\begin{aligned}
\vec{U}^{n+1}_s&=\vec{U}^{*}_s,\\
f^{**}_{s,ij}&=f^{n}_{s,ij},
\end{aligned}
\end{equation}
and the gas phase and particle phase are decoupled.
In the other dusty gas limit with $\tau_{st}\to 0$, Eq.\eqref{ugks-w2} and Eq.\eqref{ugks-f2} converge to
\begin{equation}
\begin{aligned}
\vec{U}^{n+1}_s&=\vec{U}^{n+1}_g,\\
f^{**}_{s,ij}&=\delta(\vec{v}-\vec{U}^{n+1}_g),
\end{aligned}
\end{equation}
where $\delta(\vec{x})$ is the Dirac delta function
\begin{equation}
  \delta(\vec{x})=\left\{
  \begin{aligned}
    1 \quad   &\vec{x}=0,\\
    0 \quad  &\vec{x}\neq0.
  \end{aligned}
  \right.
\end{equation}
Therefore in the dusty gas regime, solid particle shares same speed with the gas flow.

\section{Numerical test cases}
\begin{figure}
\centering
\includegraphics[width=0.45\textwidth]{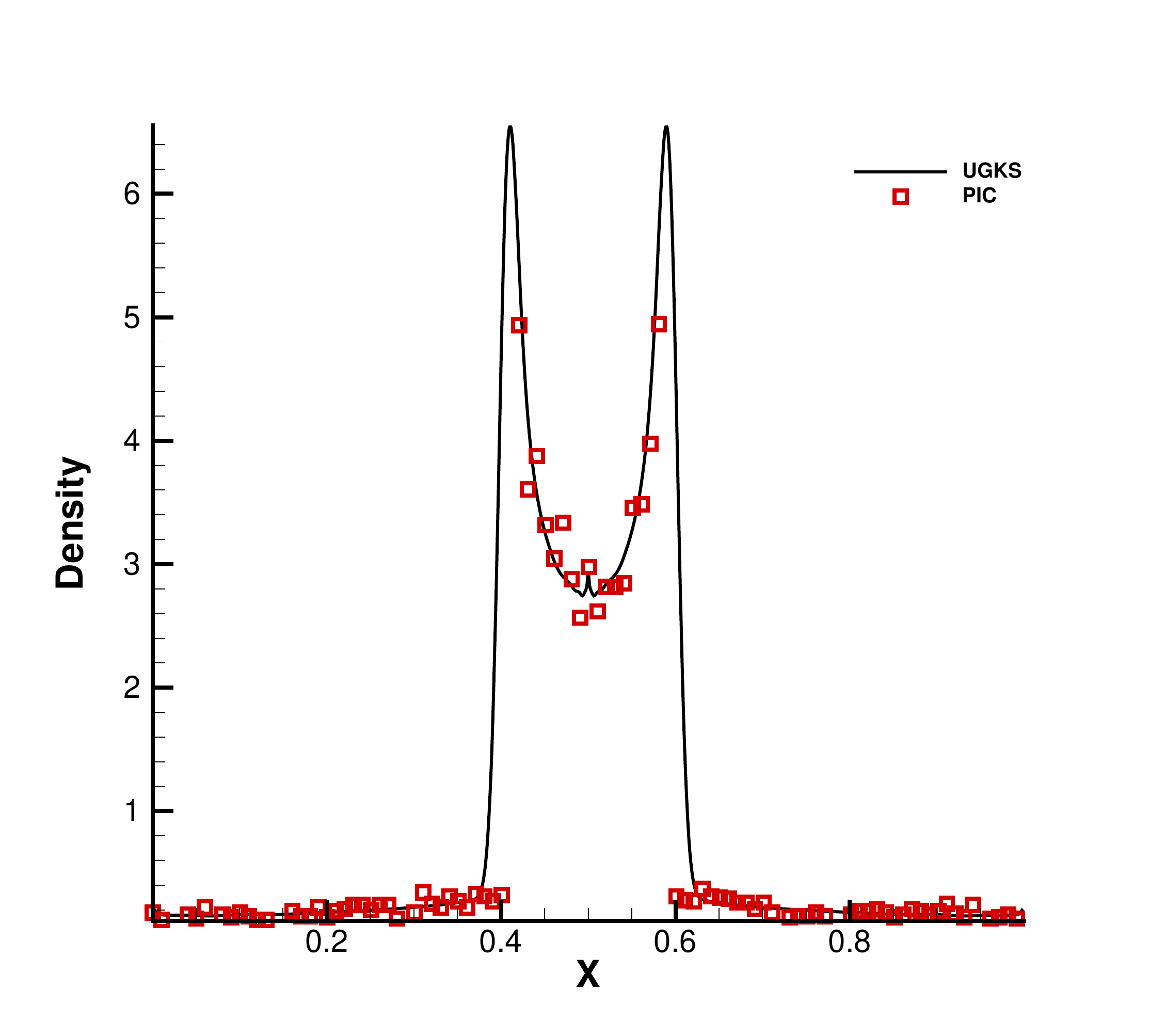}{a}
\includegraphics[width=0.45\textwidth]{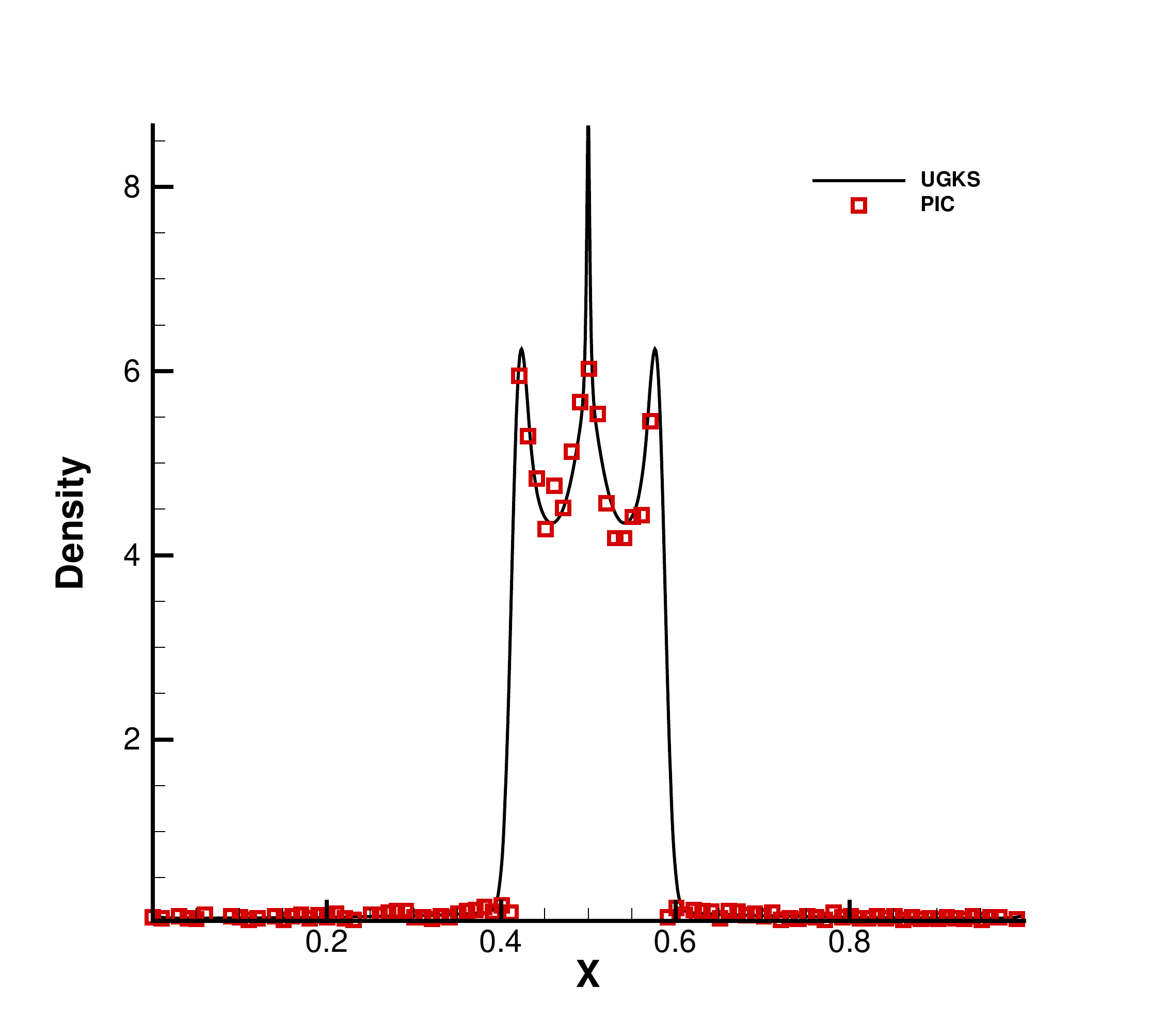}{b}\\
\includegraphics[width=0.45\textwidth]{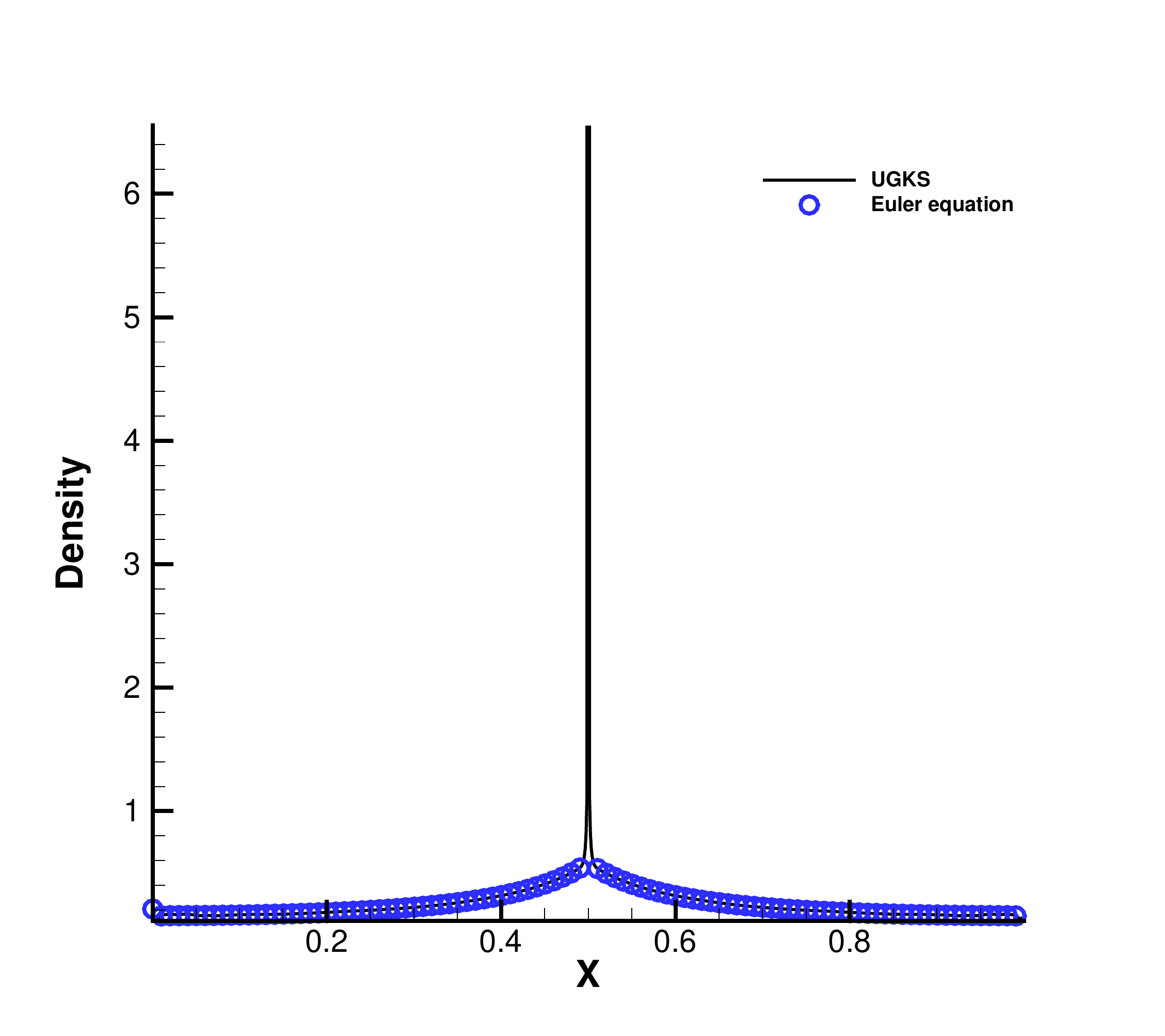}{c}
\includegraphics[width=0.45\textwidth]{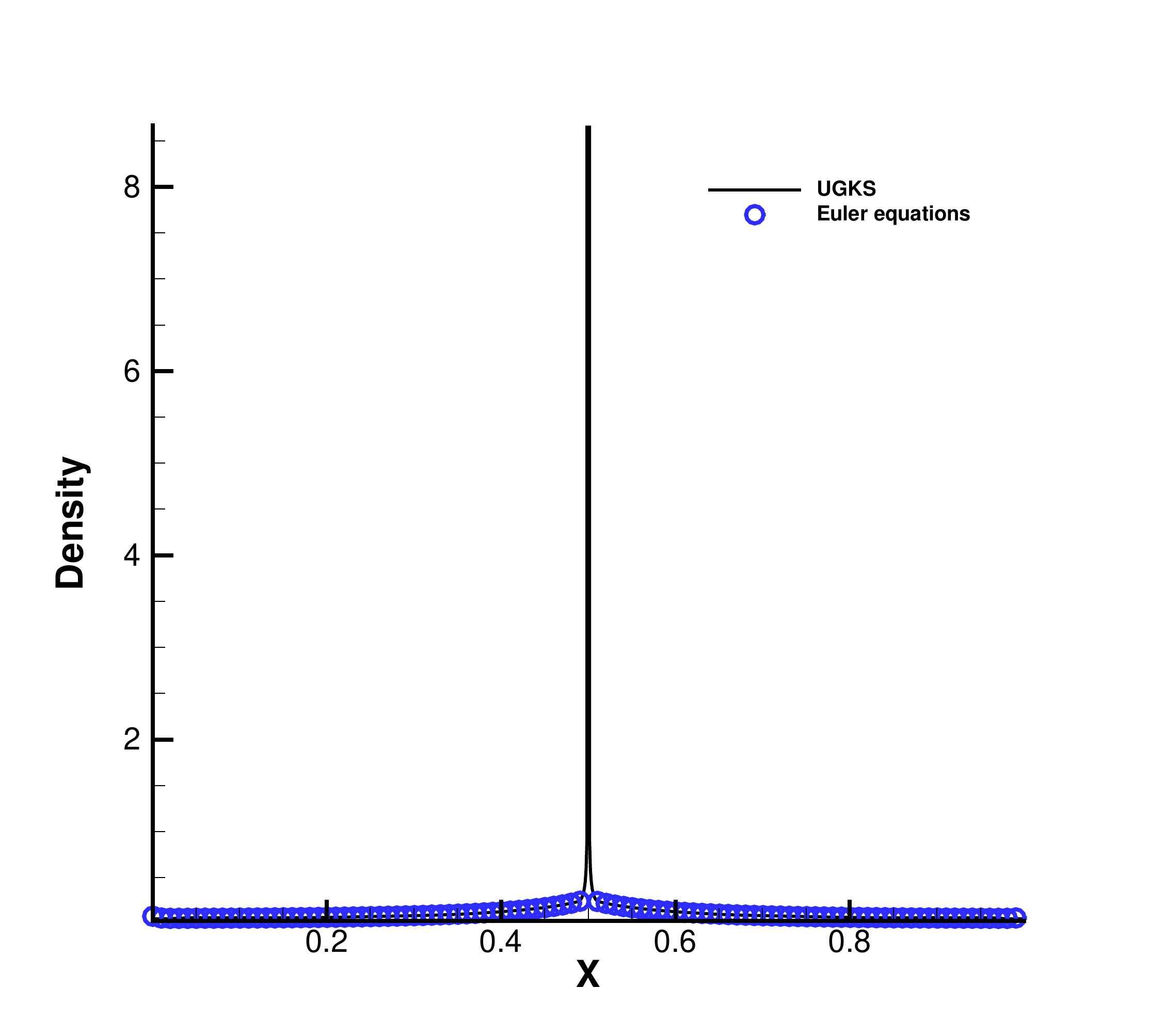}{d}
\caption{Apparent density distribution of particle phase for the one dimensional particle concentration test.
(a) the density distribution at $t=1$ with $Kn=10^{4}$ and $\tau_{st}=0.3$; solid line is the UGKS-M solution and square shows the PIC result.
(b) the density distribution at $t=1.5$ with $Kn=10^{4}$ and $\tau_{st}=0.3$; solid line is the UGKS-M solution and square shows the PIC result.
(c) the density distribution at $t=1$ with $Kn=10^{-4}$ and $\tau_{st}=0.3$; solid line is the UGKS-M solution and circle shows the Pressureless Euler solution.
(d) the density distribution at $t=1.5$ with $Kn=10^{-4}$ and $\tau_{st}=0.3$; solid line is the UGKS-M solution and circle shows the Pressureless Euler solution.}
\label{1taylor1}
\end{figure}

\begin{figure}
\centering
\includegraphics[width=0.45\textwidth]{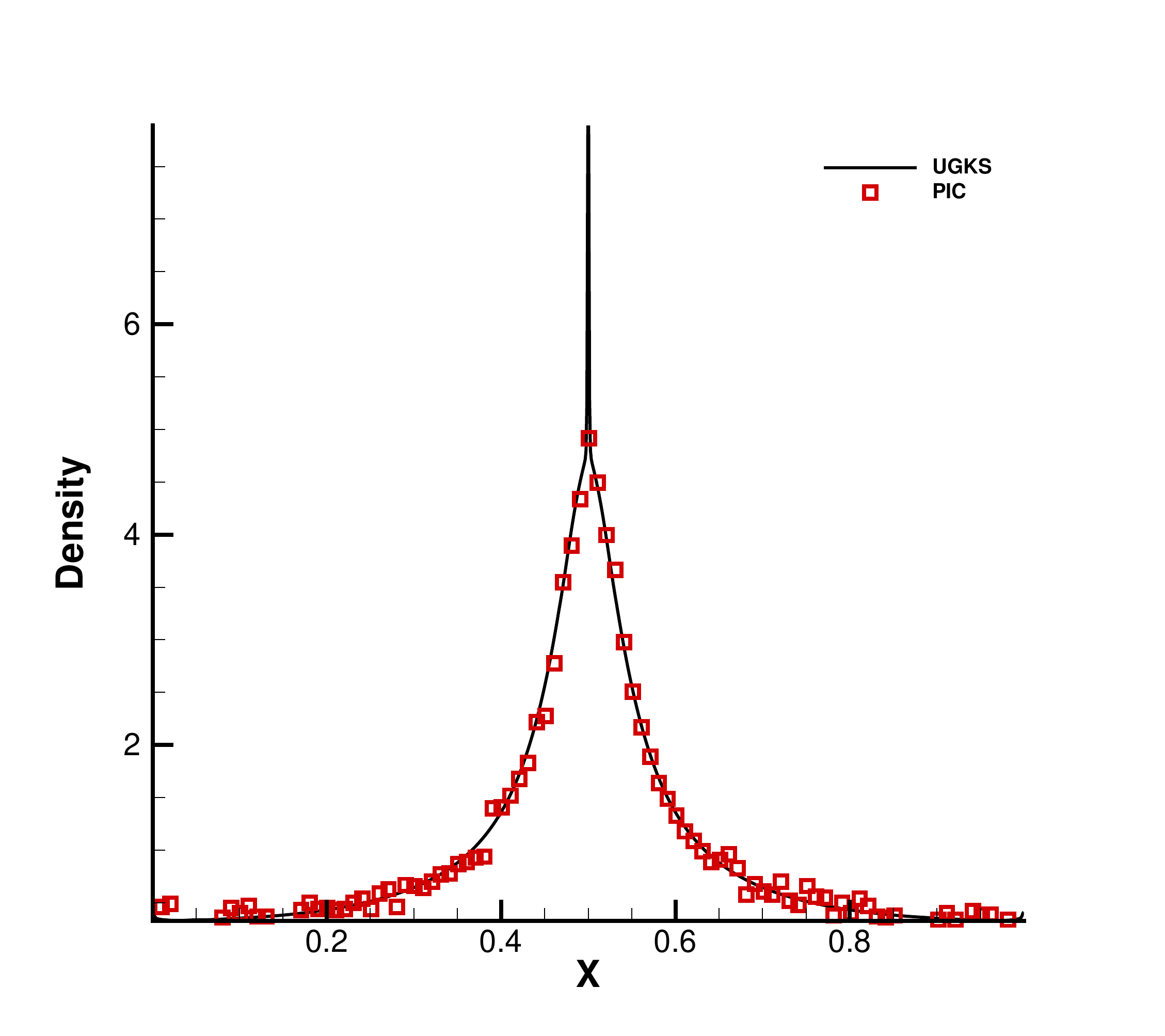}{a}
\includegraphics[width=0.45\textwidth]{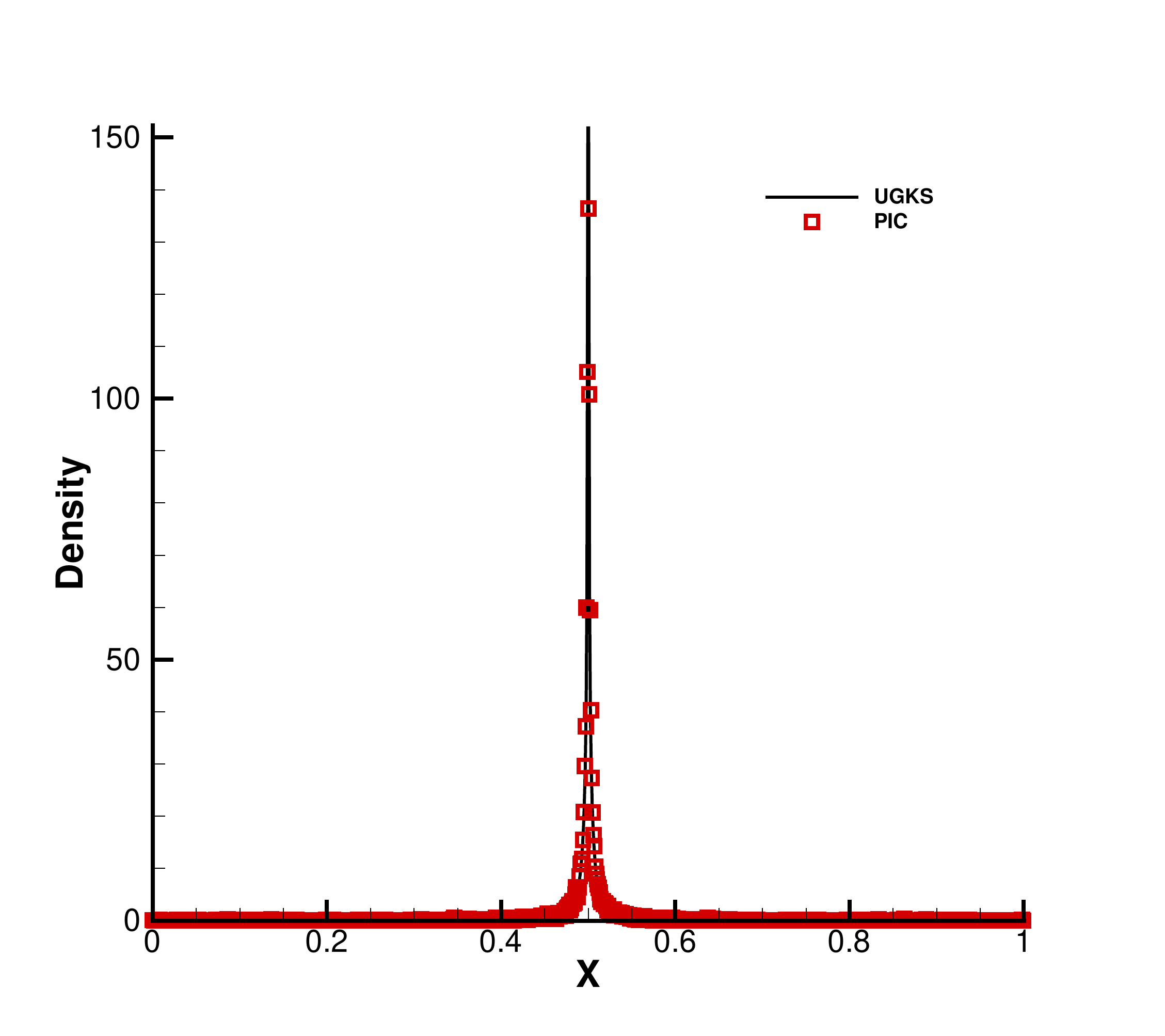}{b}\\
\includegraphics[width=0.45\textwidth]{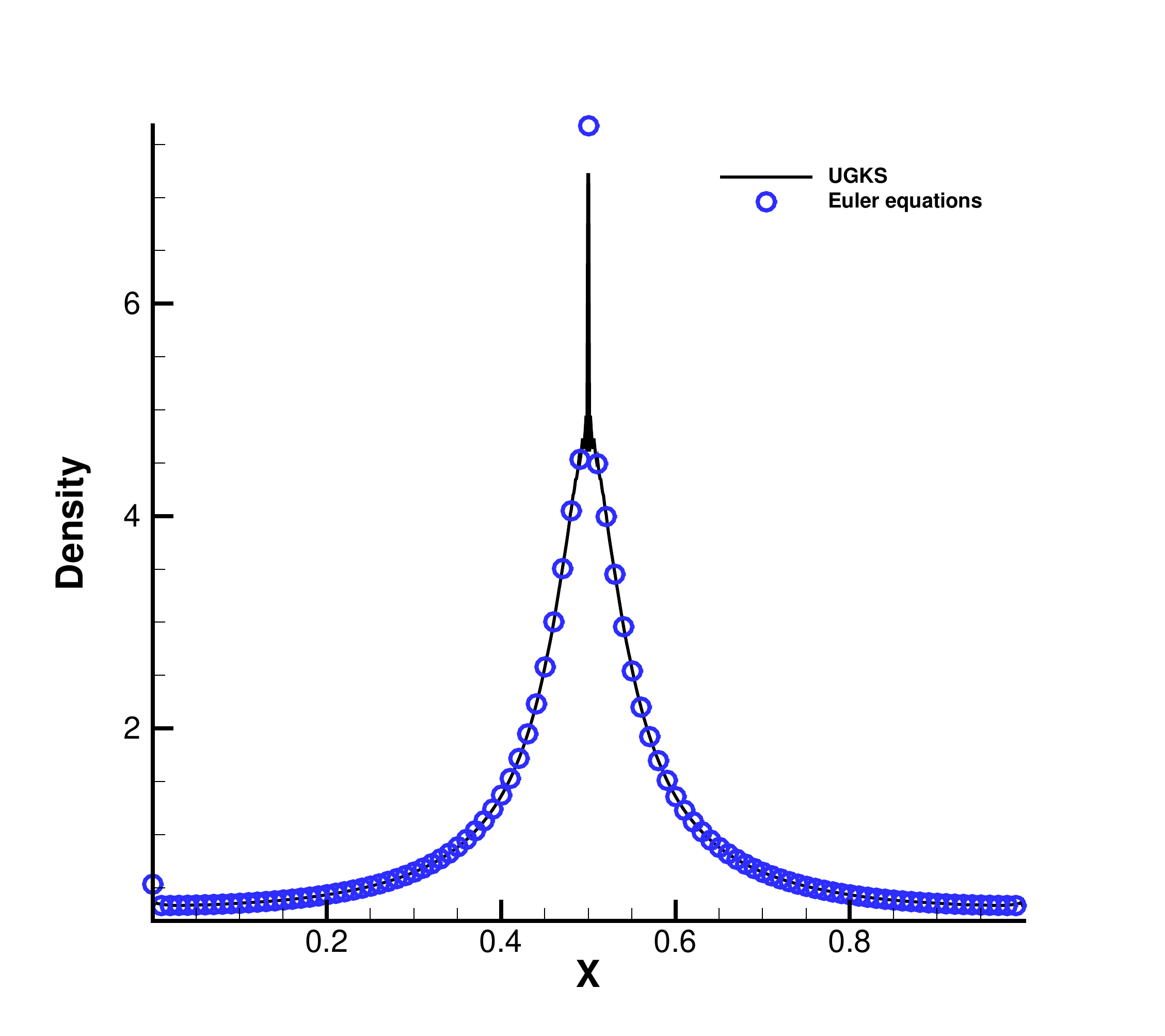}{c}
\includegraphics[width=0.45\textwidth]{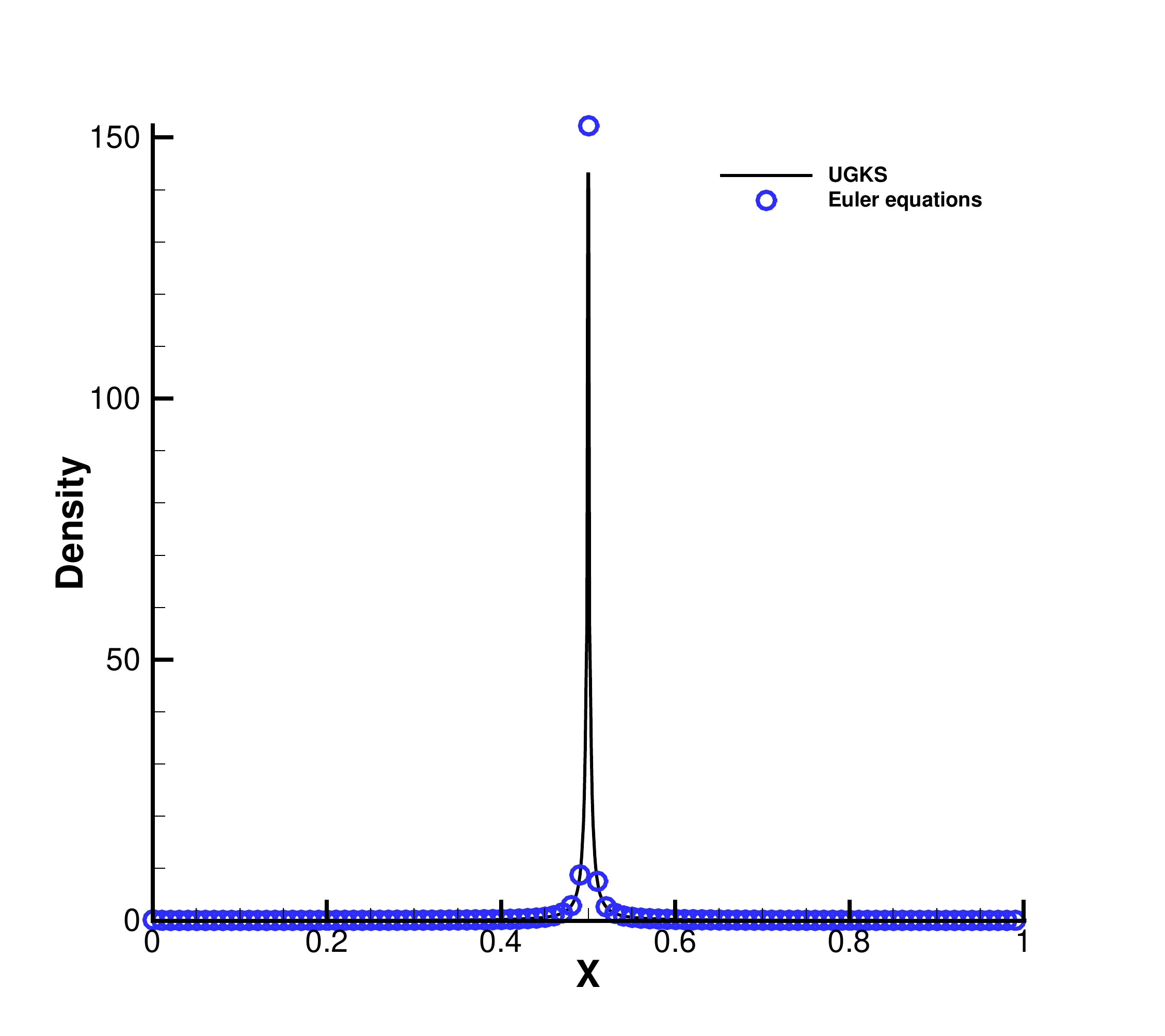}{d}
\caption{Apparent density distribution of particle phase for the one dimensional particle concentration test.
(a) the density distribution at $t=1$ with $Kn=10^{4}$ and $\tau_{st}=0.03$; solid line is the UGKS-M solution and square shows the PIC result.
(b) the density distribution at $t=1.5$ with $Kn=10^{4}$ and $\tau_{st}=0.03$; solid line is the UGKS-M solution and square shows the PIC result.
(c)  the density distribution at $t=1$ with $Kn=10^{-4}$ and $\tau_{st}=0.03$; solid line is the UGKS-M solution and circle shows the Pressureless Euler solution.
(d) the density distribution at $t=1.5$ with $Kn=10^{-4}$ and $\tau_{st}=0.03$; solid line is the UGKS-M solution and circle shows the Pressureless Euler solution.}
\label{1taylor2}
\end{figure}

In this section, we apply the UGKS-M to five numerical test cases to demonstrate the multiscale property of the numerical scheme
and its ability to capture the non-equilibrium phenomenon.
The test cases cover a wide range of particle Knudsen number and particle Stokes number.
In current calculations, non-adaptive velocity space is used.
{
The range of the velocity space is chosen to be $[U_{s,\min}-5\sqrt{2k_BT_{s,\max}/m},U_{s,\max}+5\sqrt{2k_BT_{s,\max}/m}],$
where $U_{s,\min}$, $U_{s,\max}$ are the pre-estimated lower and upper bound of solid phase macroscopic velocities,
and $T_{\max}$ is the pre-estimated highest temperature in the computational domain.
The grid size of the velocity is chosen to be $\sqrt{2k_BT_{s,\min}/m}/3$,
where $T_{s,\min}$ is the pre-estimated lowest temperature in the computational domain.
For current test cases, if $T_{x,\min}=0$, we set the maximum velocity cell numbers to be 200, which gives satisfactory resolution.
The technique of velocity adaptation will be implemented in UGKS-M in our future work, which can greatly reduce the computational cost.}
For the one-dimensional particle segmentation problem, the UGKS-M recovers the solution of collisionless Boltzmann equation in the collisionless regime and converges to the pressureless Euler solution in the continuum regime.
For one dimensional shock tube test case, the results show that at the small particle Knudsen number, the UGKS-M recovers the two-fluid NS solution,
while at large particle Knudsen number, the UGKS-M provides consistent solution with the solution of Boltzmann equation.
In the two dimensional calculations of particle jets impinging problem, physical consistent solutions are obtained with
different Knudsen number and restitution coefficient, such as the particle trajectory crossing (PTC), and particle wall rebounding.
The calculation of particle motion in a Taylor-Green flow shows the capability of UGKS-M in simulating the flow dynamics
over a wide range of Stokes number. Lastly the experiment of shock induced fluidization of particle bed is calculated by UGKS-M
and the solution is compared with the experimental measurement.
{\subsection{One dimensional particle concentration under a harmonic oscillatory flow}}
Firstly, we study the solid particle concentration in a one dimensional gas flow to test the multiscale property of UGKS-M.
This test case is a one way coupling flow with a steady gas field with velocity distribution $U_g(x)=\sin(2\pi x)$. Since the gas field is fixed, we leave out the particle material temperature and only consider the particle motion including the particle velocity distribution as well as the granular temperature in this calculation. The normalization is done according to the following reference parameters: the computational domain as the reference length,
the initial solid phase apparent density as the reference density, the maximum of the gas velocity as the reference velocity,
and the initial gas temperature as the reference temperature. The initial condition is set as $\epsilon_s\rho_s=1$, $U_s=\sin(2\pi x)$, $T_s=10^{-8}$; the boundary condition is set to be periodic and the particle collision is assumed to be fully inelastic collision with $r=0$. The computational domain in physical space is $[0,1]$ equally discretized into 1000 cells, and the velocity space is $[-1.5,1.5]$ with $32$ velocity cells.
This test case is characterized by two important parameters, the Knudsen number and the Stokes number.
Four limiting flow regimes are considered:
(i)   Large stokes number collisionless regime with $Kn=10^{4}$ and $\tau_{st}=0.3$;
(ii)  Large stokes number continuum regime with $Kn=10^{-4}$ and $\tau_{st}=0.3$;
(iii) Small stokes number collisionless regime $Kn=10^{4}$ and $\tau_{st}=0.03$;
(iv)  Small stokes number continuum regime with $Kn=10^{-4}$ and $\tau_{st}=0.03$.
{
In the collisionless regime, the reference solution is obtained by solving the collisionless Boltzmann using the particle in cell (PIC) method,
while the pressureless Euler equation is solved by GKS and serves as the reference solution in the continuum regime.
For all flow regimes, the comparison between UGKS-M solution and the reference solution at $t=1$ and $t=1.5$ is shown in Fig. \ref{1taylor1}-\ref{1taylor2}.
For large stokes regime collisionless regime, the interspecies fraction is not enough to dissipate the particle kinetic energy, and the particles will oscillate in the gas field for a while before reaching the same speed with gas field. In such regime, good agreement between UGKS-M and PIC solution can be observed in Fig. \ref{1taylor1}(a) and \ref{1taylor1}(b). When the Knudsen number decreases, the intense inelastic collision dissipates the particle kinetic energy, and the particles show a tendency of concentration. The UGKS-M well recovers the pressureless Euler solution in such regime. When reducing the Stokes number to $\tau_{st}=0.03$, the oscillatory behavior of particles will be suppressed by a strong interspecies fraction.
In both collisionless and continuum regimes, the UGKS-M shows good agreements with reference solutions in the highly different flow regimes.
The one dimensional particle concentration test shows the capability of UGKS-M in predicting the behavior of particles in continuum and rarefied flow regime with different Stokes numbers.}

\subsection{Wind-sand shock tube}
\begin{figure}
\centering
\includegraphics[width=0.45\textwidth]{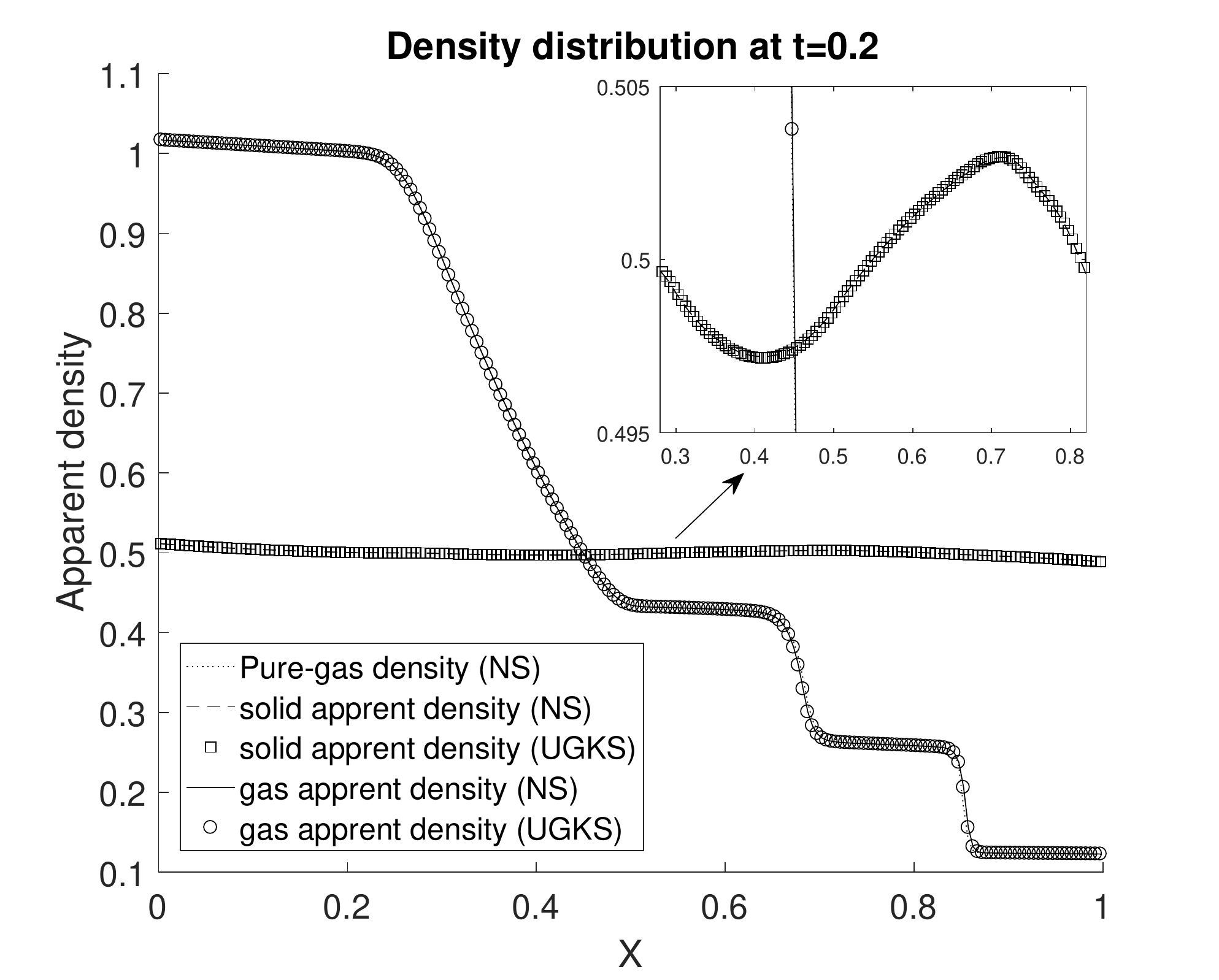}{a}
\includegraphics[width=0.45\textwidth]{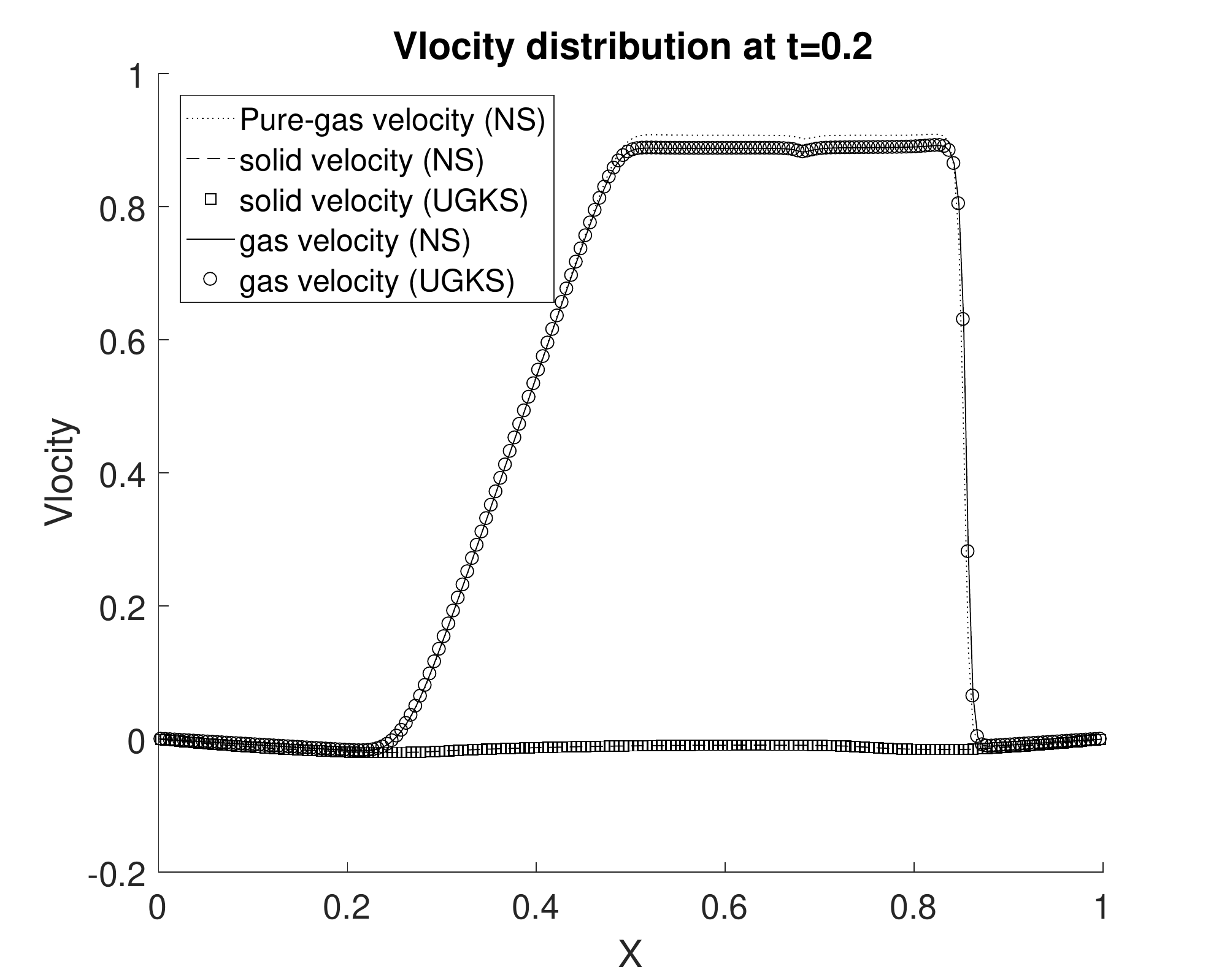}{b}\\
\includegraphics[width=0.45\textwidth]{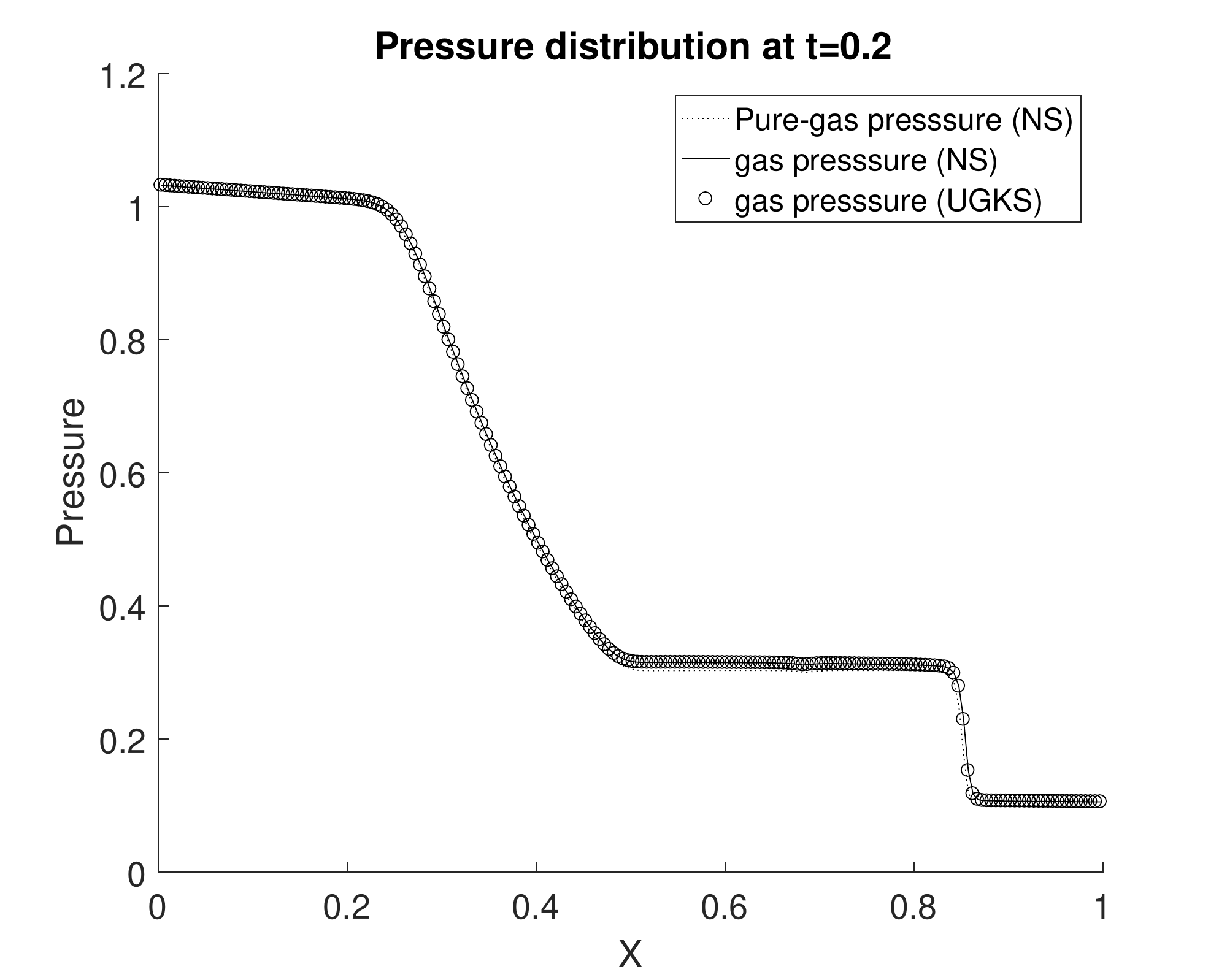}{c}
\includegraphics[width=0.45\textwidth]{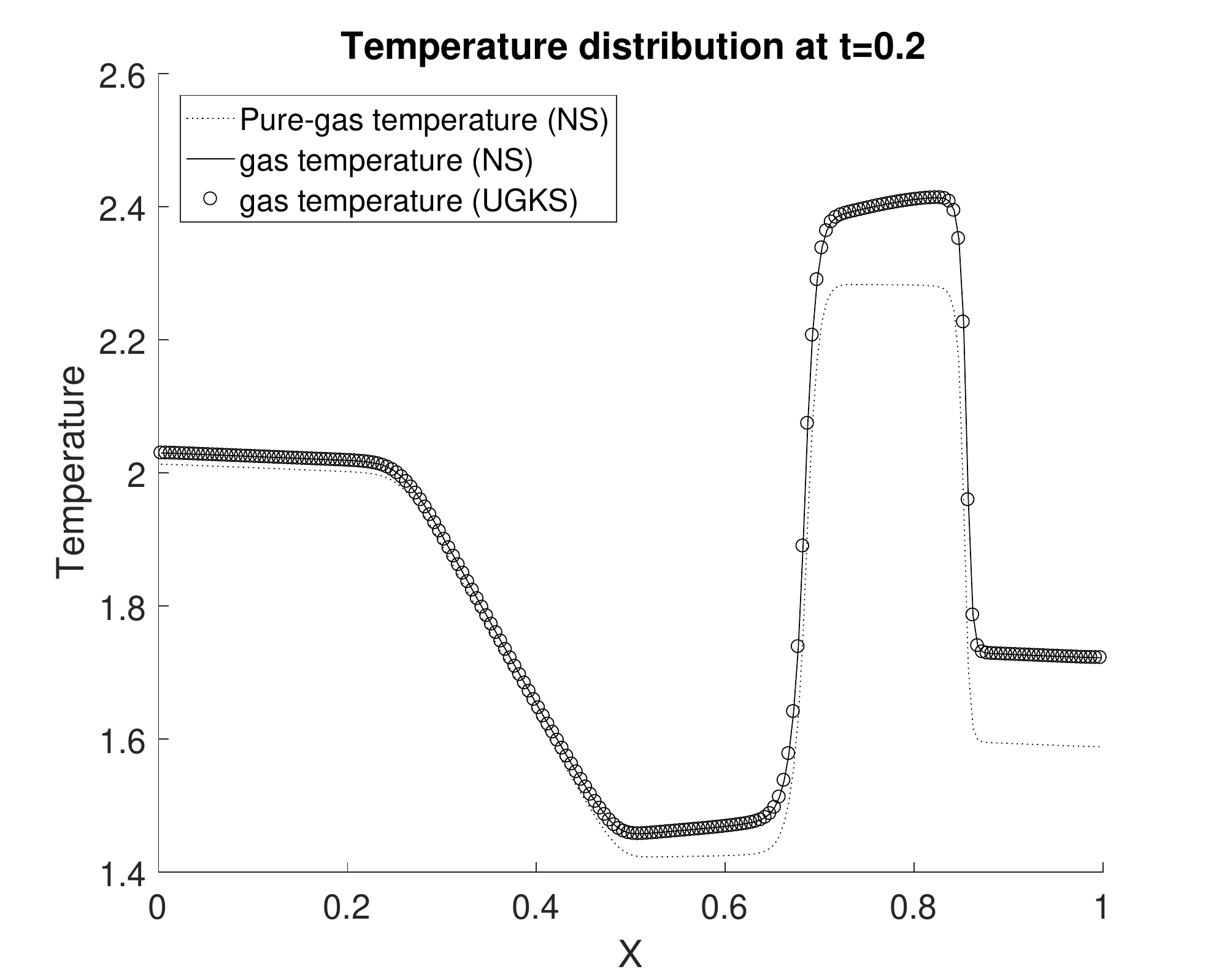}{d}\\
\includegraphics[width=0.45\textwidth]{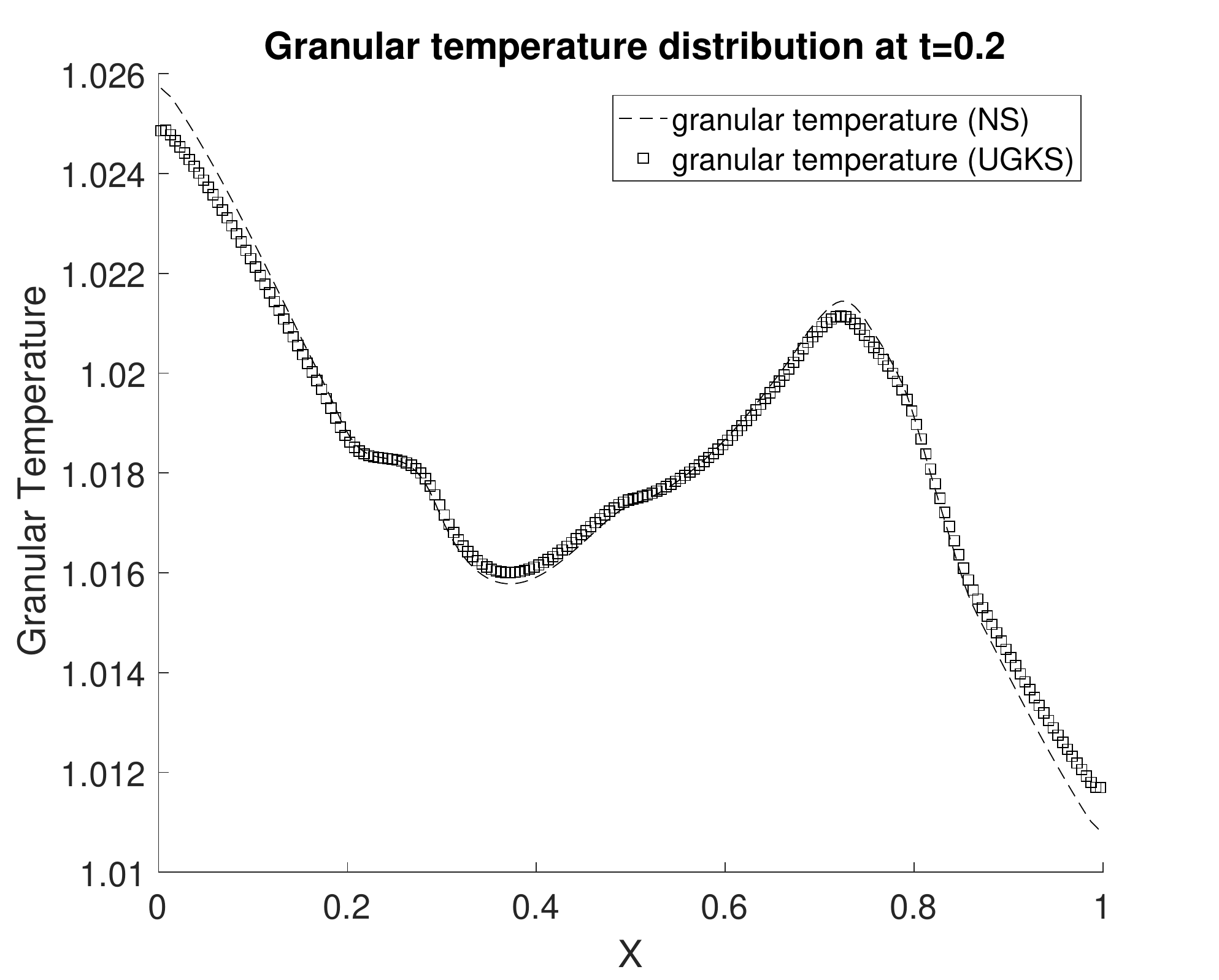}{c}
\includegraphics[width=0.45\textwidth]{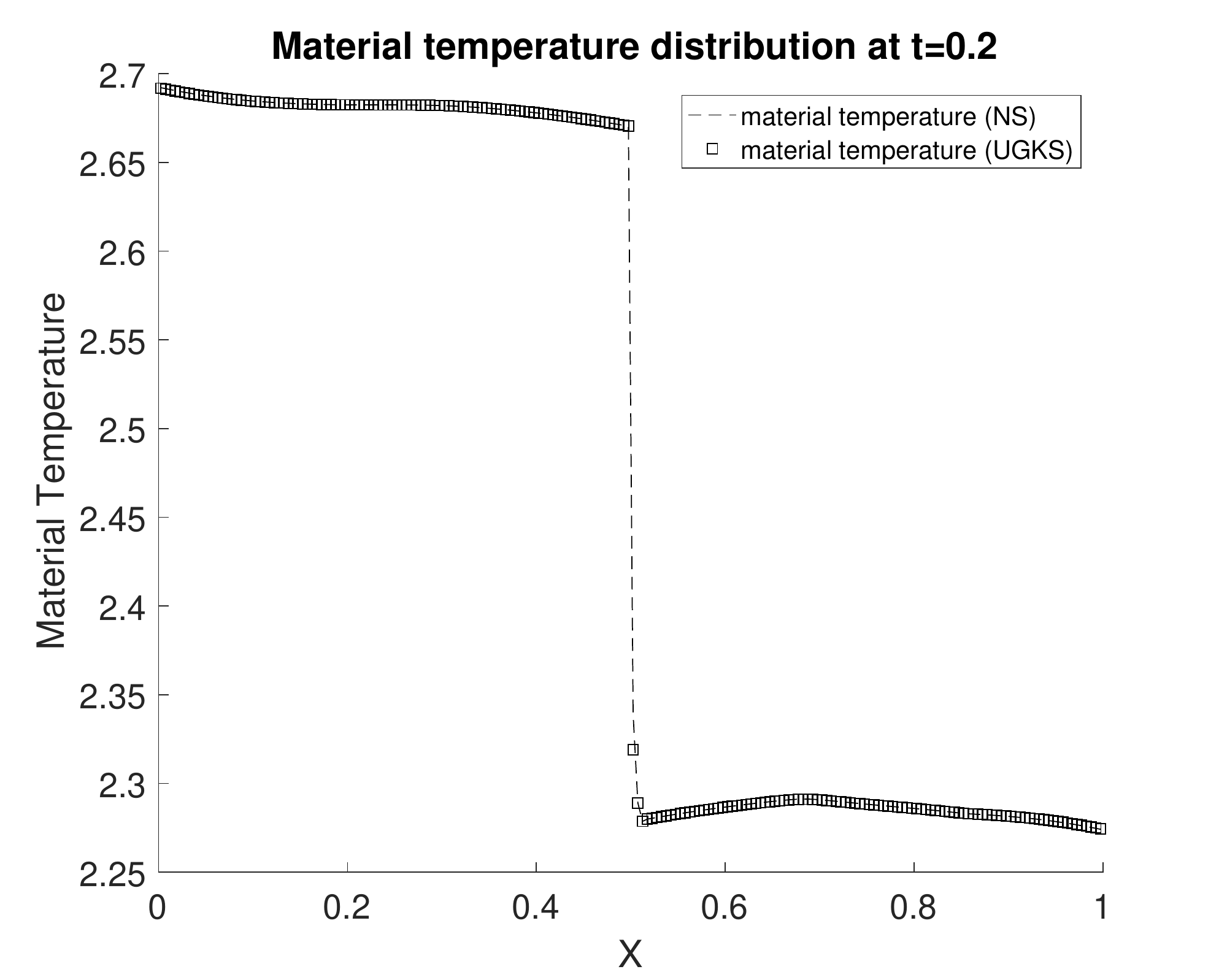}{d}
\caption{Results of the wind-sand shock tube problem at $t=0.2$ with $Kn_s=10^{-4}$, $\tau_{st}=10$, and $\tau_{T}=10$. (a) apparent density, (b) velocity, (c) gas pressure, (d) gas temperature, (e) particle granular temperature, (f) particle material temperature. The solutions of UGKS-M are shown in symbols (circle for gas phase and square for solid phase), and the solutions of two-fluid NS system are shown in lines (solid for gas phase and dashed for solid phase). The pure gas solutions are shown in dotted lines for reference.}
\label{shocktube1}
\end{figure}

\begin{figure}
\centering
\includegraphics[width=0.45\textwidth]{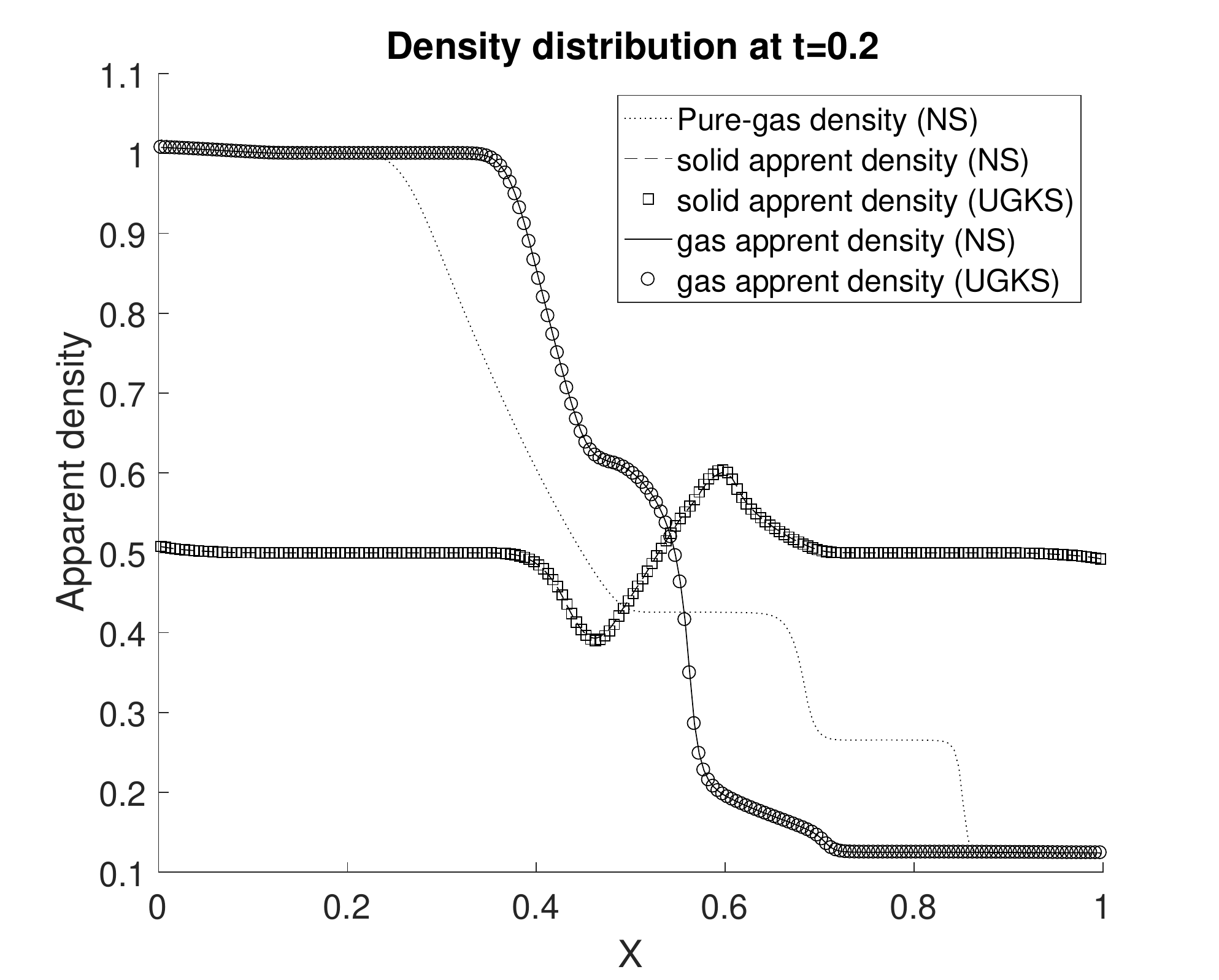}{a}
\includegraphics[width=0.45\textwidth]{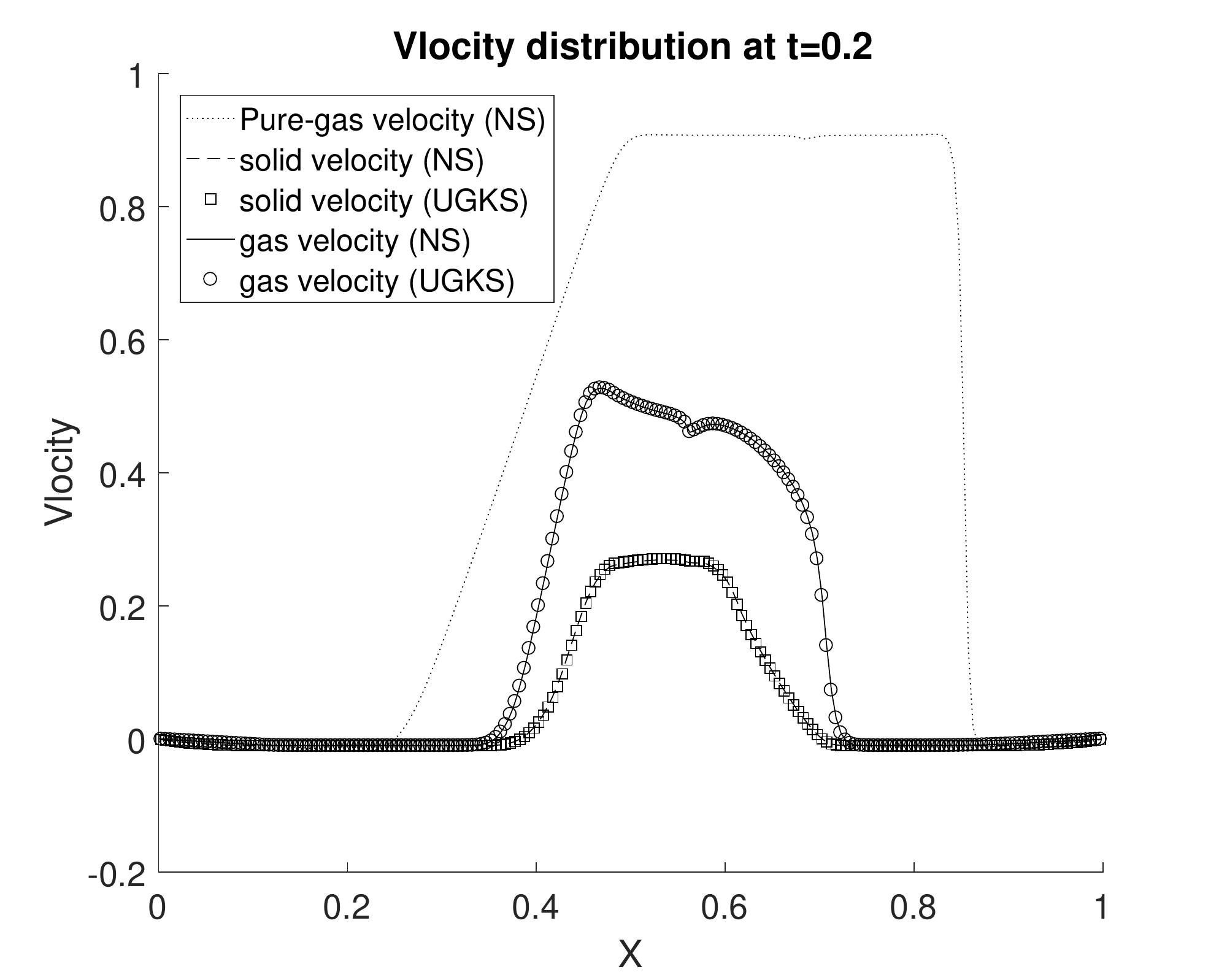}{b}\\
\includegraphics[width=0.45\textwidth]{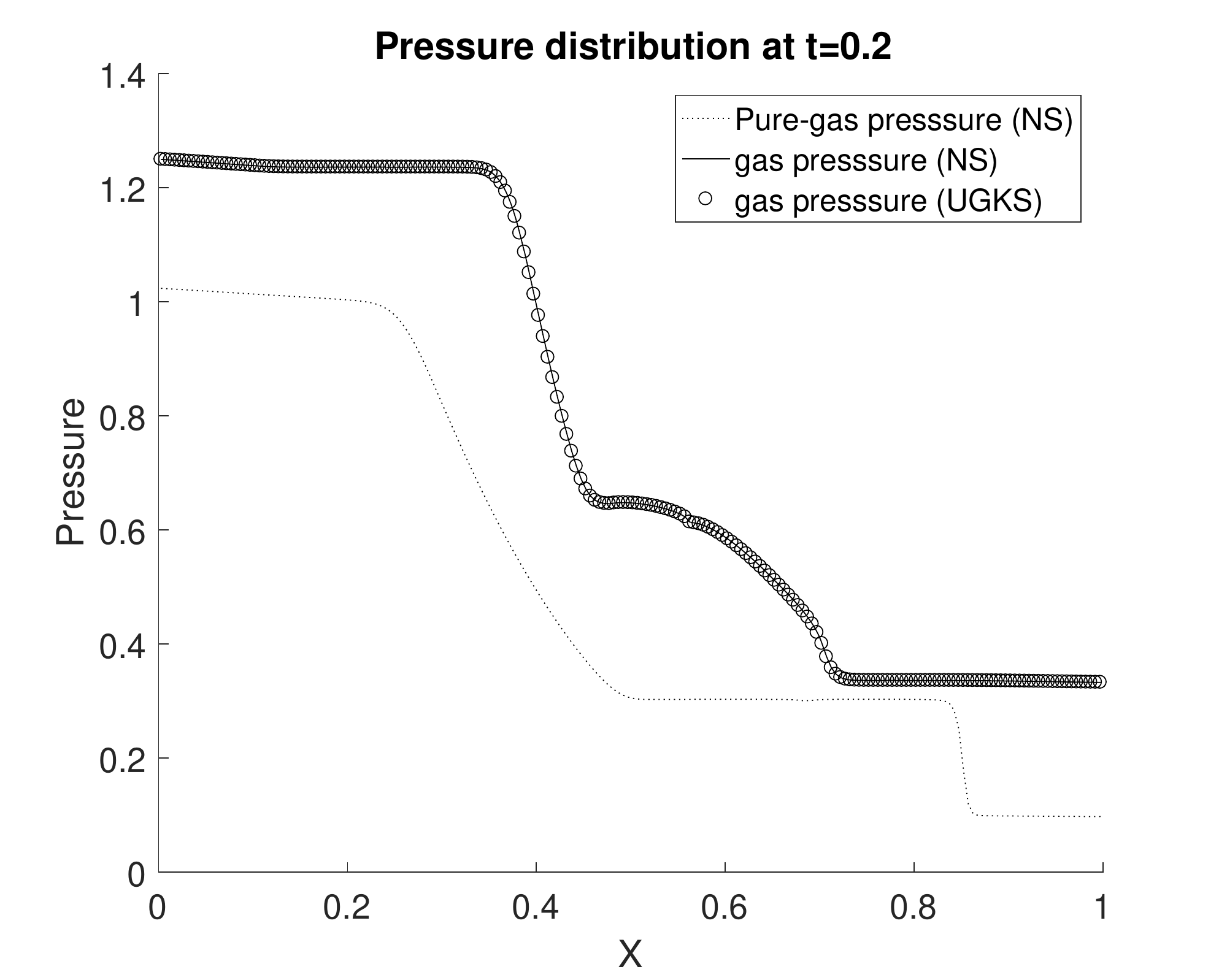}{c}
\includegraphics[width=0.45\textwidth]{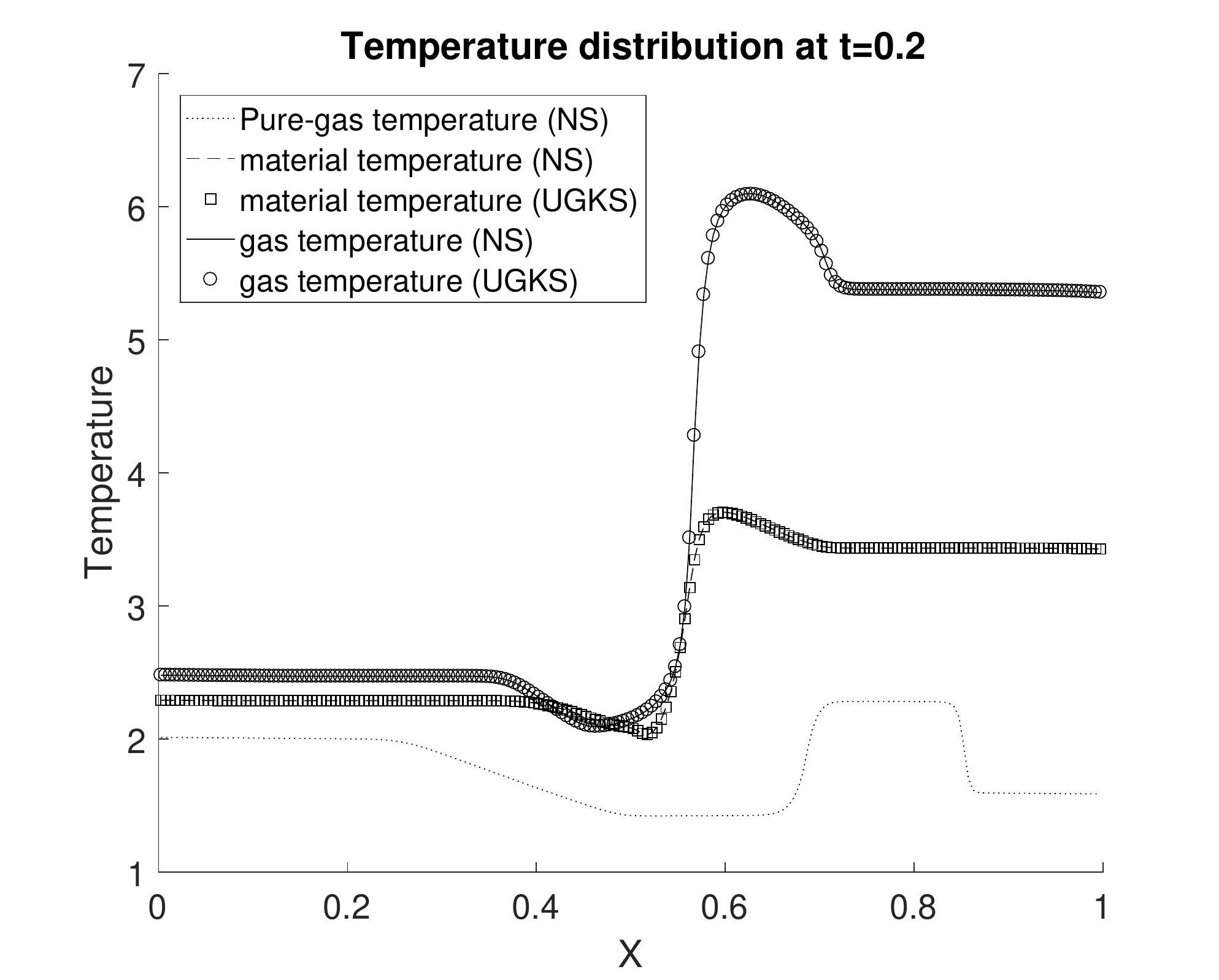}{d}\\
\includegraphics[width=0.45\textwidth]{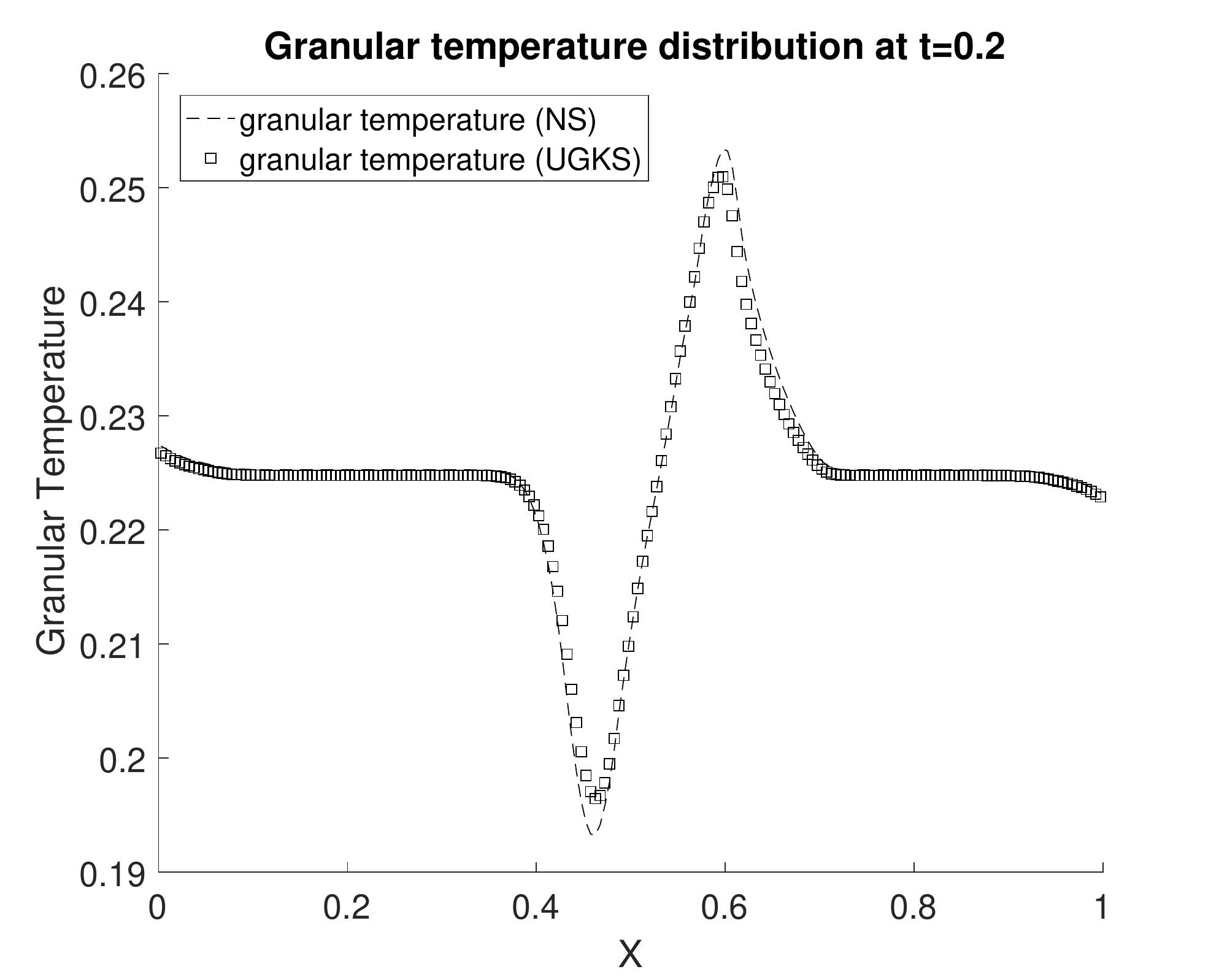}{c}
\includegraphics[width=0.45\textwidth]{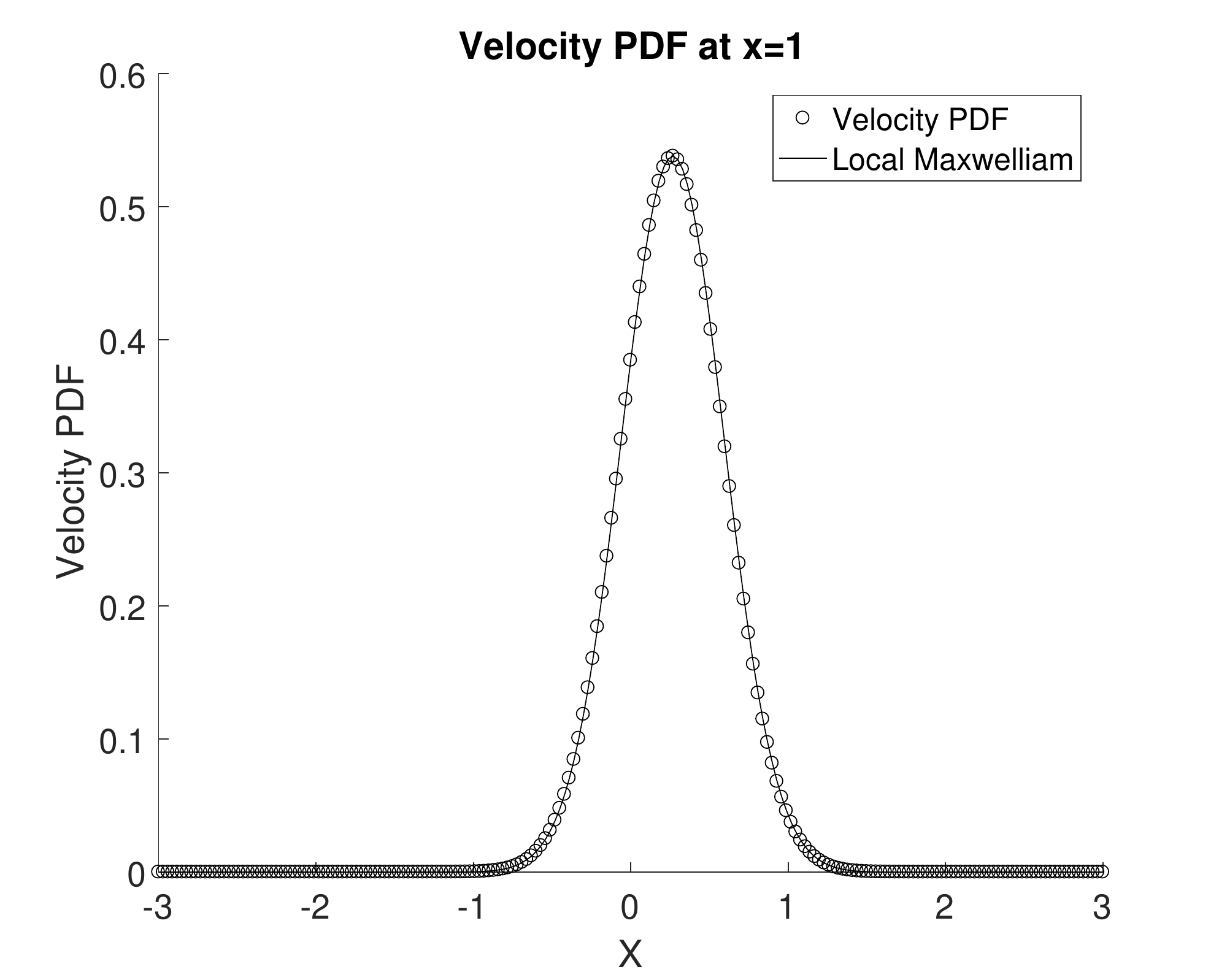}{d}
\caption{Results of the wind-sand shock tube problem at $t=0.1$ with $Kn_s=10^{-4}$, $\tau_{st}=0.1$, and $\tau_{T}=0.1$. (a) apparent density, (b) velocity, (c) gas pressure, (d) gas temperature and particle material temperature, (e) particle granular temperature, (f) solid particle velocity distribution function (circle) and the local Maxwellian distribution (line) at $x=0.5$. For (a)-(e), the solutions of UGKS-M are shown in symbols (circle for gas phase and square for solid phase), and the solutions of two-fluid NS system are shown in lines (solid for gas phase and dashed for solid phase). The pure gas solutions are shown in dotted lines for reference.}
\label{shocktube2}
\end{figure}

\begin{figure}
\centering
\includegraphics[width=0.45\textwidth]{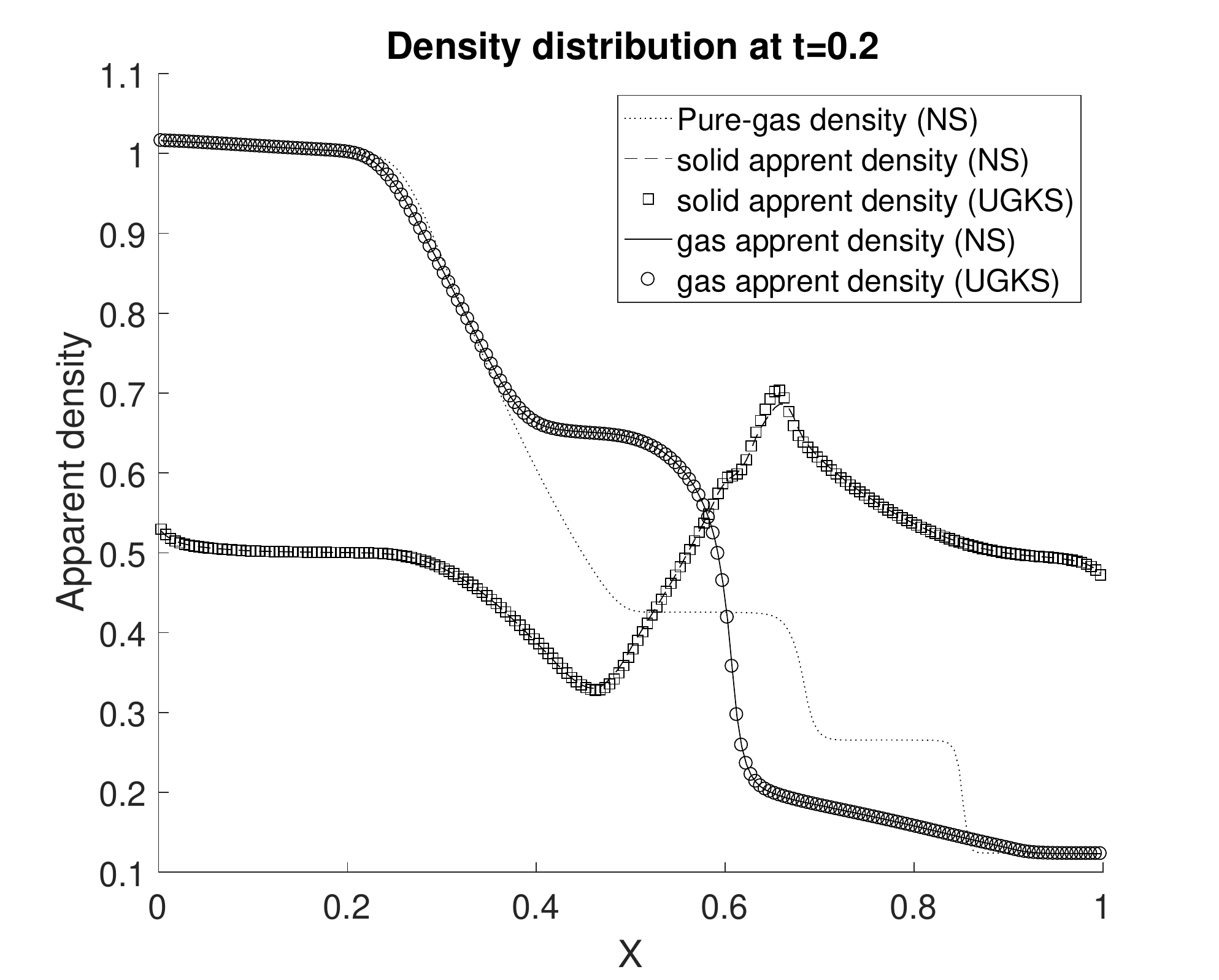}{a}
\includegraphics[width=0.45\textwidth]{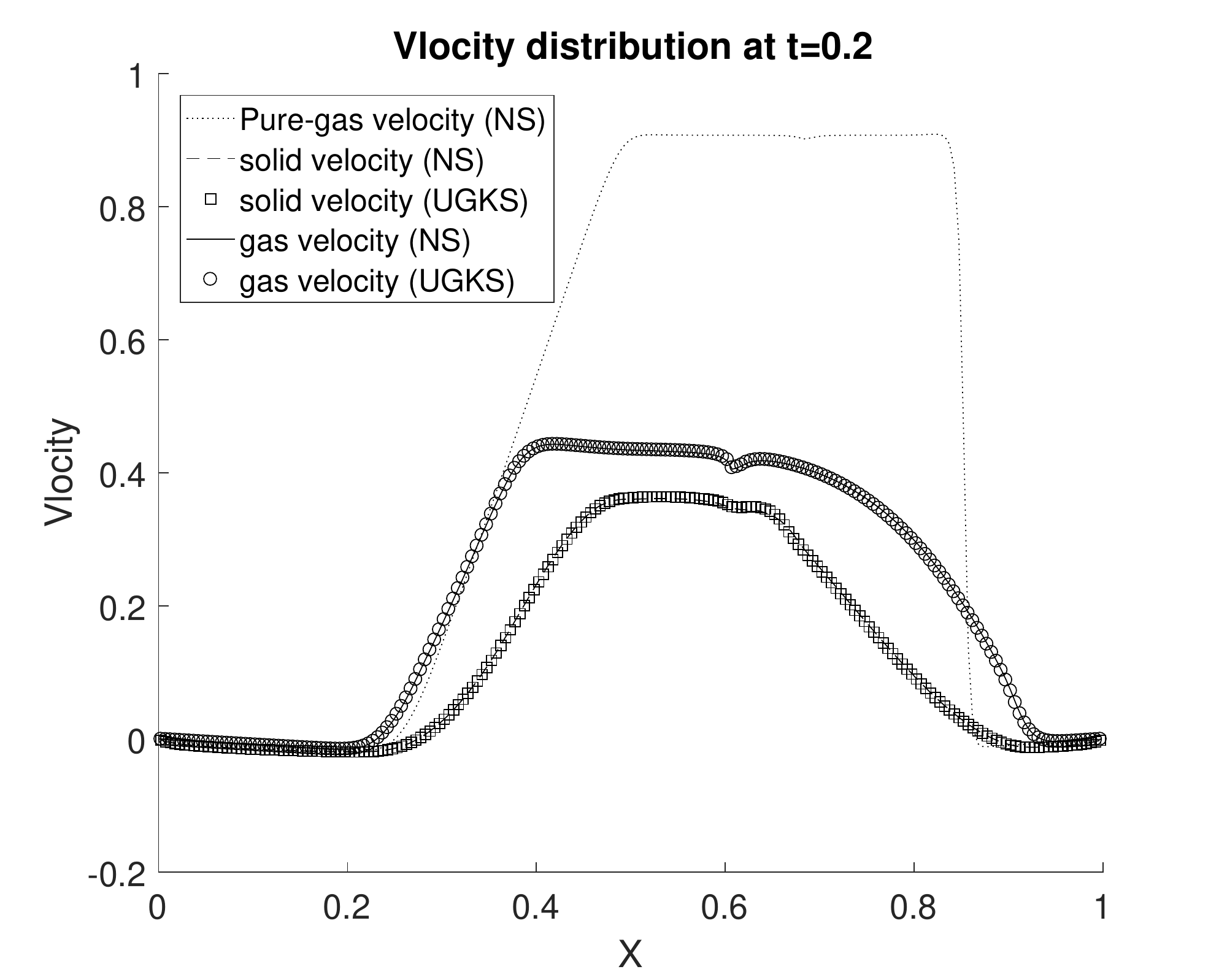}{b}\\
\includegraphics[width=0.45\textwidth]{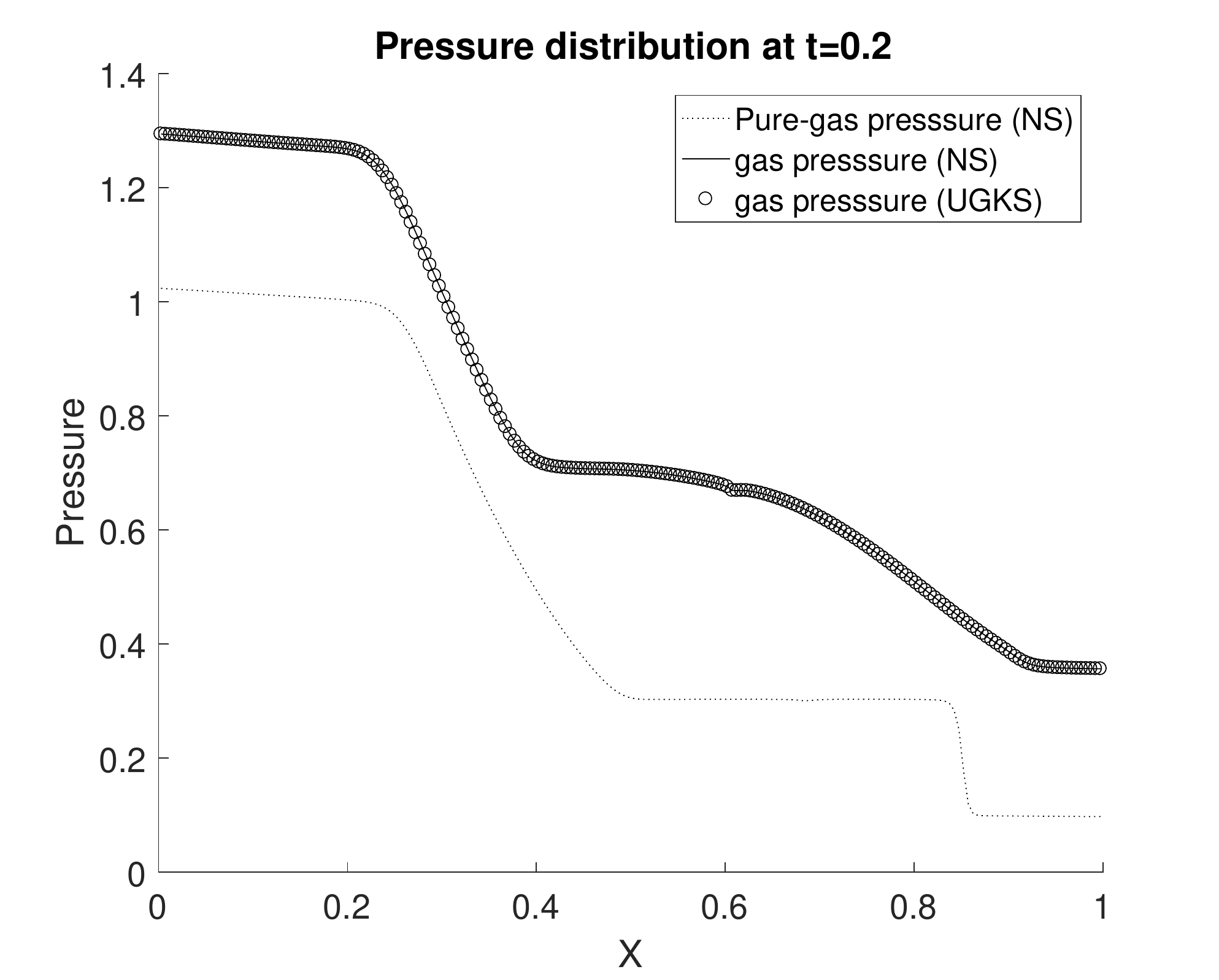}{c}
\includegraphics[width=0.45\textwidth]{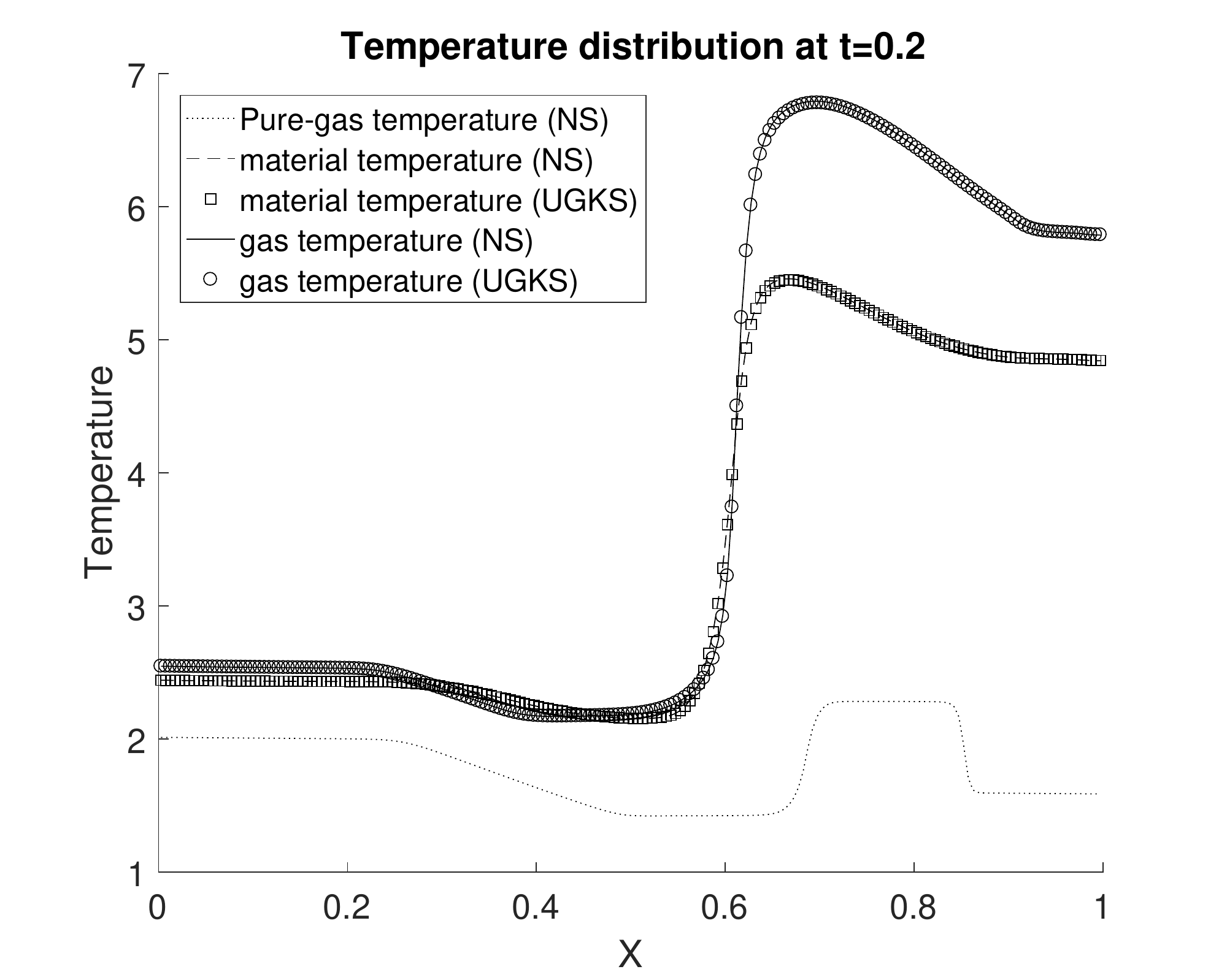}{d}\\
\includegraphics[width=0.45\textwidth]{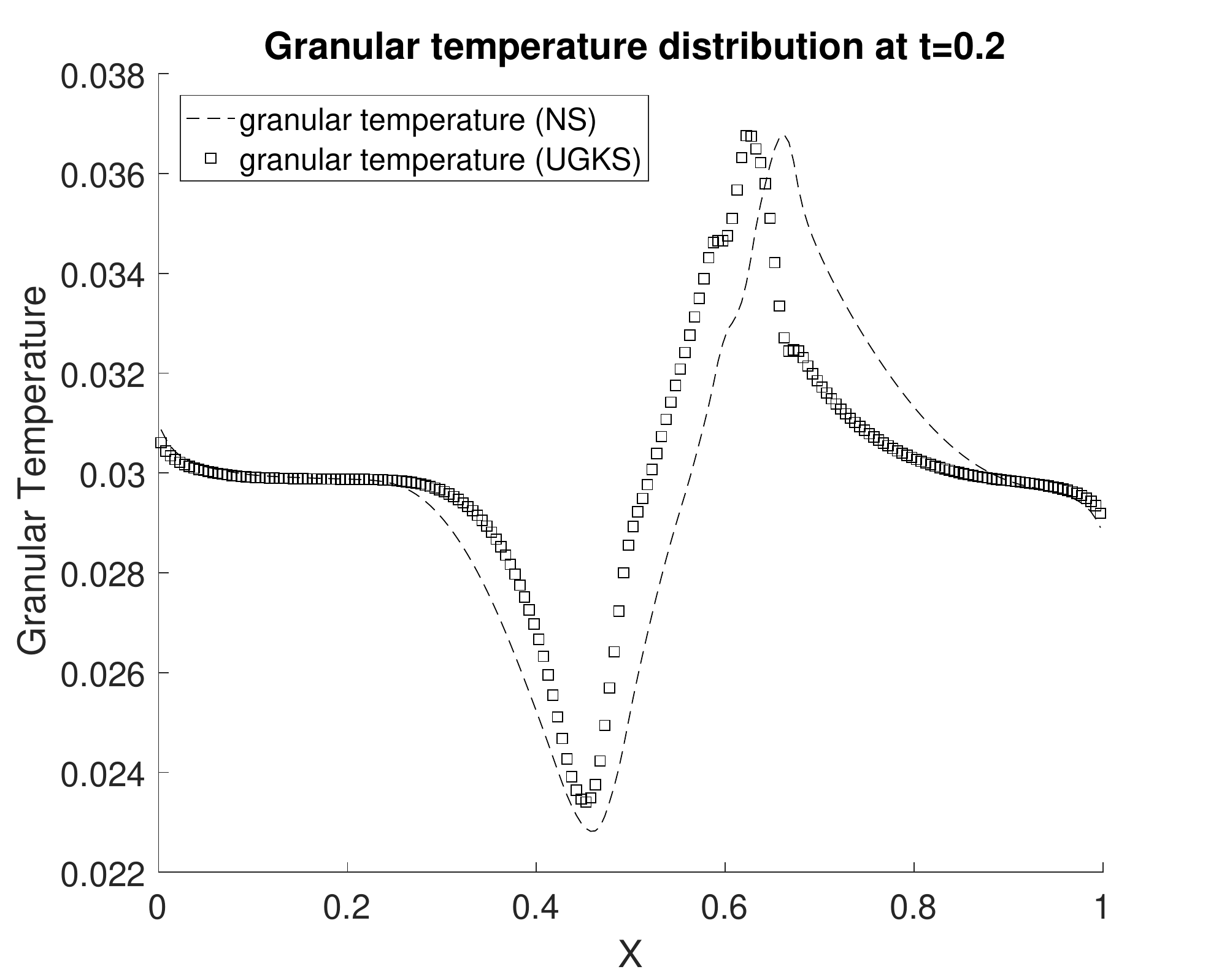}{c}
\includegraphics[width=0.45\textwidth]{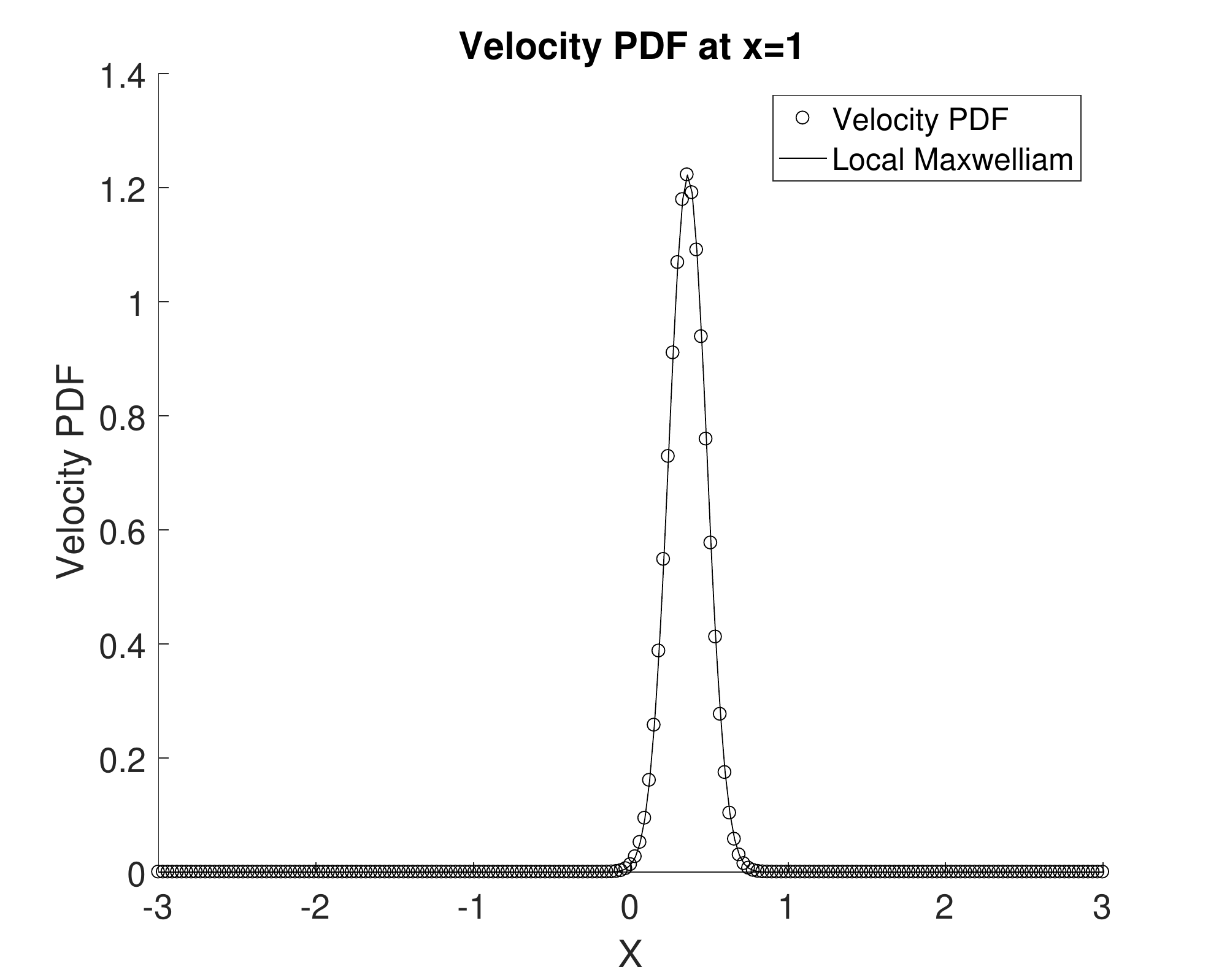}{d}
\caption{Results of the wind-sand shock tube problem at $t=0.2$ with $Kn_s=10^{-4}$, $\tau_{st}=0.1$, and $\tau_{T}=0.1$. (a) apparent density, (b) velocity, (c) gas pressure, (d) gas temperature and particle material temperature, (e) particle granular temperature, (f) solid particle velocity distribution function (circle) and the local Maxwellian distribution (line) at $x=0.5$. For (a)-(e), the solutions of UGKS-M are shown in symbols (circle for gas phase and square for solid phase), and the solutions of two-fluid NS system are shown in lines (solid for gas phase and dashed for solid phase). The pure gas solutions are  shown in dotted lines for reference.}
\label{shocktube3}
\end{figure}

\begin{figure}
\centering
\includegraphics[width=0.45\textwidth]{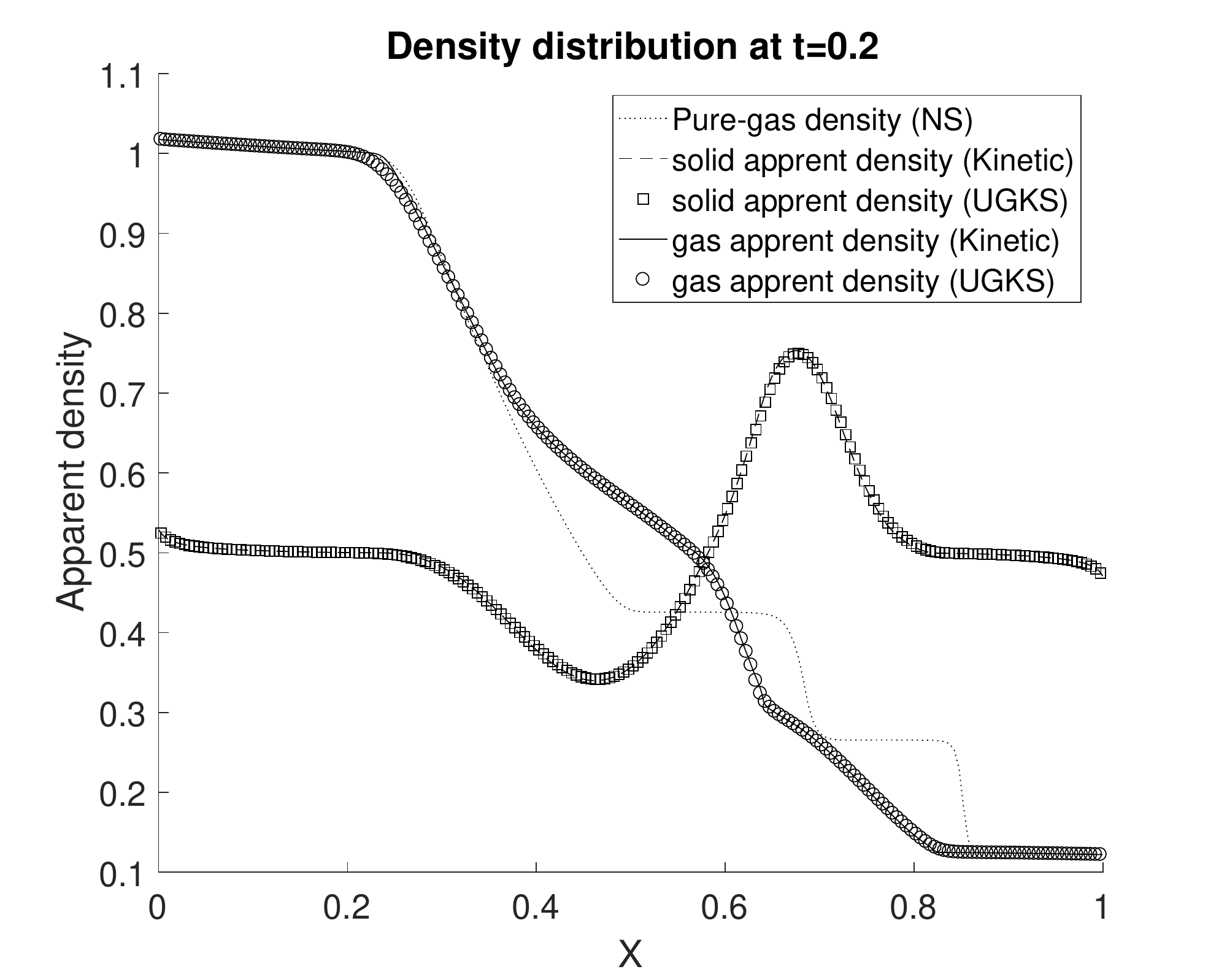}{a}
\includegraphics[width=0.45\textwidth]{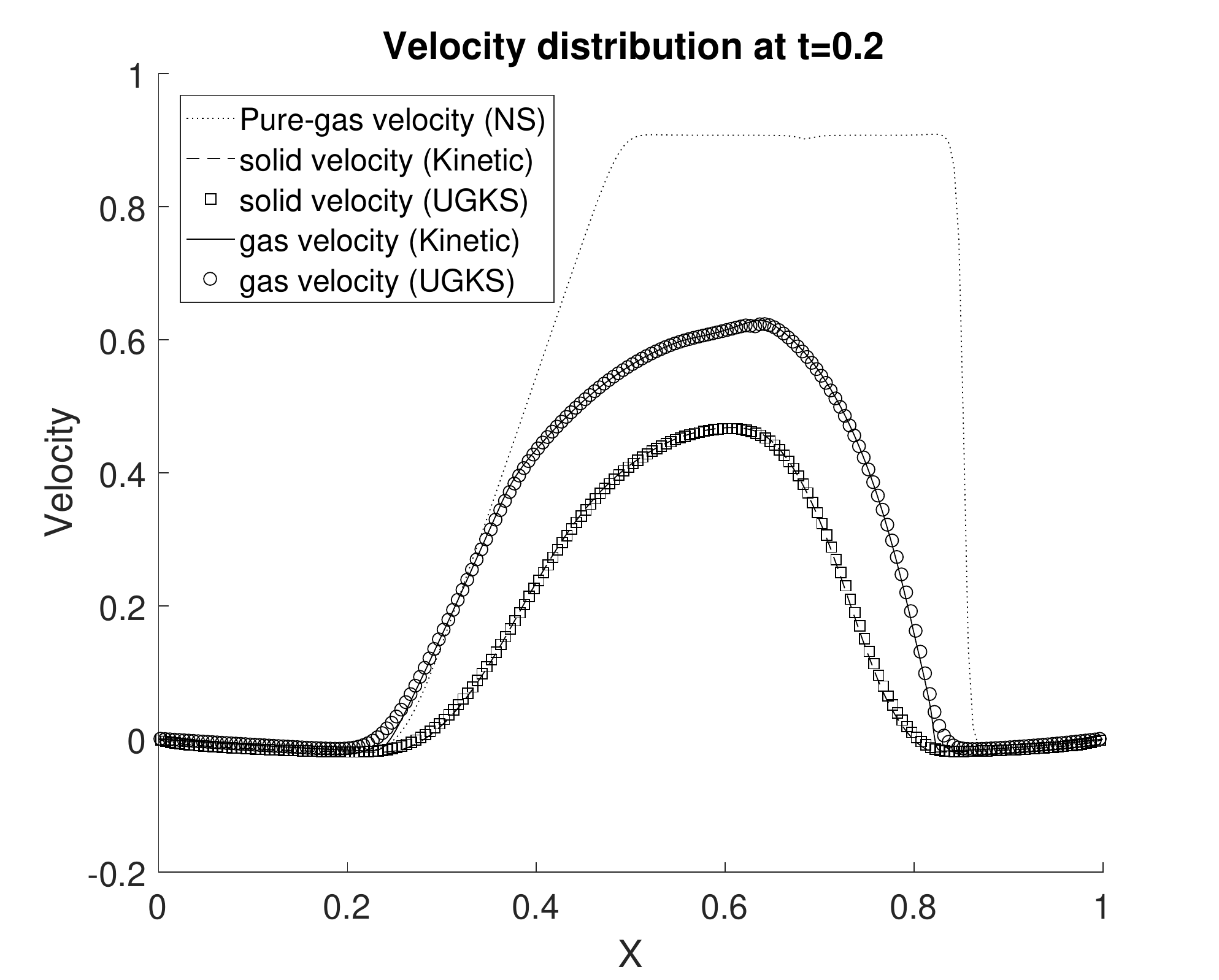}{b}\\
\includegraphics[width=0.45\textwidth]{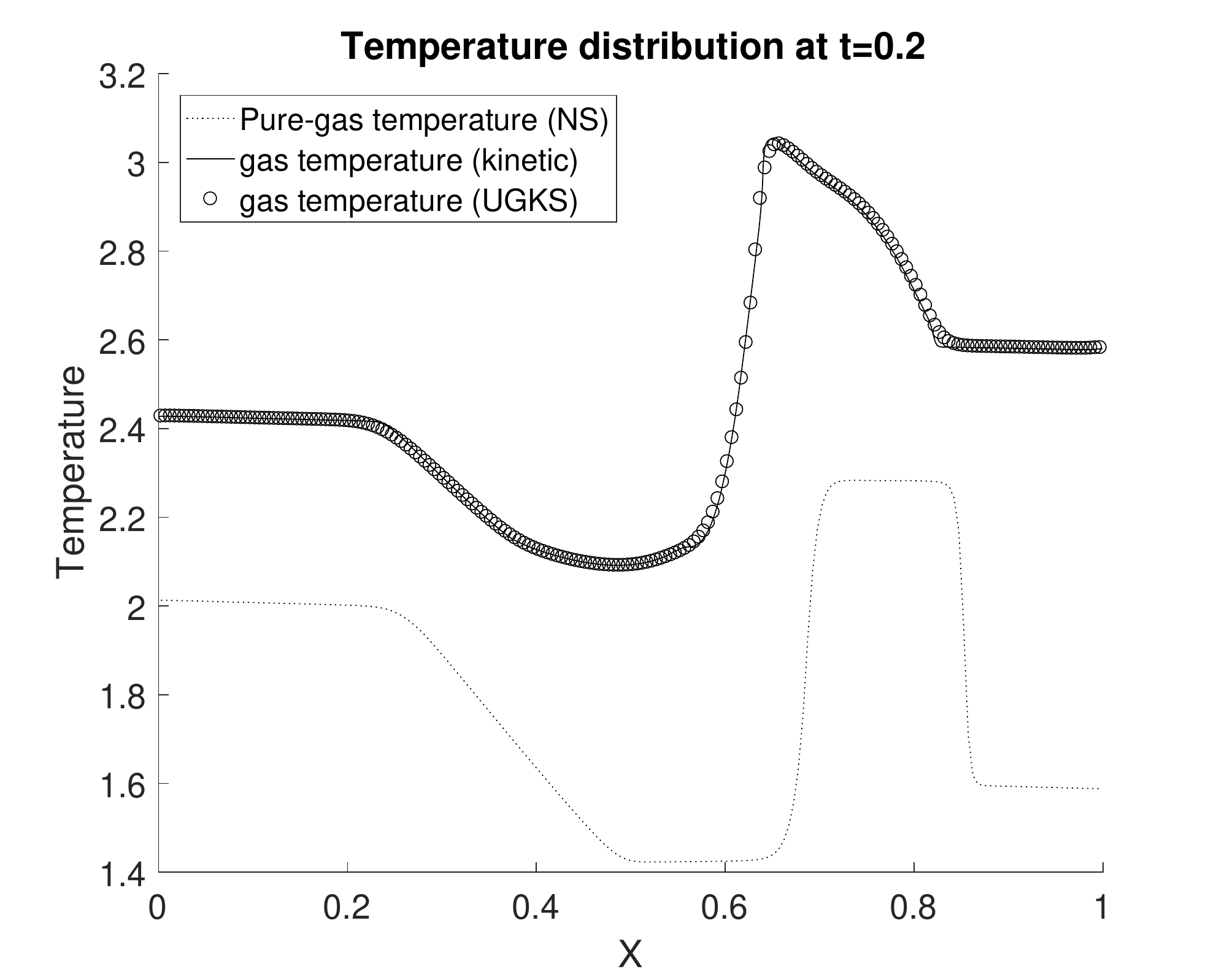}{c}
\includegraphics[width=0.45\textwidth]{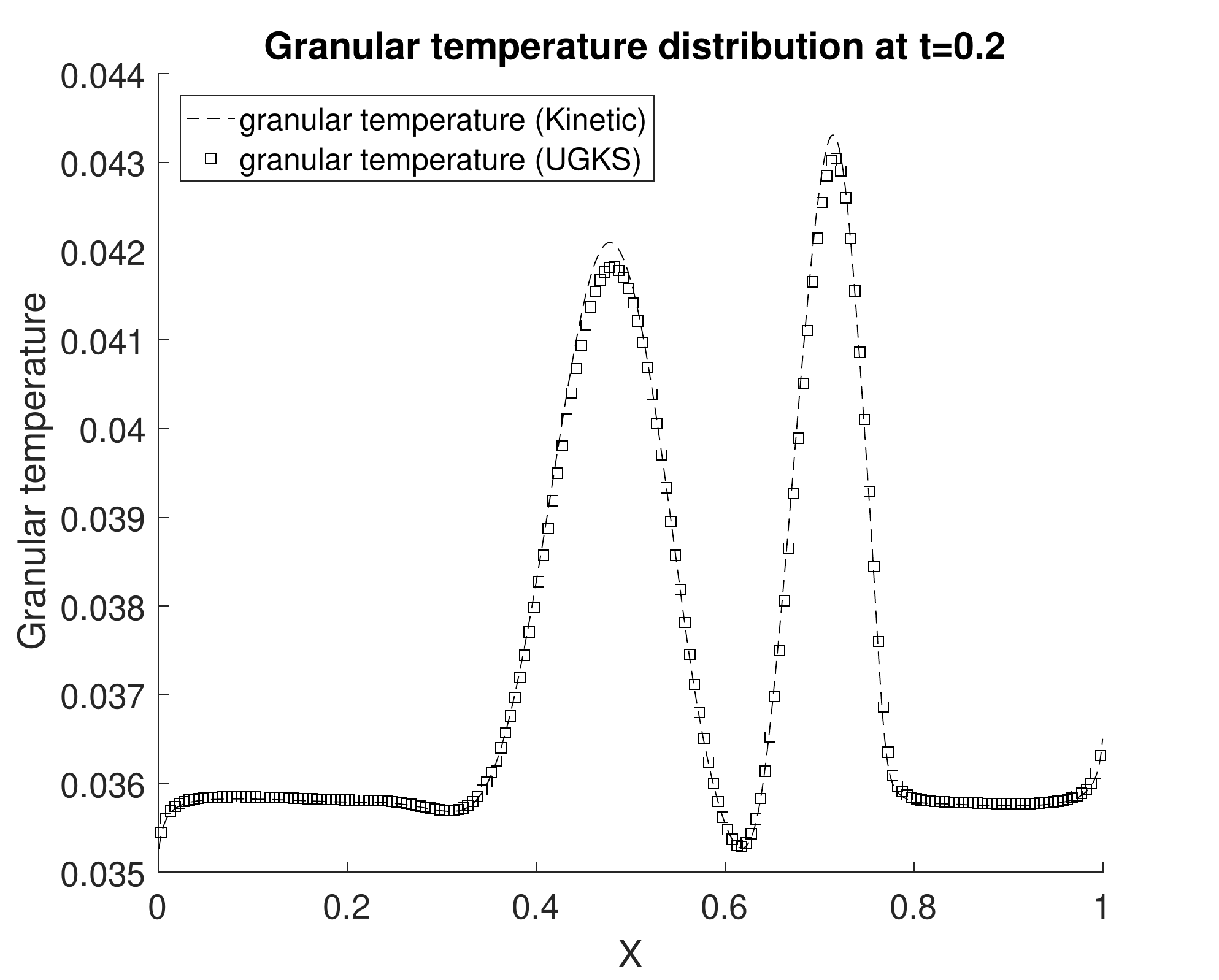}{d}\\
\includegraphics[width=0.45\textwidth]{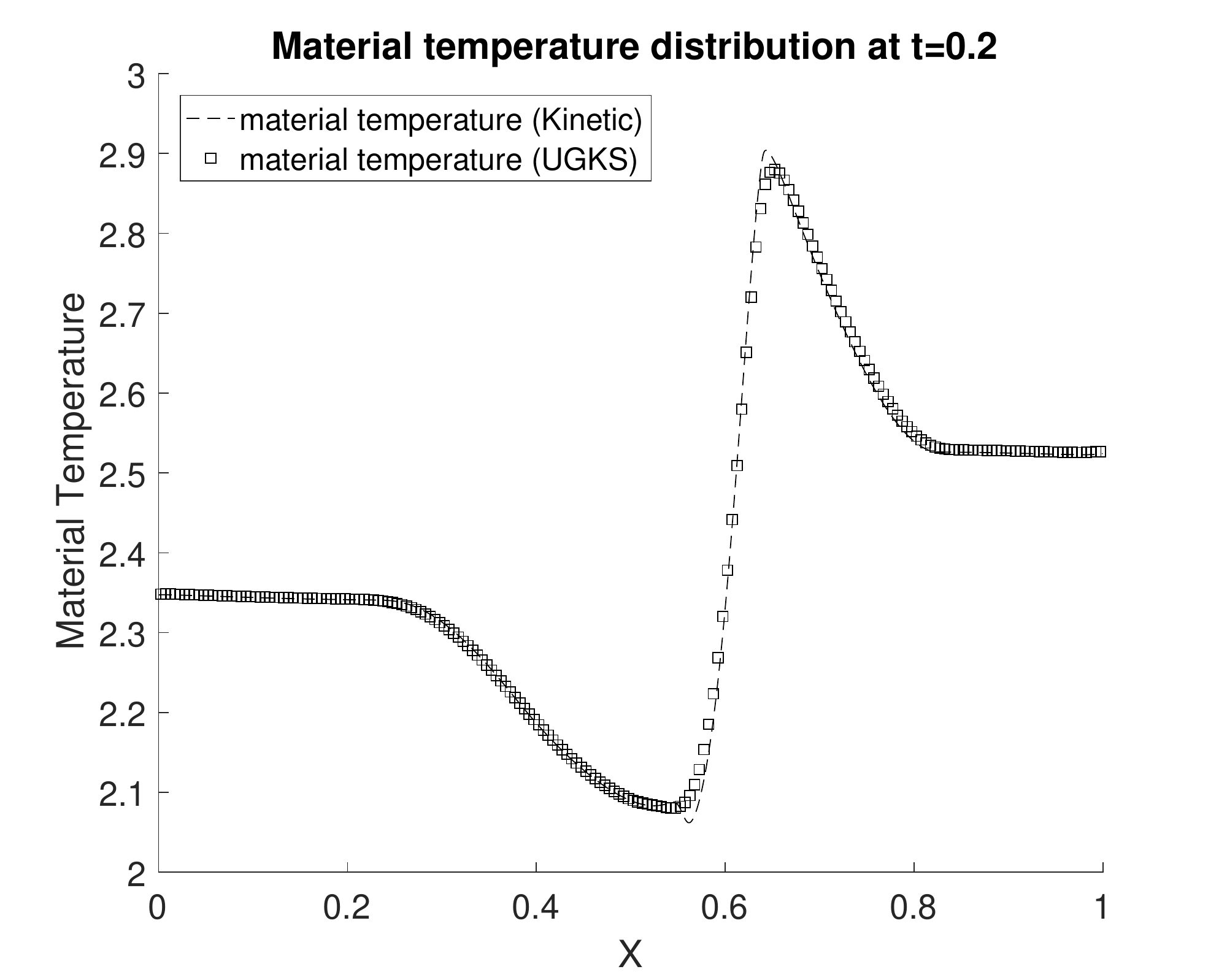}{c}
\includegraphics[width=0.45\textwidth]{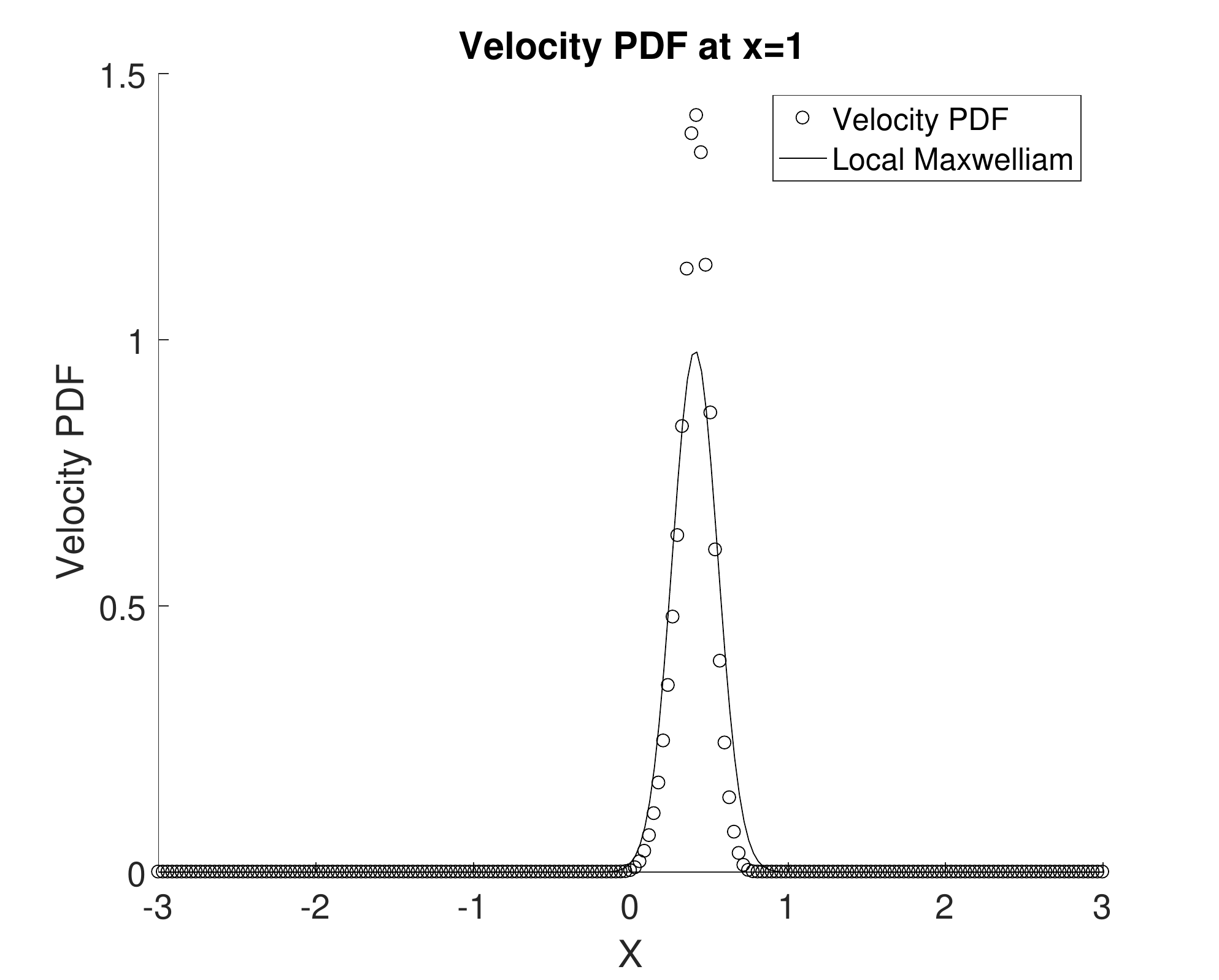}{d}
\caption{Results of the wind-sand shock tube problem at $t=0.2$ with $Kn_s=1$, $\tau_{st}=0.1$, and $\tau_{T}=0.1$. (a) apparent density, (b) velocity, (c) gas pressure, (d) gas temperature and particle material temperature, (e) particle granular temperature, (f) solid particle velocity distribution function (circle) and the local Maxwellian distribution (line) at $x=0.5$. For (a)-(e), the solutions of UGKS-M are shown in symbols (circle for gas phase and square for solid phase), and the solutions of kinetic equation are shown in lines (solid for gas phase and dashed for solid phase). The pure gas solutions are shown in dotted lines for reference.}
\label{shocktube4}
\end{figure}

{We calculate the one-dimensional wind-sand shock tube problem similar to the numerical calculations done by T. Saito \cite{saito2002numerical}, T. Saito et al. \cite{saito2003numerical}, but with a simplified drag force formulation Eq.\eqref{simplified-drag}.}
Initially the solid phase is uniformly distributed in the computational domain $x\in[0,1]$.
With the evolution of time, the solid phase will be driven by gas due to friction.
{
We use the nondimensional parameters with respect to the following reference parameters:
$$\rho_{ref}=\rho_{g,L}, \quad U_{ref}=\sqrt{\frac{\gamma p_{g,L}}{\rho_{g,L}}},\quad T_{ref}=0.5T_{g,L},$$
and the reference length is the computational domain.}
The initial condition is shown in table \ref{initial-1d}.
\begin{table}[!h]
  \centering
  \caption{Initial condition for wind-sand shock tube problem}
  \vspace{3mm}
\begin{tabular}{|l |c c c c| c c c c|}
  \hline
  Phase  & $\rho_{g,L}/\epsilon_{s,L}\rho_{s,L}$ & $U_L$ & $p_L$ &$T_{g,L}/T_{M,L}$ & $\rho_{g,R}/\epsilon_{s,R}\rho_{s,R} $ & $ U_R $ & $ p_R$  &$T_{g,R}/T_{M,R}$\\ \hline
  Gas & 1.0 & 0 & 1.0 & 2.0 &0.125 & 0 & 0.1 & 1.6 \\
  Solid & 0.5 & 0 & 0.5 & 2.0 &0.5 & 0 & 0.5 & 1.6\\
  \hline
\end{tabular}
\label{initial-1d}
\end{table}
The nondimensional gravitational acceleration is $\vec{g}=-0.1\vec{x}$; the nondimensional restitution coefficient is $r=0.9999$; the nondimensional solid heat capacity is $C_s=0.1$, and the inelastic collision energy transfer coefficient is $r_{Tm}=1.0$.
The physical space is divided into $500$ cells and the particle velocity space is $[-3.0,3.0]$ divided into $80$ cells.
The Knudsen number of gas phase is $Kn_g=10^{-4}$, and two Knudsen numbers are considered for solid phase, namely $Kn_s=10^{-4}$, and $Kn_s=1.0$.
{
For the continuum regime with $Kn_s=10^{-4}$, the two-fluid Navier-Stokes system is calculated by GKS \cite{xu2001}, which serves as reference solution.
For large Knudsen number, the kinetic equation is solved by discrete ordinate method (DOM) under a fine mesh, which provides the cell converged kinetic solution.}
The solutions of UGKS-M compared with reference solutions in Fig. \ref{shocktube1}-\ref{shocktube4}.
For this test case, other than the Knudesn number, two more important parameters are
the particle response time $\tau_{st}$ and the heat conduction time scale $\tau_T$.
Firstly, we set the parameters $Kn_s=10^{-4}, \tau_{st}=10, \tau_T=10$, and compare the density, velocity, pressure, gas temperature, solid granular temperature, and solid material temperature with the two-fluid NS solutions at $t=0.2$. As shown in Fig. \ref{shocktube1}, the UGKS-M solutions are shown in symbols and the reference solution in lines, and good agreements are shown in such flow regime.
{It can be observed comparing with the pure gas solution (dotted lines), the solid phase is slightly driven by the gas phase. Due to the interspecies fraction, the gas is heated as the gas temperature is higher than the pure gas case, while the solid phase granular temperature decreases. The material temperature of solid phase doesn't change much, as the gas-particle heat conduction is weak.}
Secondly, we keep $Kn_s=10^{-4}$ and decrease the Stokes number to $\tau_{st}=0.1$ and $\tau_{T}=0.1$, and compare the density, velocity, pressure, gas temperature, solid granular temperature and solid material temperature with the two-fluid NS solutions at $t=0.1,0.2$. Good agreement can be observed in Fig. \ref{shocktube2} and \ref{shocktube3}.
{In such case, the momentum and energy transfer between species is enhanced, which leads to a lower granular temperature. The decrease of granular temperature is on one hand due to the fraction, and on the other due to the inelastic collision of solid particles. In this test case, the energy loss in the collision process purely increases the material temperature of solid particle.} We also compare the velocity distribution of solid particle with the local Maxwellian distribution at $x=0.5$. In such regime, two distribution functions agree well and the two-fluid model holds.
Lastly, we keep $\tau_{st}=0.1, \tau_{T}=0.1$, and increase the Knudsen number $Kn_s=1.0$, and
{compare with the kinetic solution with 2000 cells}. As shown in Fig. \ref{shocktube4}, the UGKS-M performs well in the rarefied regime. It is shown that velocity distribution of solid phase is a leptokurtic distribution and deviates from the local Maxwellian distribution. Since the close to equilibrium assumption is violated, the hydrodynamic two-fluid modeling will not properly describe the flow dynamics.

\subsection{Particle jets impinging problem}
\begin{figure}
\centering
\includegraphics[width=0.45\textwidth]{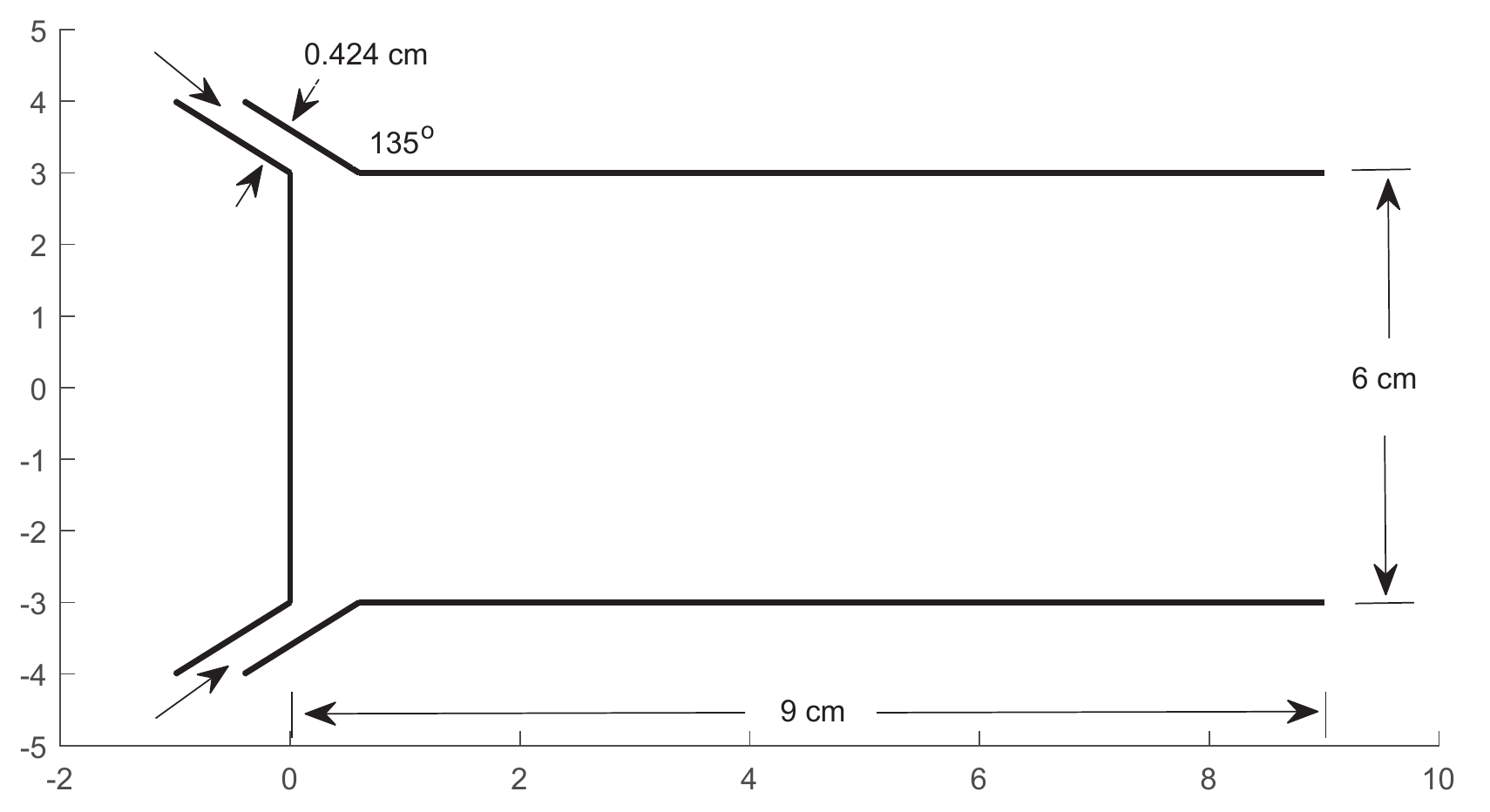}
\includegraphics[width=0.5\textwidth]{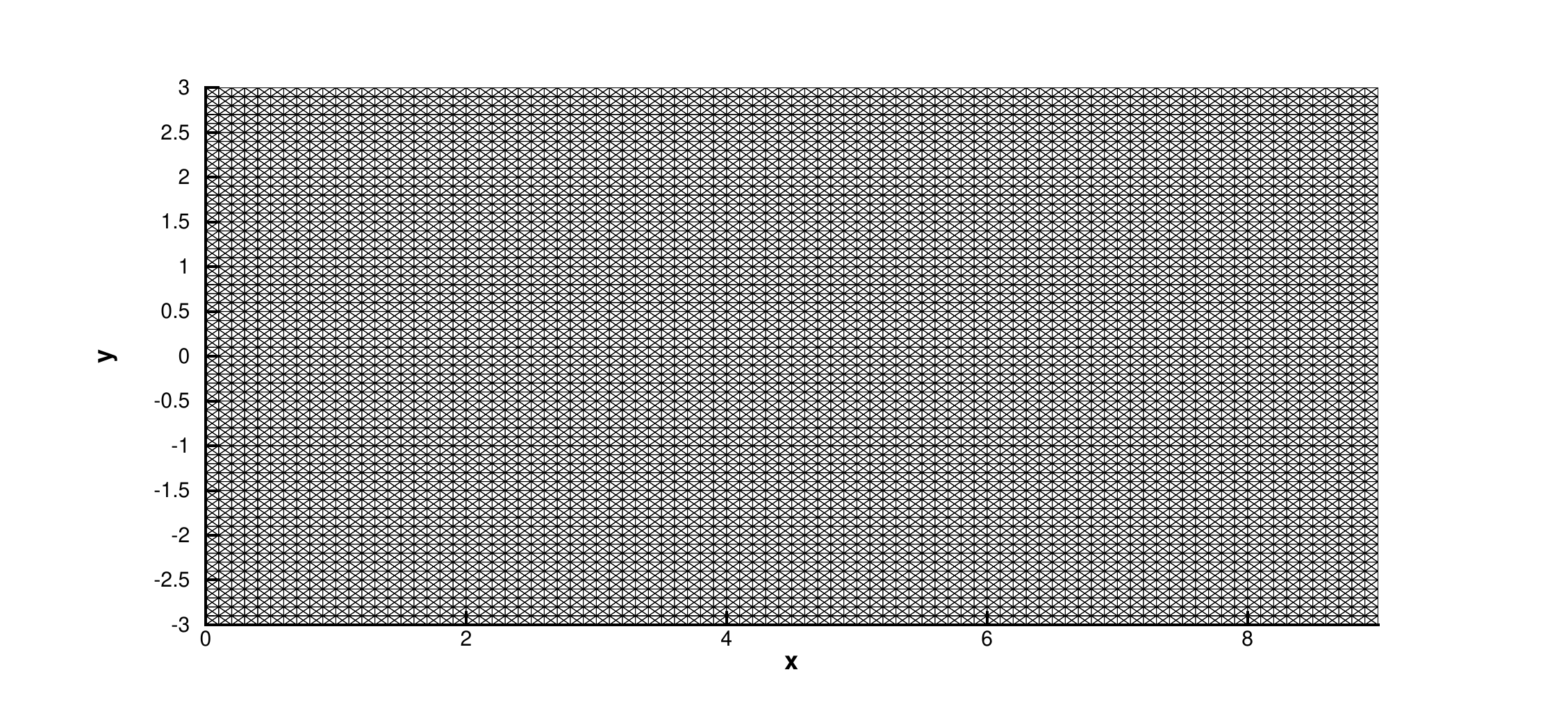}
\caption{Channel geometry for the calculation of two impinging particle jets.}
\label{jets-initial}
\end{figure}

\begin{figure}
\centering
\includegraphics[width=0.45\textwidth]{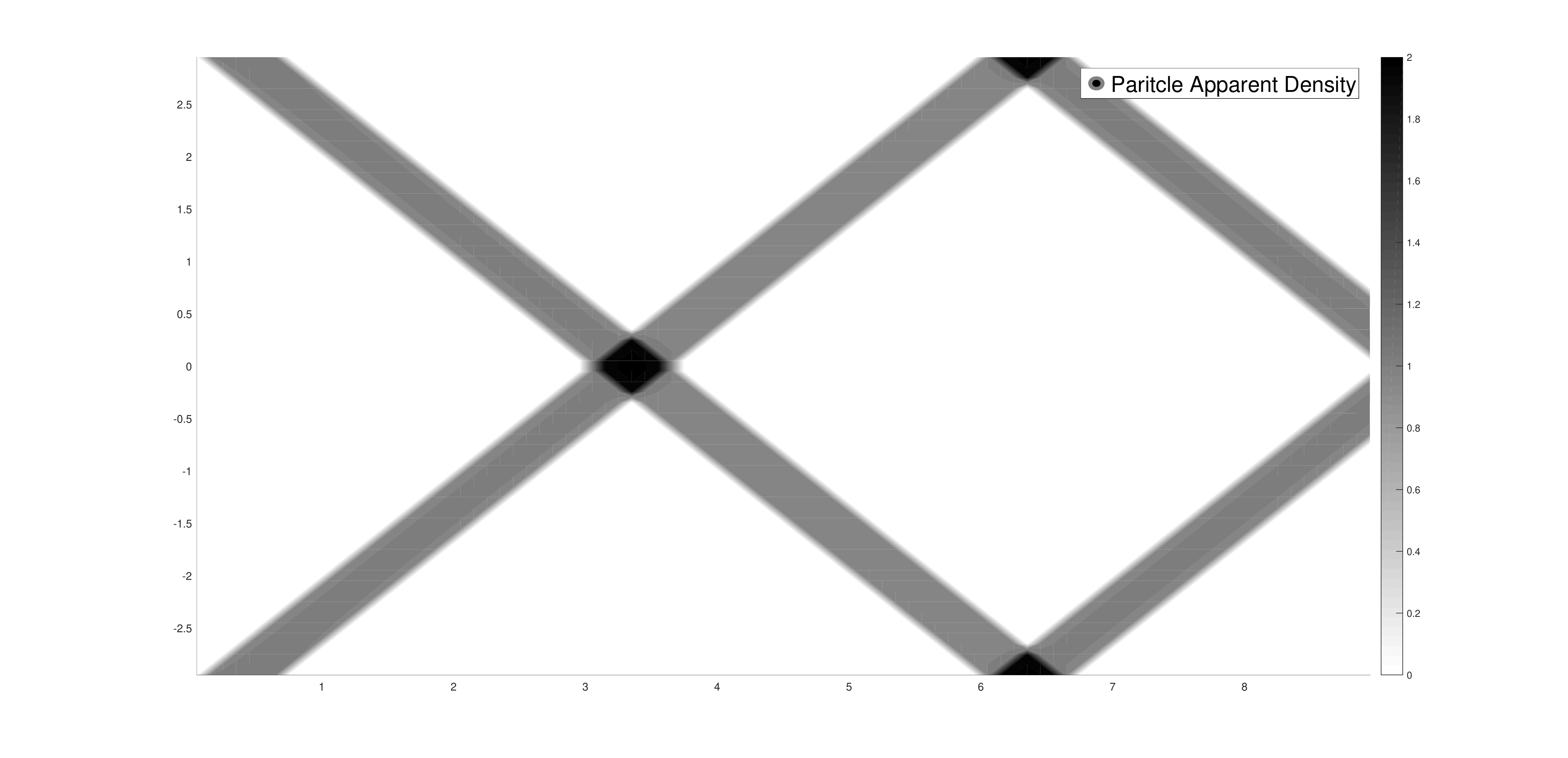}{a}
\includegraphics[width=0.48\textwidth]{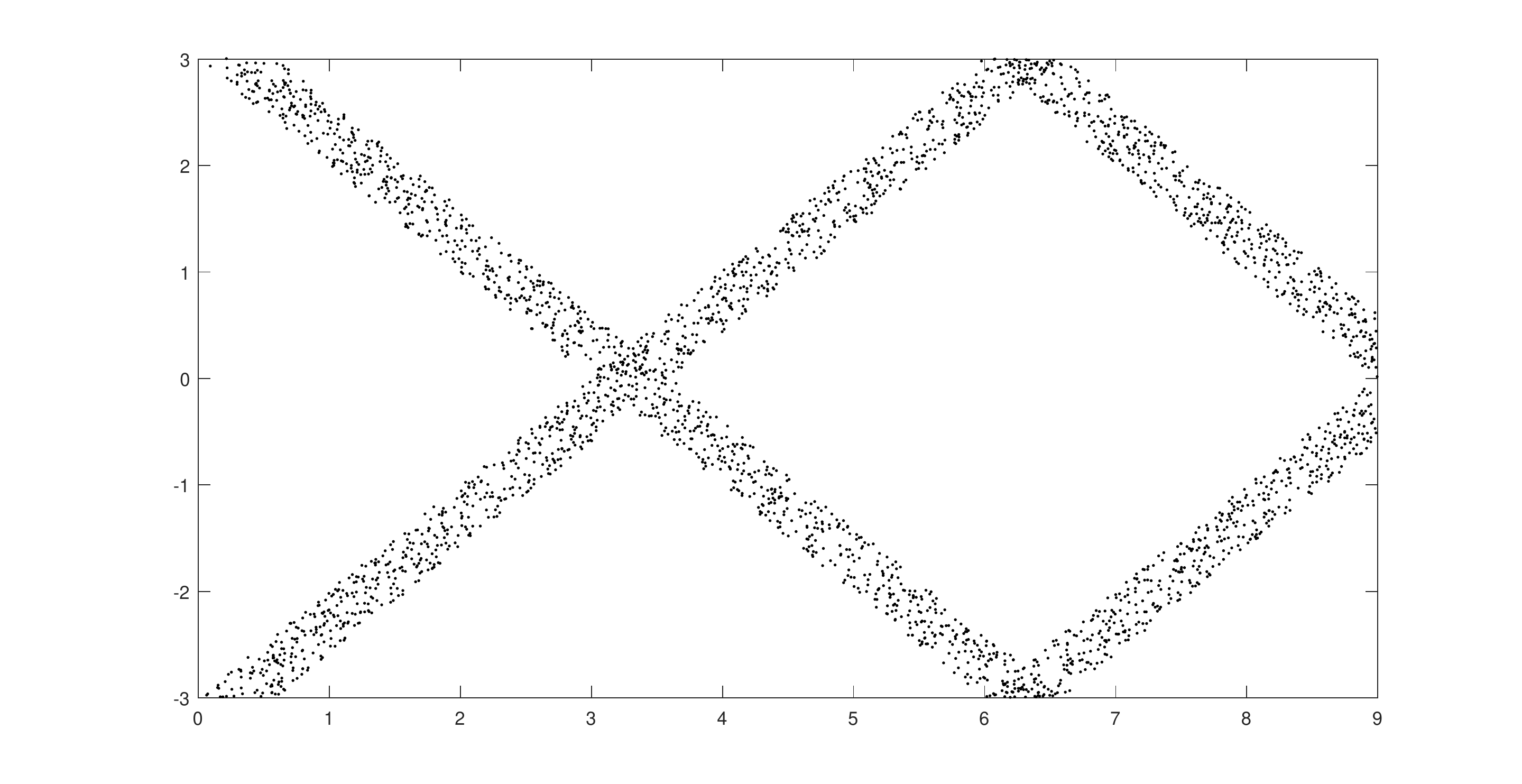}{b}
\caption{Particle number density at t=20 for collisionless regime: (a) UGKS-M result, and (b) PIC result.}
\label{jets1}
\end{figure}

\begin{figure}
\centering
\includegraphics[width=0.45\textwidth]{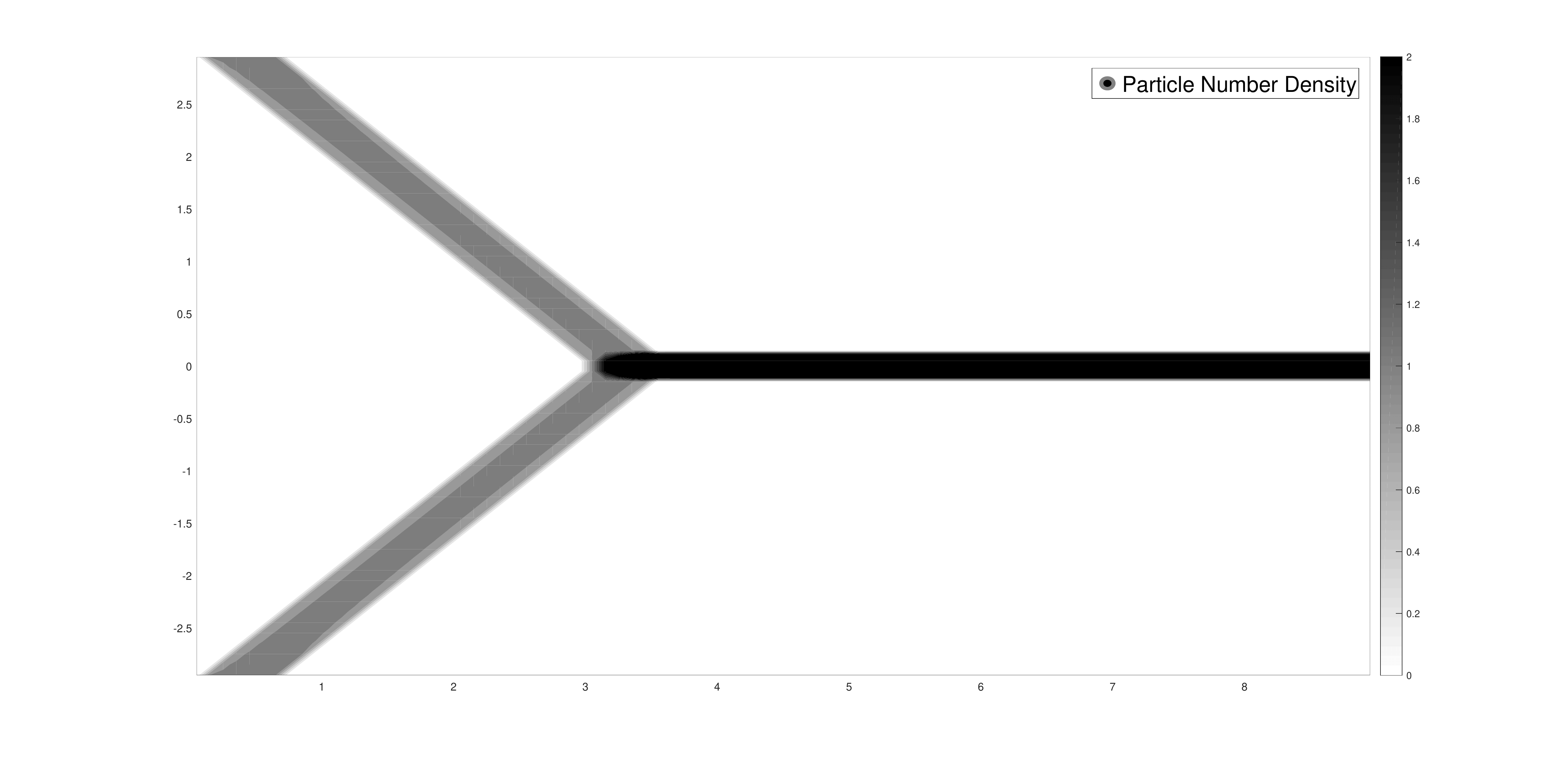}{a}
\includegraphics[width=0.48\textwidth]{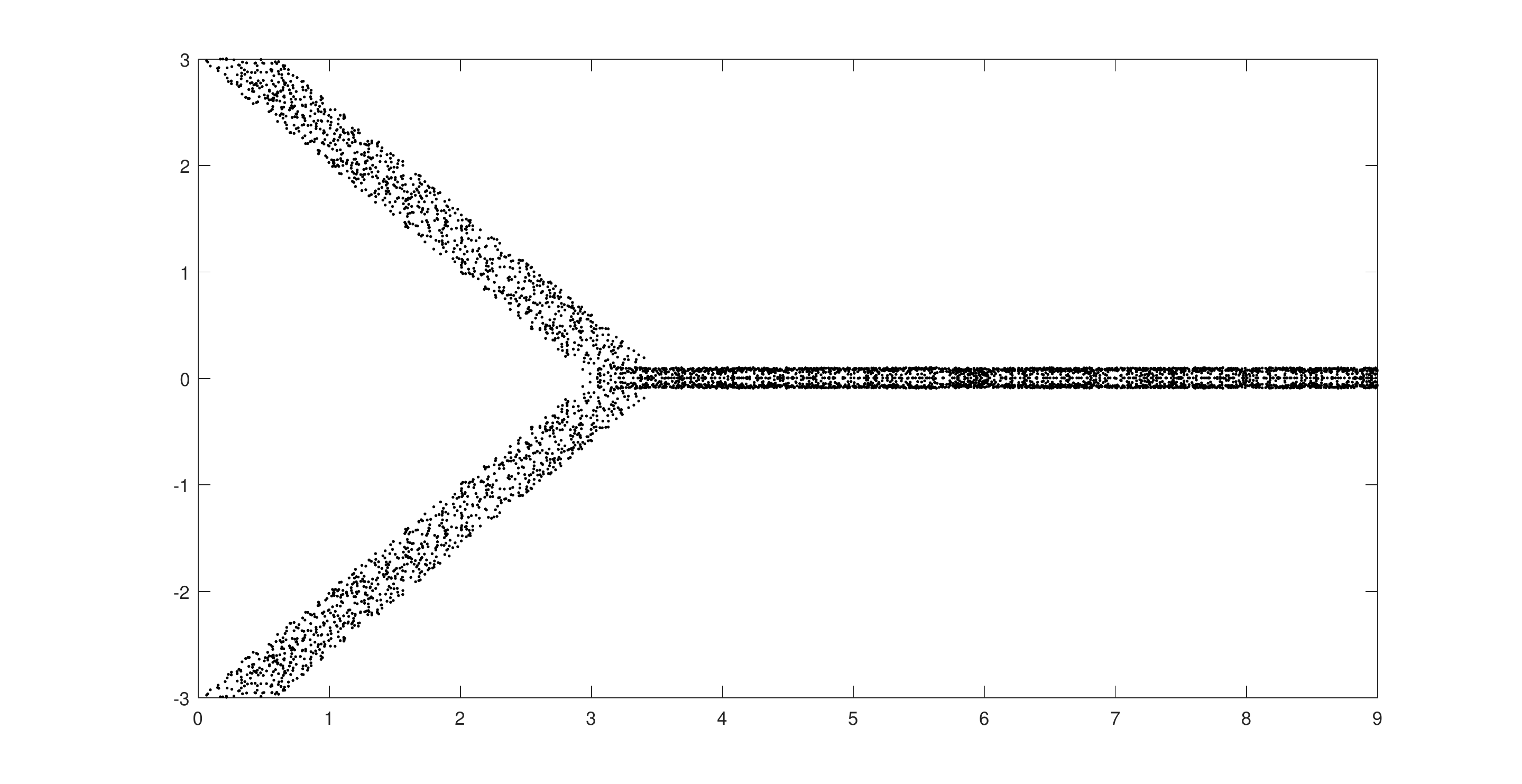}{b}
\caption{Particle number density at t=20 for highly collision regime at $Kn_s=1.0\times 10^{-4}$ and $r=0$ : (a) UGKS-M result, and (b) PIC result.}
\label{jets2}
\end{figure}

\begin{figure}
\begin{minipage}[t]{0.48\textwidth}
\centering
\includegraphics[width=\textwidth]{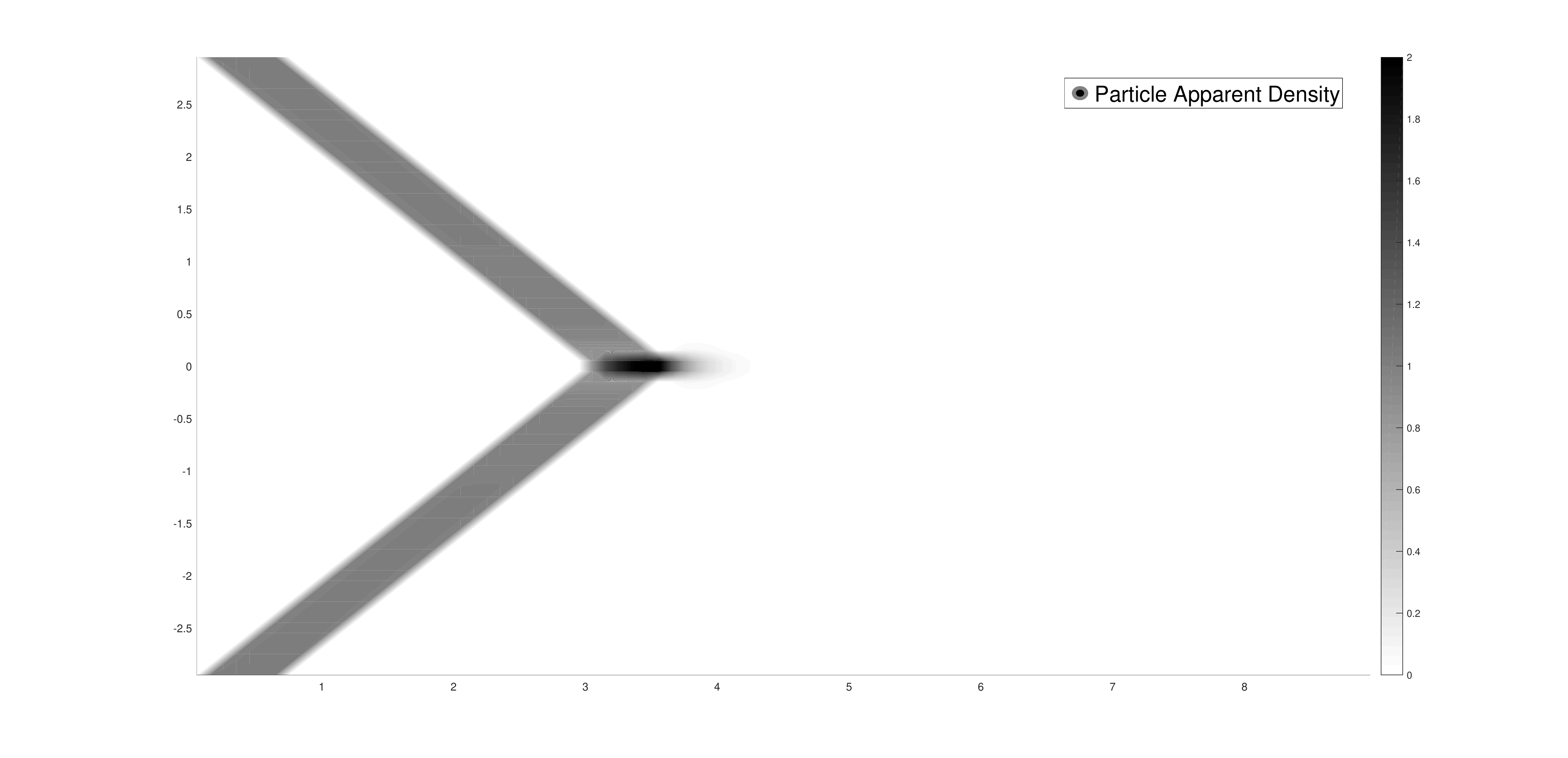}
\centerline{\footnotesize (a) t=4.4}
\end{minipage}
\hfill
\begin{minipage}[t]{0.48\textwidth}
\centering
\includegraphics[width=\textwidth]{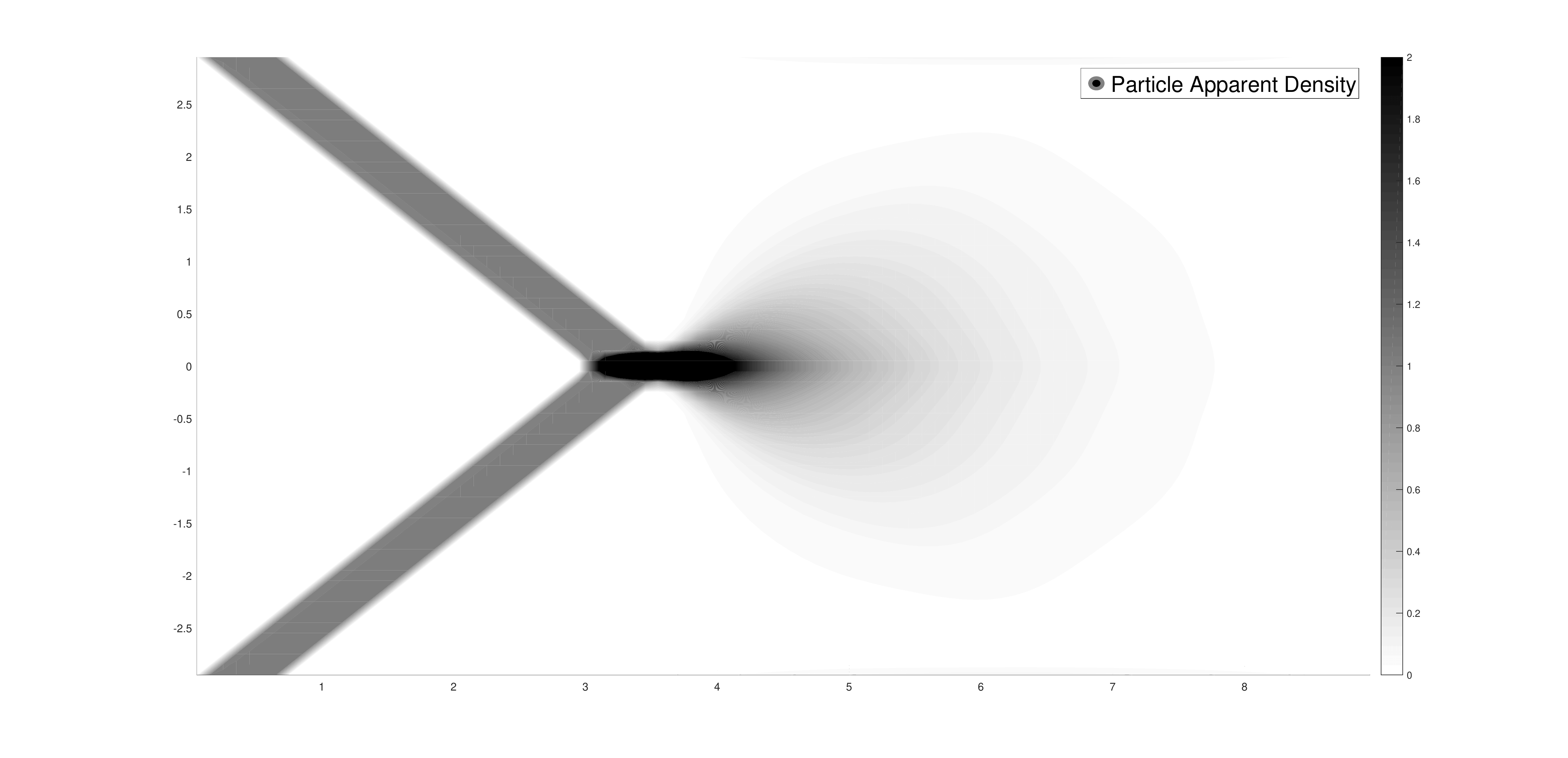}
\centerline{\footnotesize (b) t=8.8}
\end{minipage}
\vfill
\begin{minipage}[t]{0.48\textwidth}
\centering
\includegraphics[width=\textwidth]{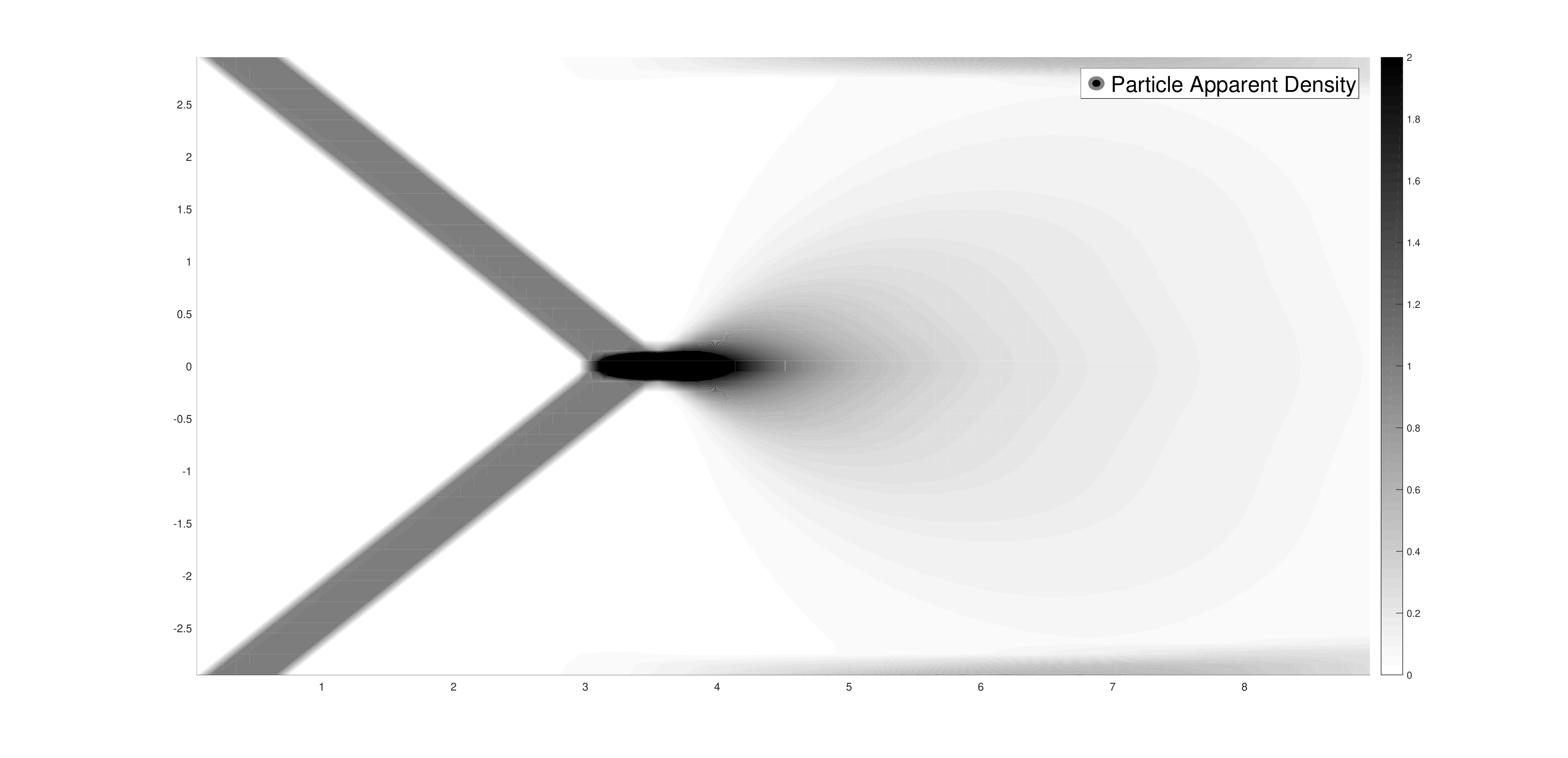}
\centerline{\footnotesize (c) t=13.2}
\end{minipage}
\hfill
\begin{minipage}[t]{0.48\textwidth}
\centering
\includegraphics[width=\textwidth]{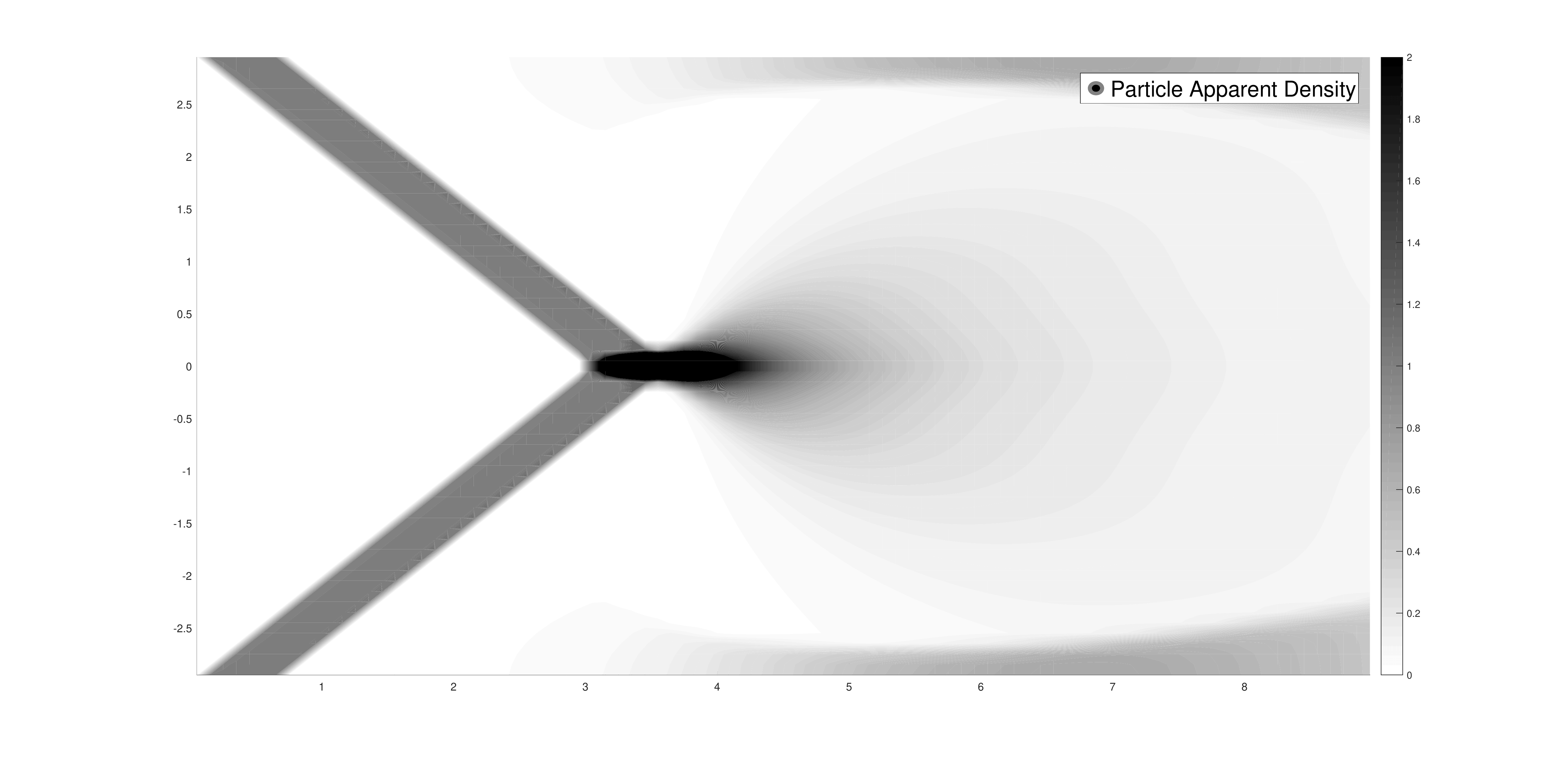}
\centerline{\footnotesize (d) t=20}
\end{minipage}
\caption{Particle number density at different output times with $Kn_s=1.0\times 10^{-4}$ and $r=0.4$.}
\label{jets3}
\end{figure}

\begin{figure}
\begin{minipage}[t]{0.48\textwidth}
\centering
\includegraphics[width=\textwidth]{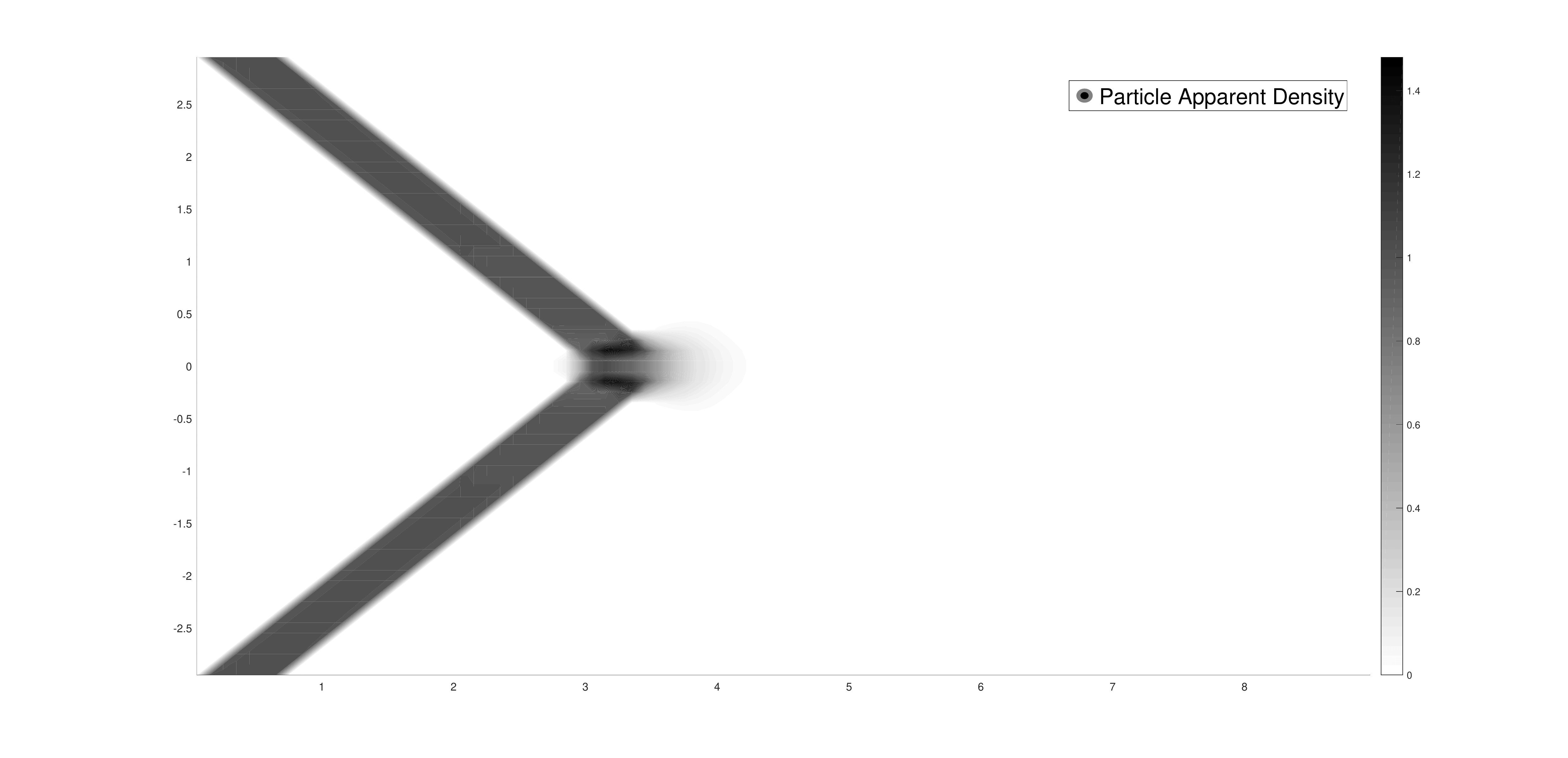}
\centerline{\footnotesize (a) t=4.4}
\end{minipage}
\hfill
\begin{minipage}[t]{0.48\textwidth}
\centering
\includegraphics[width=\textwidth]{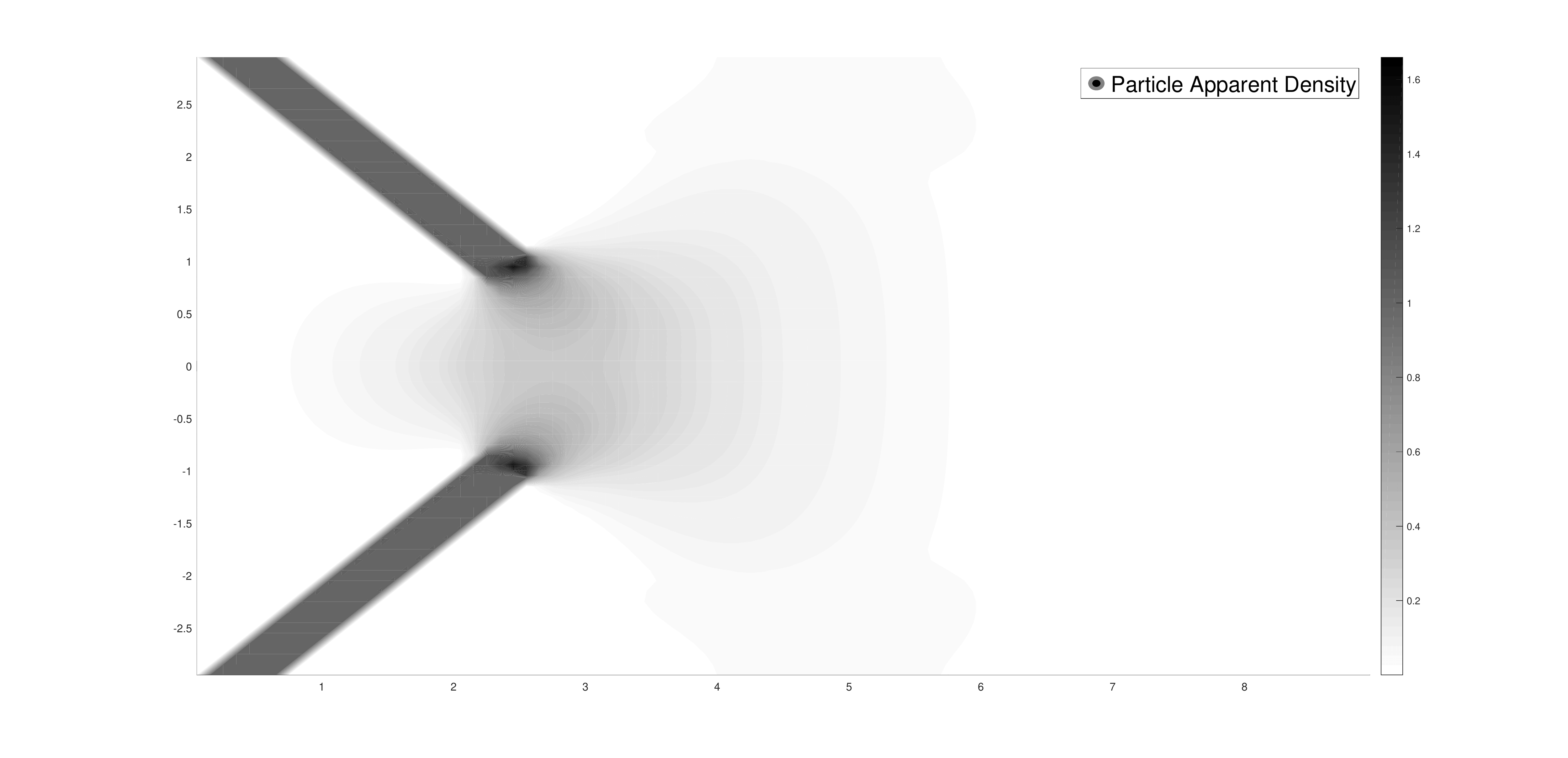}
\centerline{\footnotesize (b) t=8.8}
\end{minipage}
\vfill
\begin{minipage}[t]{0.48\textwidth}
\centering
\includegraphics[width=\textwidth]{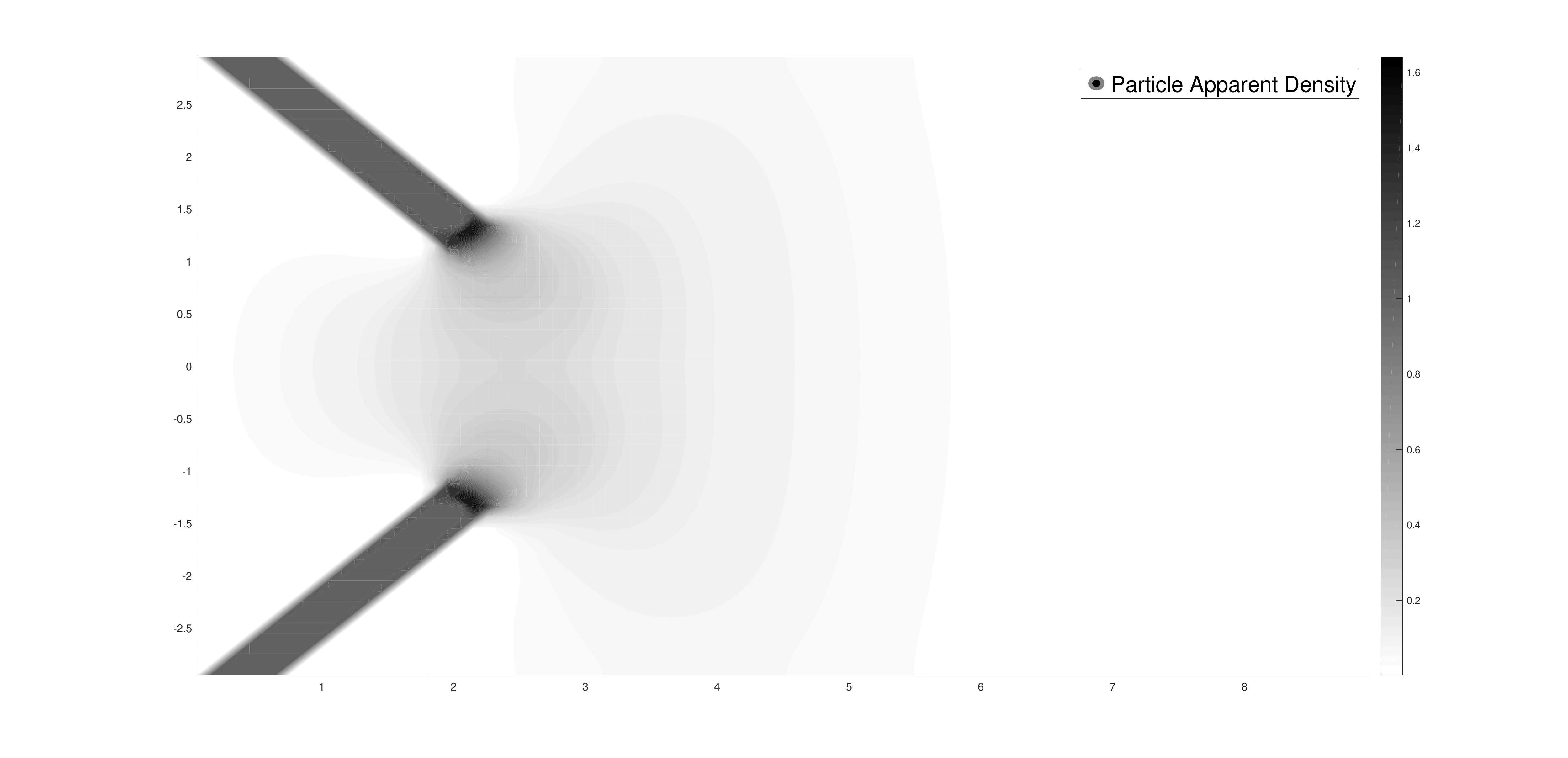}
\centerline{\footnotesize (c) t=13.2}
\end{minipage}
\hfill
\begin{minipage}[t]{0.48\textwidth}
\centering
\includegraphics[width=\textwidth]{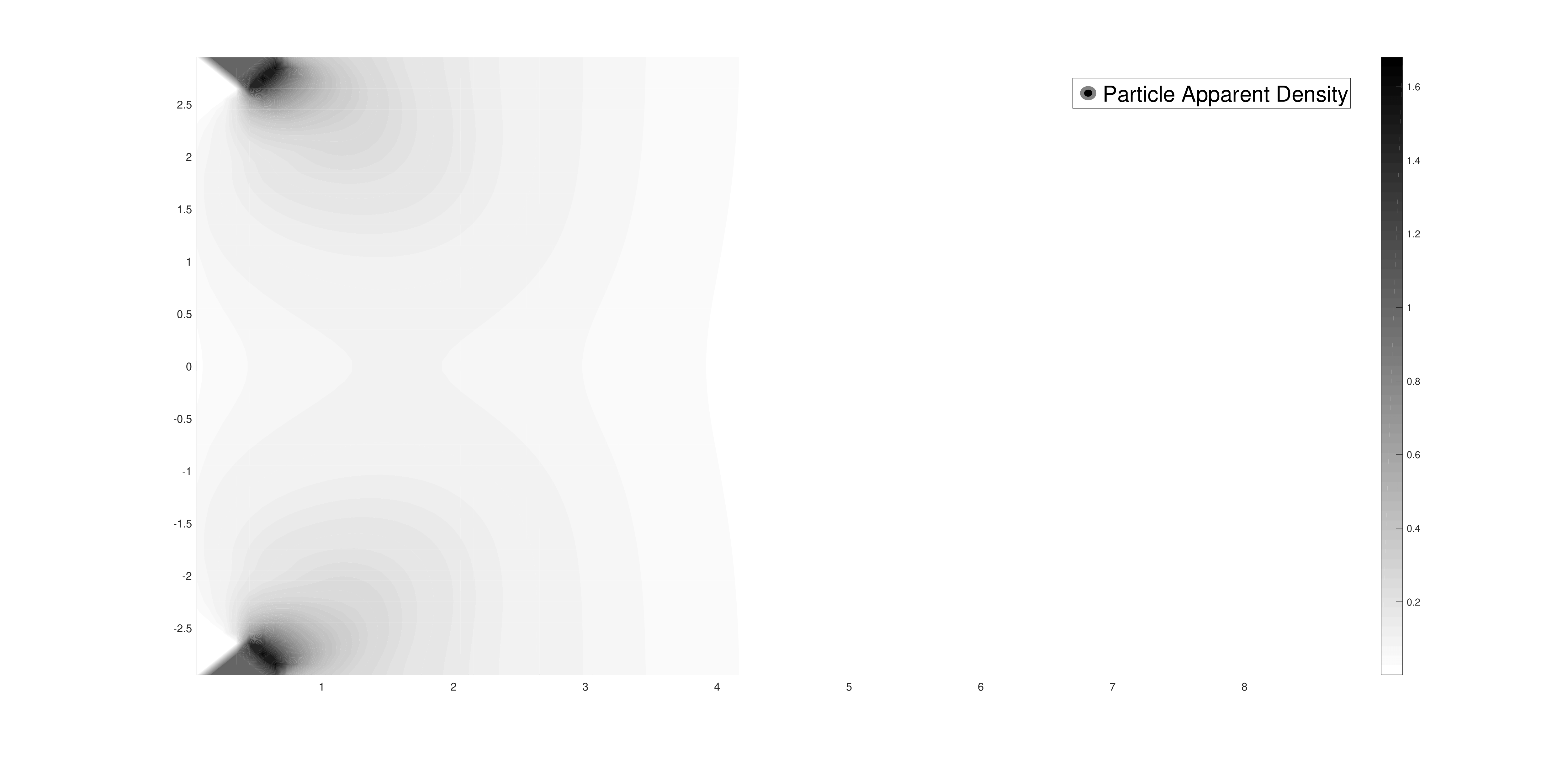}
\centerline{\footnotesize (d) t=20}
\end{minipage}
\caption{Particle number density at different output times with $Kn_s=1.0\times 10^{-4}$ and $r=1.0$.}
\label{jets4}
\end{figure}

The particle trajectory crossing (PTC) and particle wall reflecting are two important tests to show the ability of the numerical scheme in capturing the rarefied particle flow. The hydrodynamic models fail to capture these two phenomena and gives nonphysical $\delta$-shock \cite{marchisio2013computational}.
In this example, we calculate the problem of two particle jets impinging into a rectangular chamber to demonstrate the ability of the UGKS-M to capture the PTC and particle wall reflecting in two-dimensional flows.
To omit the influence of gas phase, we set the Stokes number infinity.
{The channel geometry as well as the mesh geometry is shown in Fig. \ref{jets-initial}, the mesh we use is an unstructured mesh with $\Delta x=0.1$.}
Initially, two particle jets are injected from left top and left bottom corner of a rectangular chamber with adiabatic wall.
The apparent density of the jet flow is used as the reference apparent density and set $\epsilon_s\rho_s=1$,
and the injection velocity is along $135^\text{o}$ and $225^\text{o}$ directions with respect to the positive x-axis.
The velocity magnitude of the jet is used as the reference velocity and set $|\vec{U}_s|=1.0$; and granular temperature is $T_s=0$.
The velocity space is $[-\sqrt{2},\sqrt{2}]$ divided into $17\times17$ cells.
Four sets of particle Knudsen number and restitution coefficient are calculated.
{
Firstly, we calculate the collisionless regime with infinite Knudsen number and compare with the PIC result,
the distribution of particle apparent density at $t=20$ is shown in Fig. \ref{jets1}.
It can be observed that UGKS-M recovers the physical consistent PTC and wall reflecting phenomena.
Then we decrease the particle Knudsen number to $1.0\times 10^{-4}$, and set the restitution coefficient $r=0$.
The distribution of particle apparent density at $t=20$ is shown in Fig. \ref{jets2}, compared with the PIC result.
In such a situation, two solid particles will share same speed after collision, and the two particle jets merge into a single one.}
Next, we increase the restitution coefficient to $r=0.4$ and $r=1.0$, the density contours are shown in Fig. \ref{jets3}-\ref{jets4}.
The particle scattering effect appears and the particles fill the chamber due to elastic collision.
\subsection{Particle segregation in Taylor-Green flow}
\begin{figure}
\centering
\includegraphics[width=0.48\textwidth]{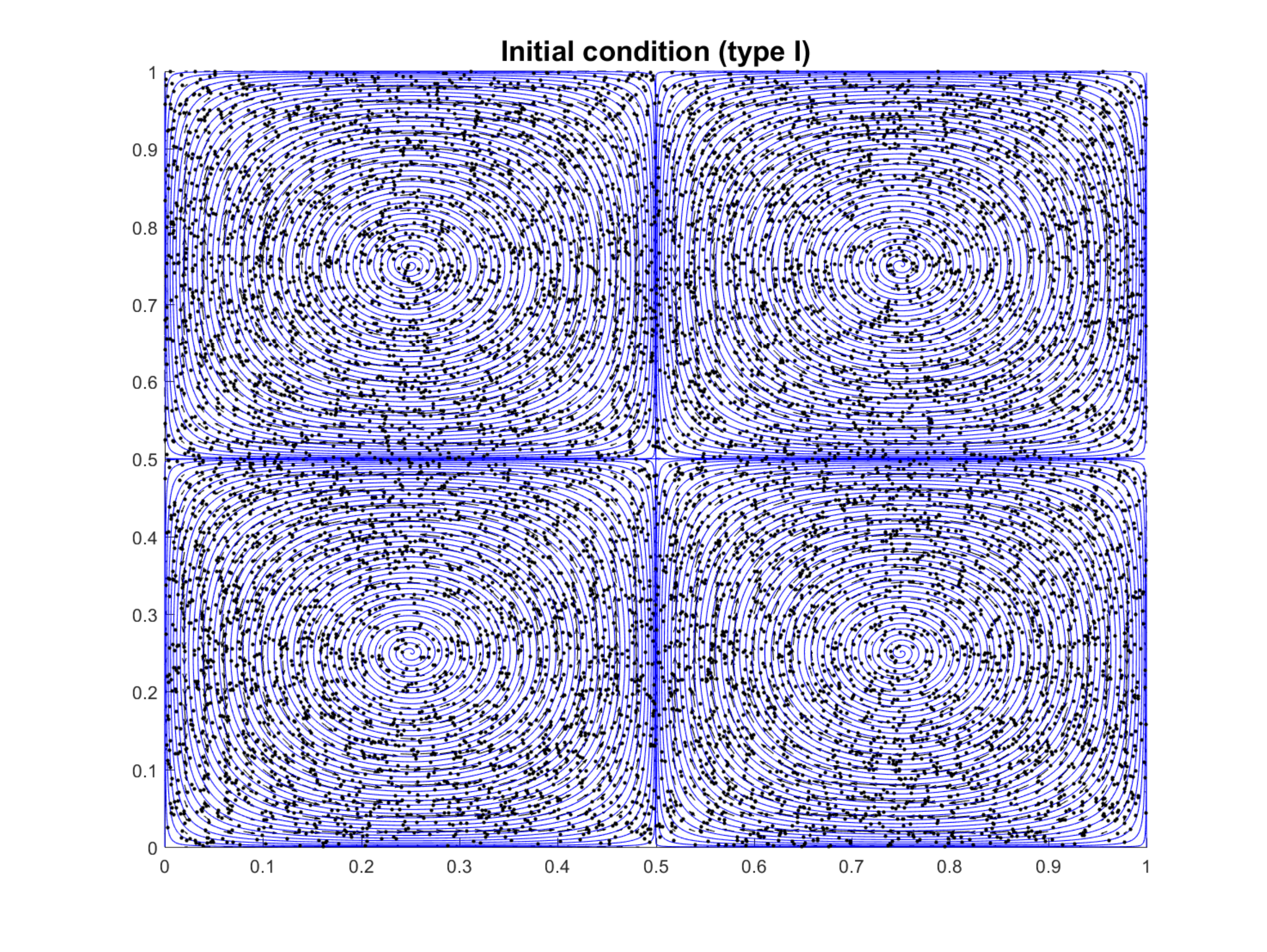}
\includegraphics[width=0.48\textwidth]{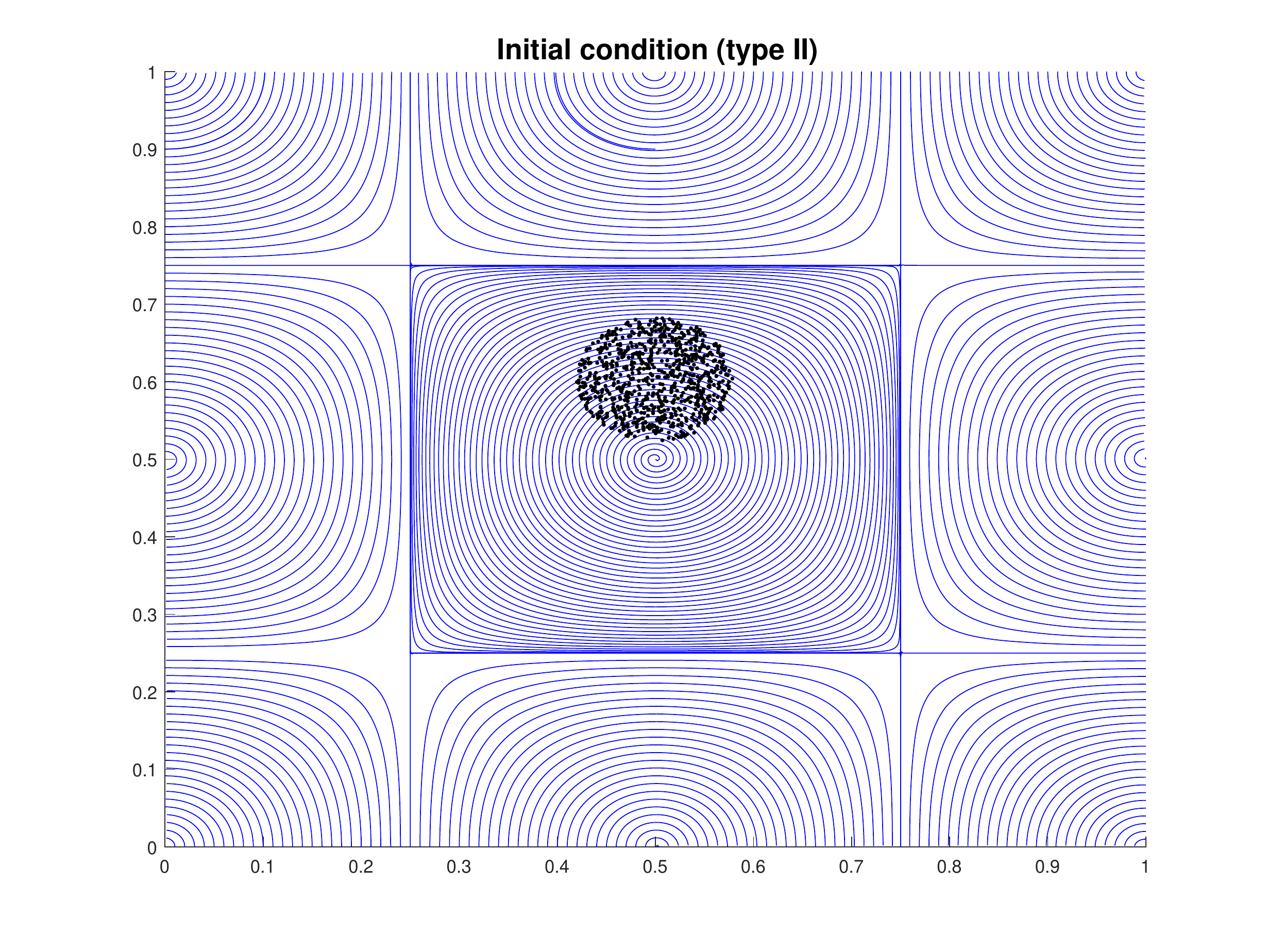}
\caption{Two initial particle apparent density and streamline used for Taylor{-}Green flow test cases.}
\label{2taylor0}
\end{figure}

\begin{figure}
\begin{minipage}[t]{0.48\textwidth}
\centering
\includegraphics[width=\textwidth]{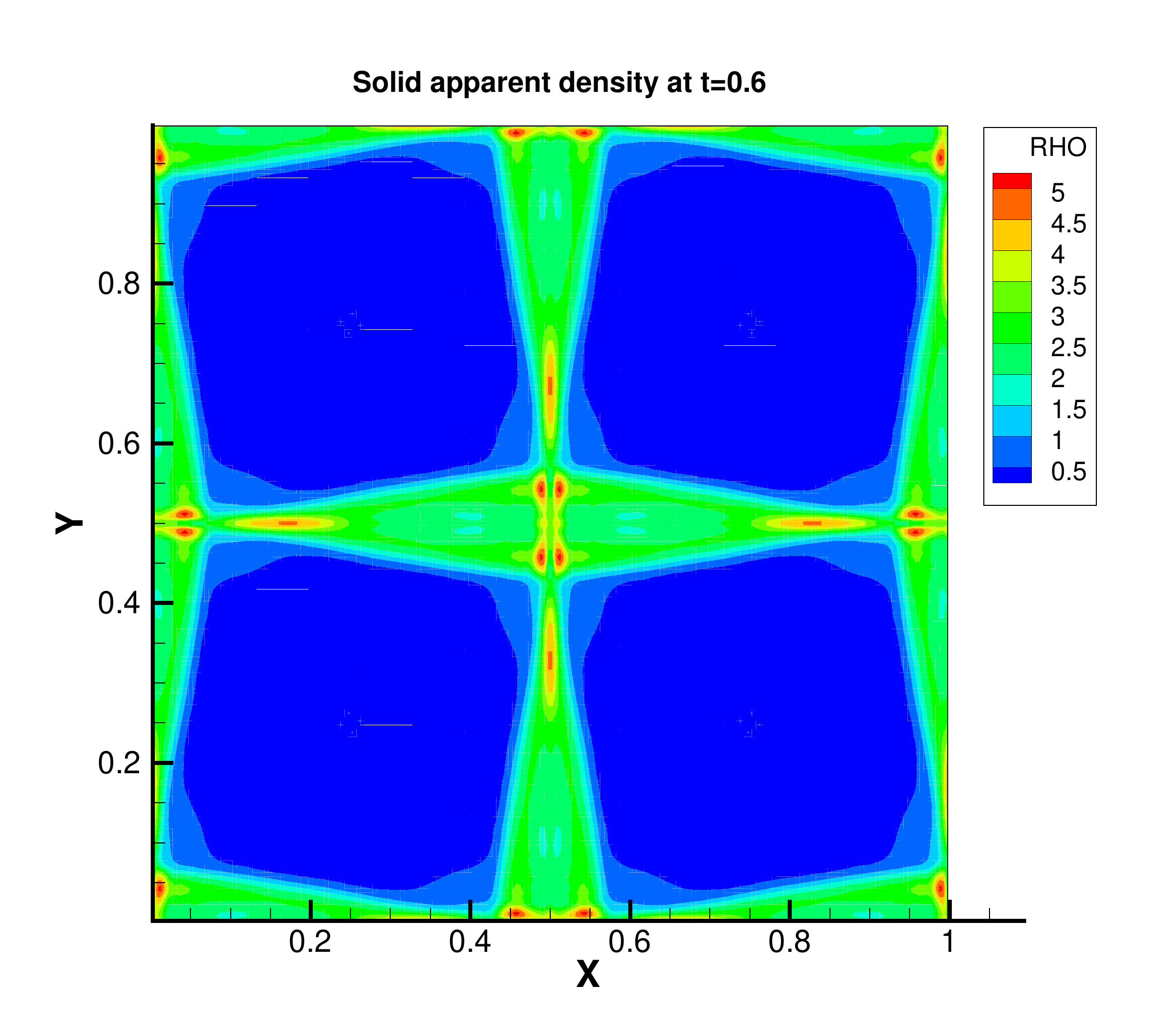}
\centerline{\footnotesize (a) UGKS-M at t=0.6}
\end{minipage}
\hfill
\begin{minipage}[t]{0.48\textwidth}
\centering
\includegraphics[width=\textwidth]{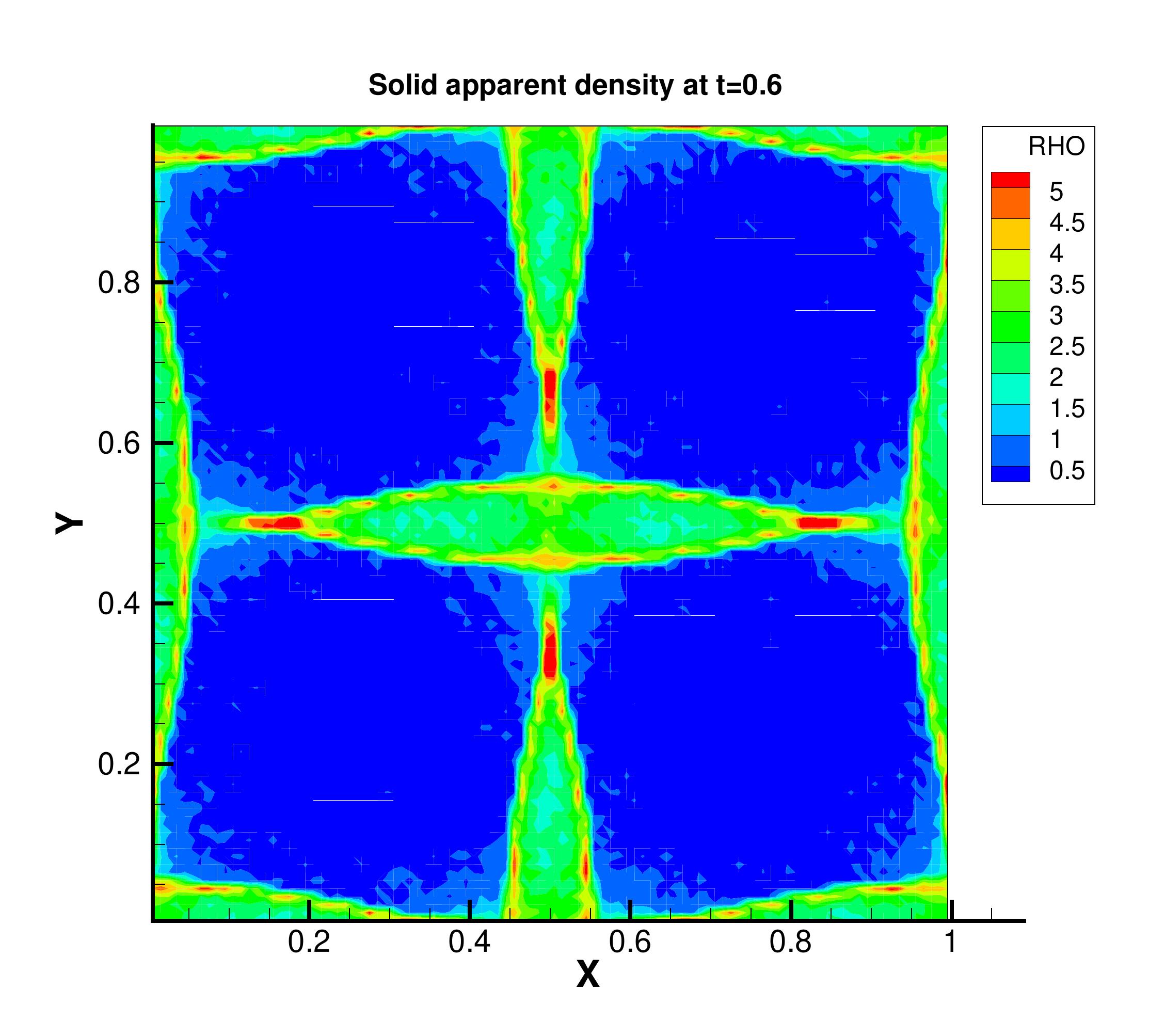}
\centerline{\footnotesize (b) PIC at t=0.6}
\end{minipage}
\vfill
\begin{minipage}[t]{0.48\textwidth}
\centering
\includegraphics[width=\textwidth]{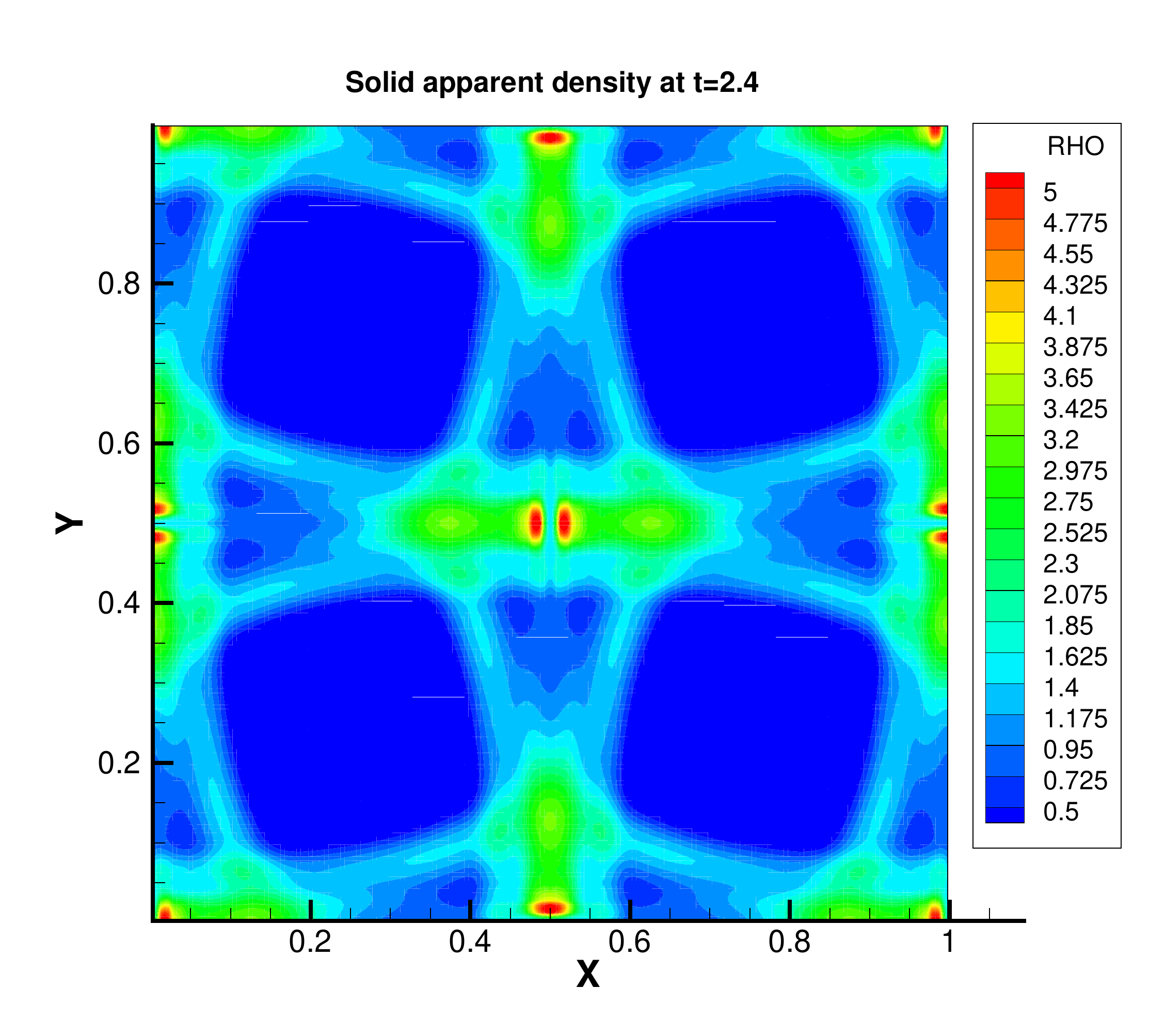}
\centerline{\footnotesize (c) UGKS-M at t=2.0}
\end{minipage}
\hfill
\begin{minipage}[t]{0.48\textwidth}
\centering
\includegraphics[width=\textwidth]{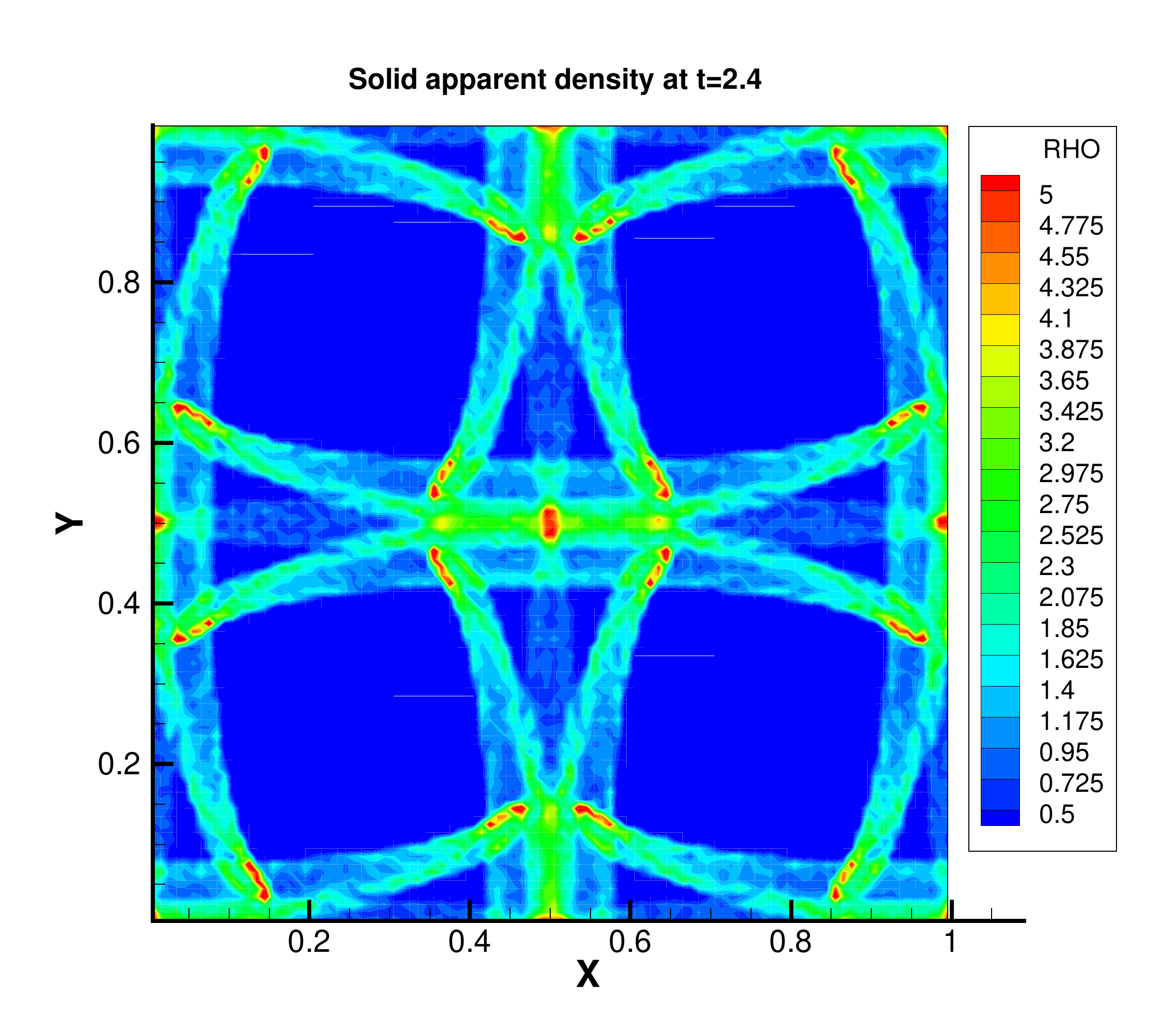}
\centerline{\footnotesize (d) PIC at t=2.0}
\end{minipage}
\caption{The UGKS-M and PIC result of particle apparent density distribution in {Taylor-Green} flow at $t=0.6$ and $t=2$, with parameter $Kn_s=10^{4}$ and $St_s=0.3$.}
\label{2taylor1}
\end{figure}

\begin{figure}
\begin{minipage}[t]{0.48\textwidth}
\centering
\includegraphics[width=\textwidth]{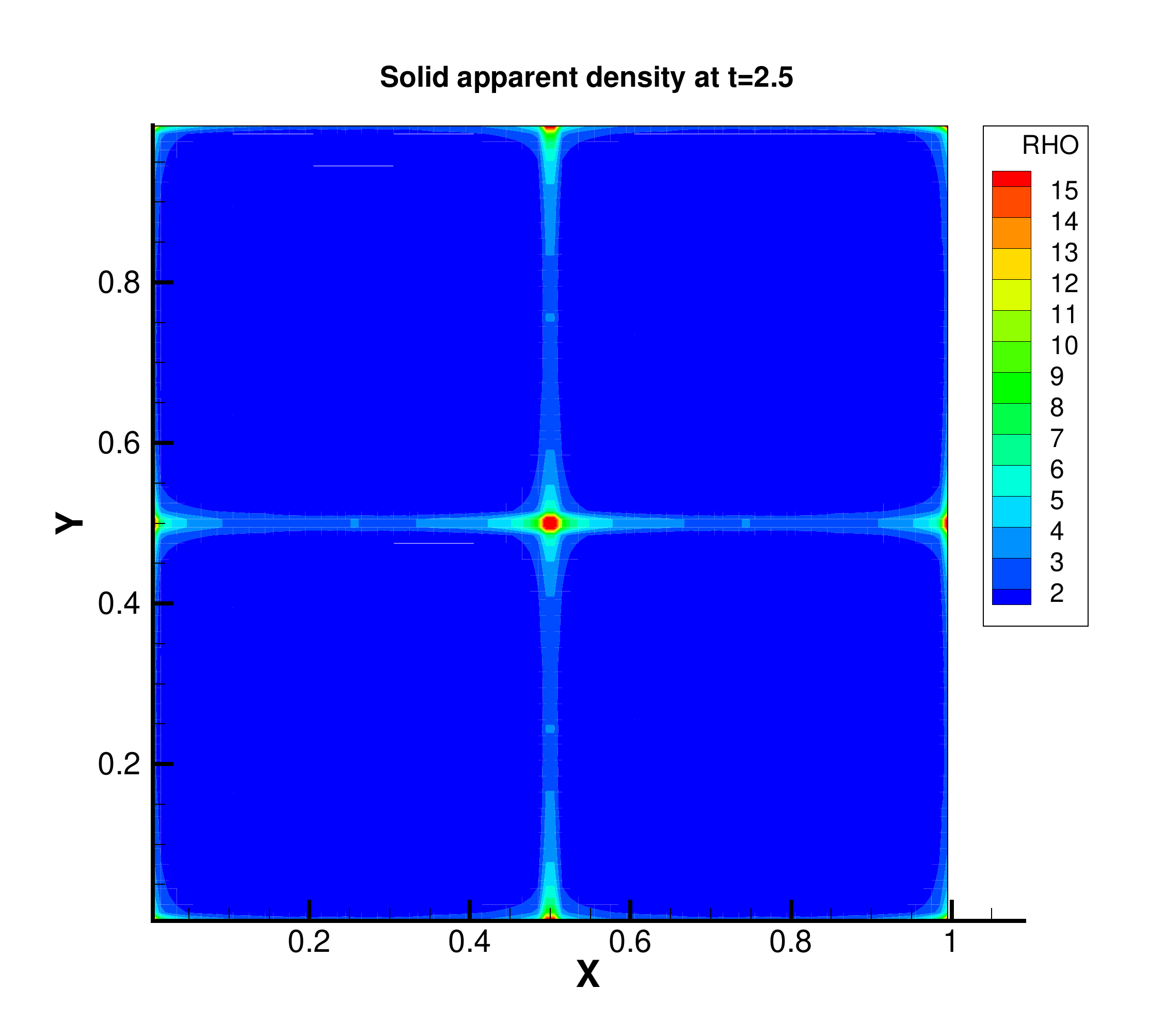}
\centerline{\footnotesize (a) UGKS-M at t=0.6}
\end{minipage}
\hfill
\begin{minipage}[t]{0.48\textwidth}
\centering
\includegraphics[width=\textwidth]{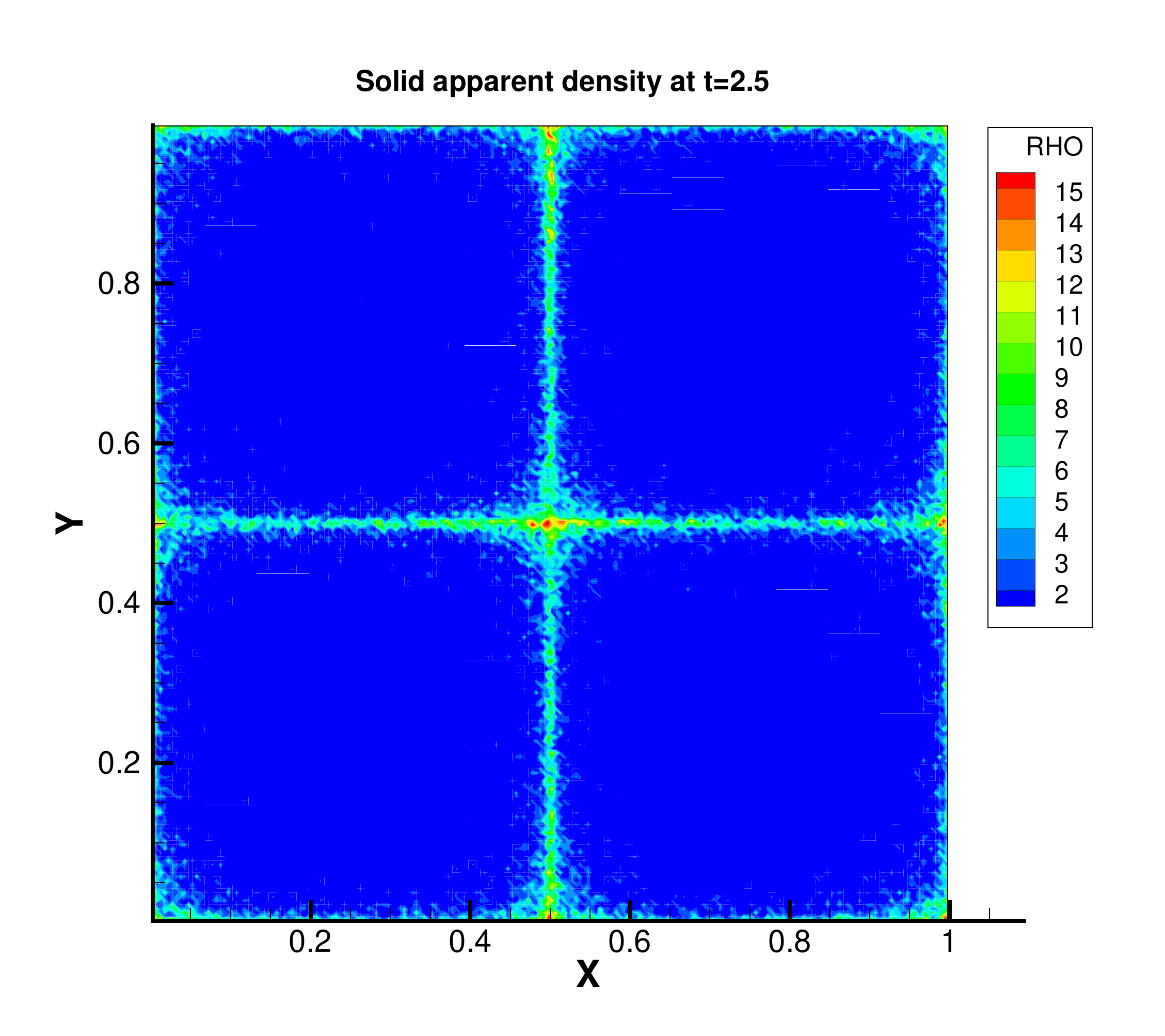}
\centerline{\footnotesize (b) PIC at t=0.6}
\end{minipage}
\caption{The UGKS-M and PIC result of particle apparent density distribution in {Taylor-Green} flow at $t=0.6$, with parameter $Kn_s=10^{4}$ and $St_s=0.03$.}
\label{2taylor2}
\end{figure}

\begin{figure}
\begin{minipage}[t]{0.48\textwidth}
\centering
\includegraphics[width=\textwidth]{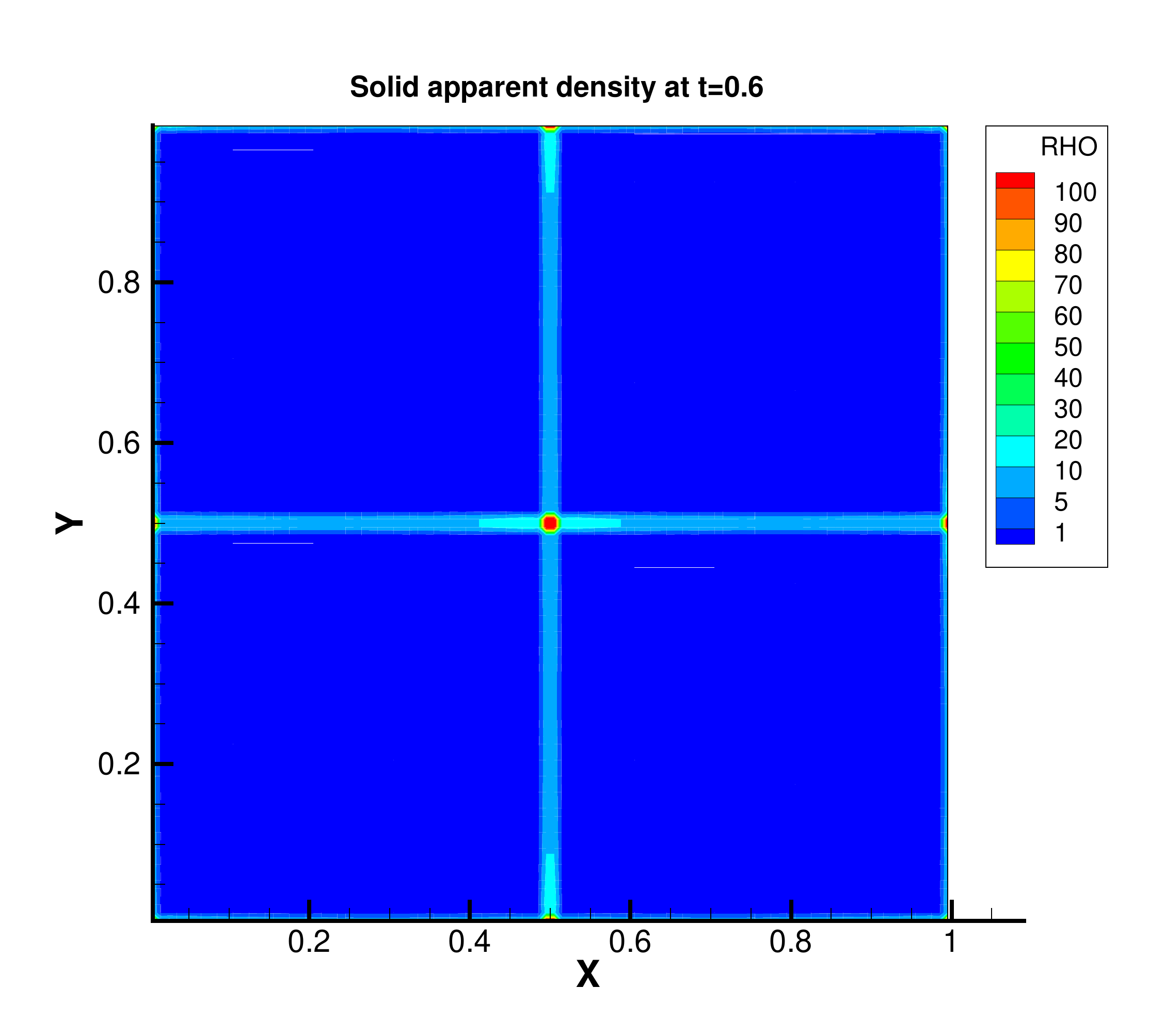}
\centerline{\footnotesize (a) UGKS-M at t=0.6}
\end{minipage}
\hfill
\begin{minipage}[t]{0.48\textwidth}
\centering
\includegraphics[width=\textwidth]{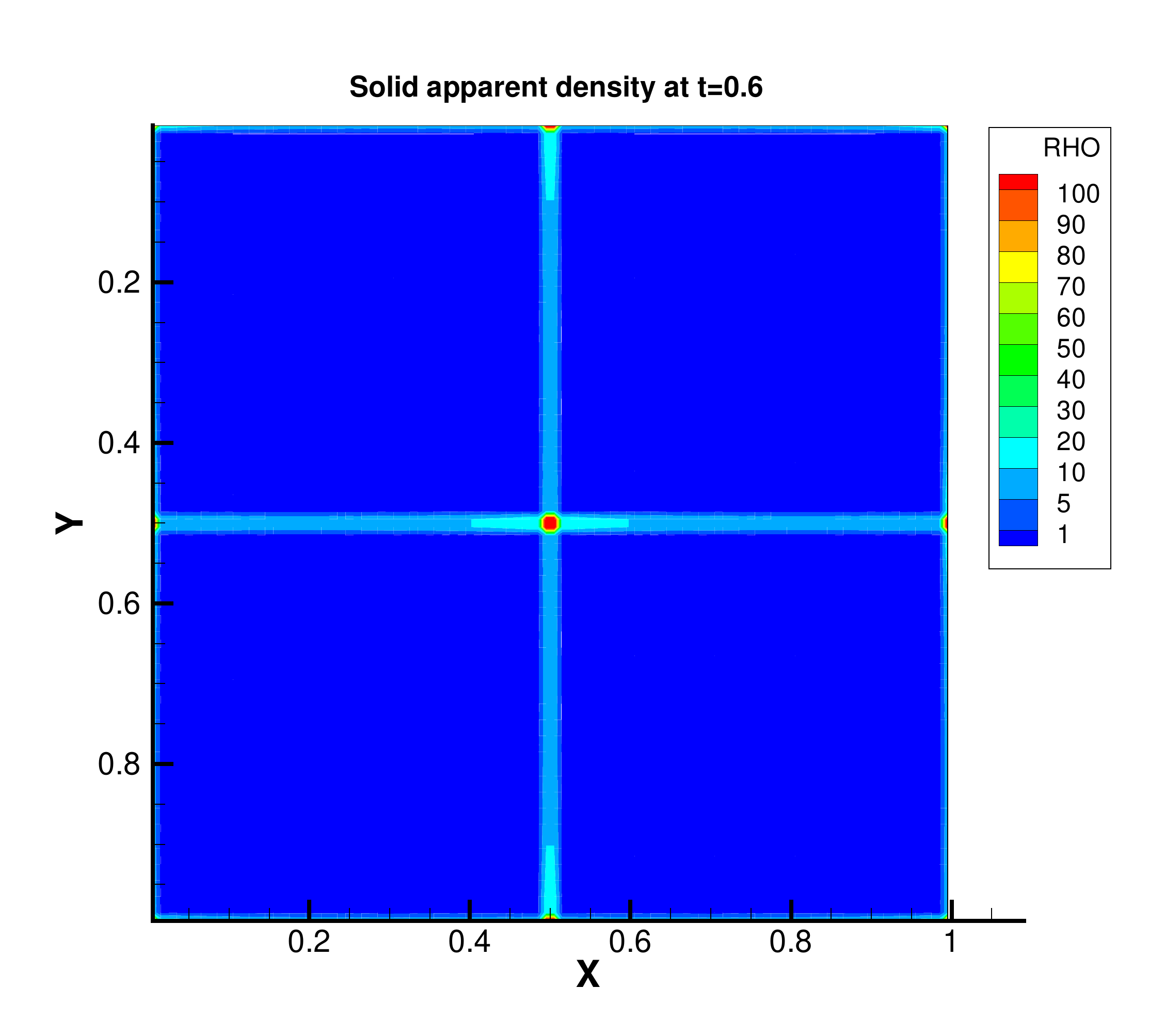}
\centerline{\footnotesize (b) Pressureless Euler at t=0.6}
\end{minipage}
\caption{The UGKS-M and pressureless Euler result of particle apparent density distribution in {Taylor-Green} flow at $t=0.6$, with parameter $Kn_s=10^{-4}$ and $St_s=0.3$.}
\label{2taylor3}
\end{figure}

\begin{figure}
\begin{minipage}[t]{0.48\textwidth}
\centering
\includegraphics[width=\textwidth]{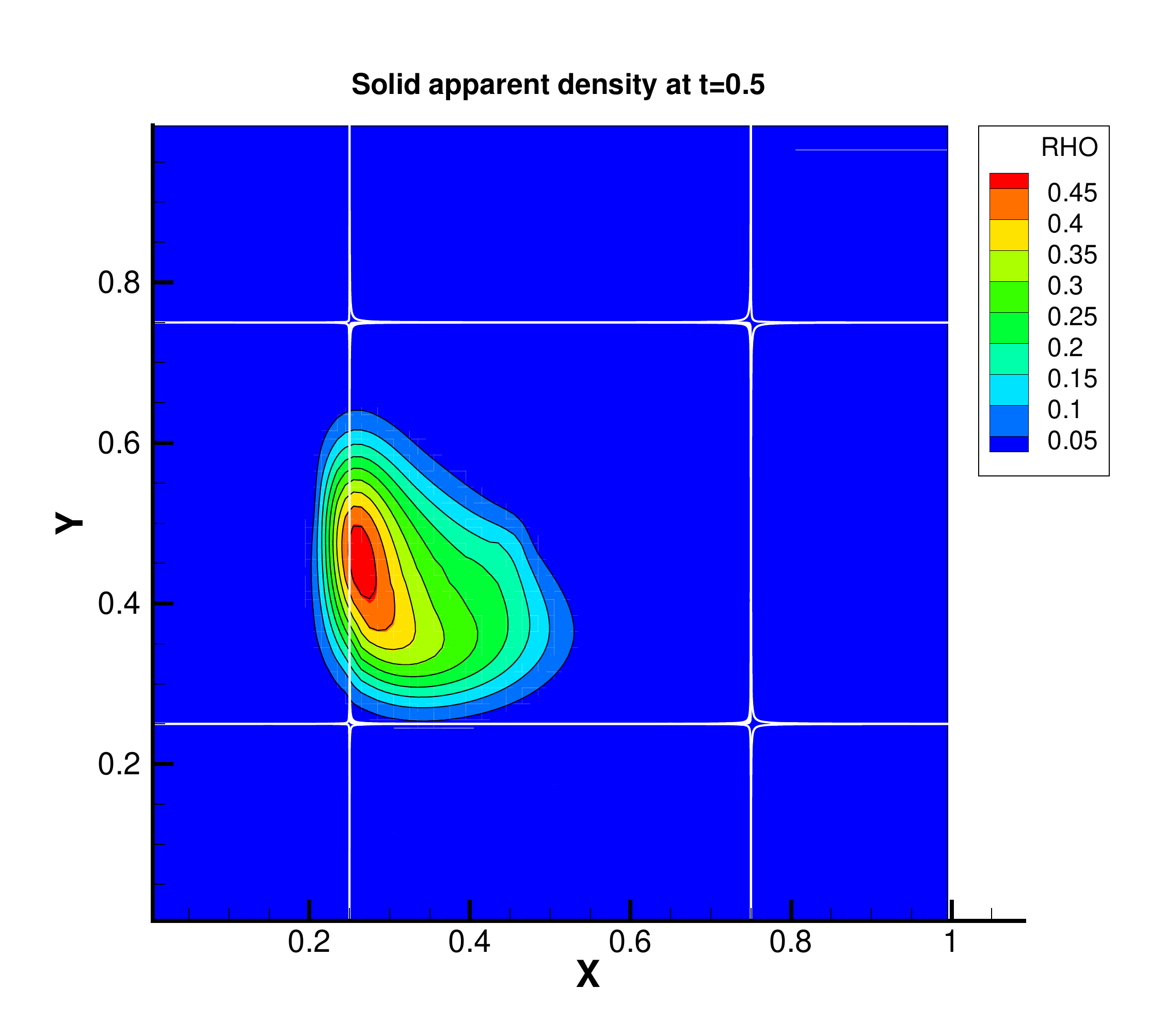}
\centerline{\footnotesize (a) t=0.6}
\end{minipage}
\hfill
\begin{minipage}[t]{0.48\textwidth}
\centering
\includegraphics[width=\textwidth]{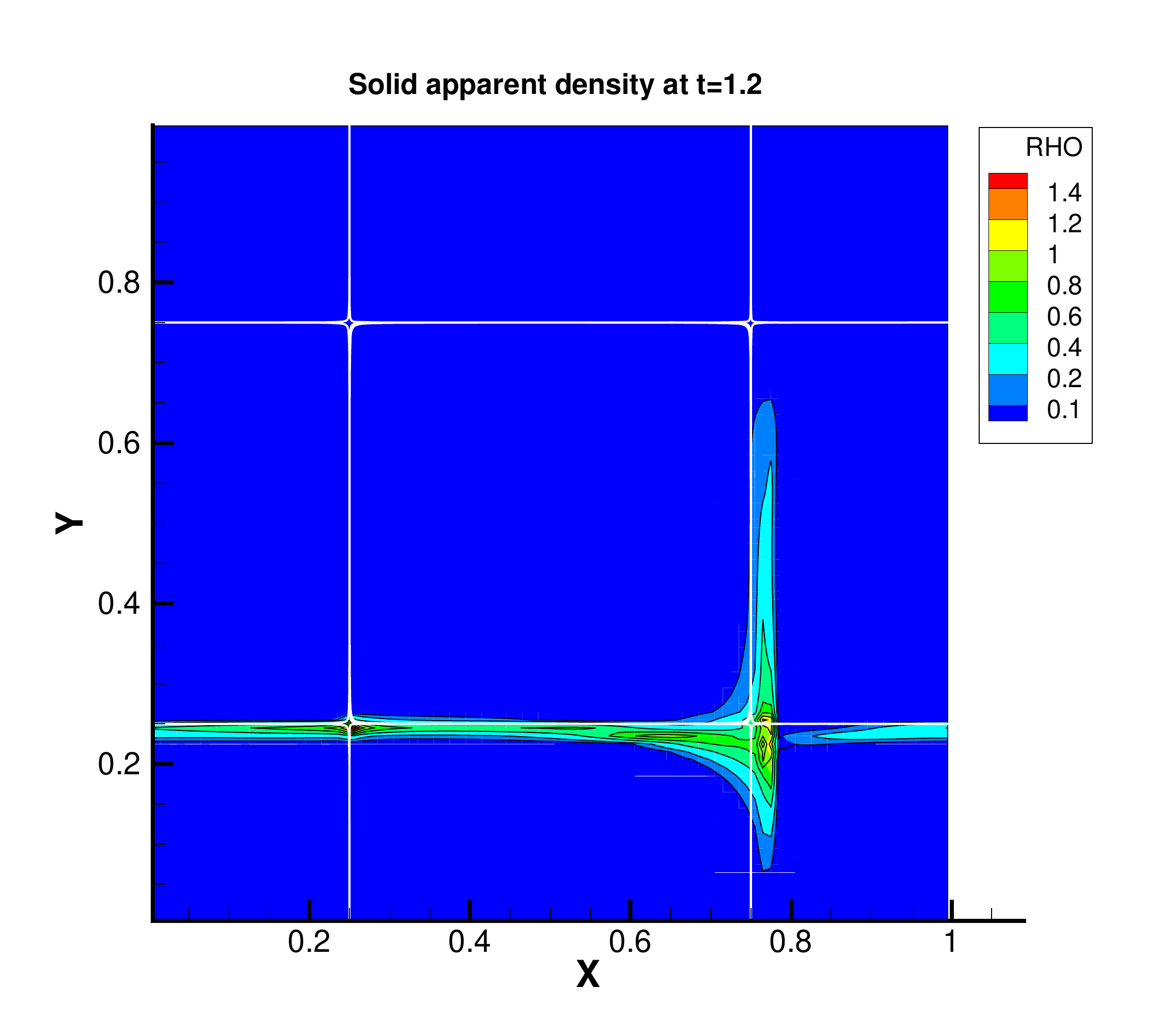}
\centerline{\footnotesize (b) t=1.2}
\end{minipage}
\caption{The UGKS-M (contour) and pressureless Euler (lines) results of particle apparent density distributions in {Taylor-Green} flow at $t=0.6$ and $t=1.2$, with parameter $Kn_s=10^{-4}$ and $St_s=0.1$.}
\label{2taylor4}
\end{figure}

\begin{figure}
\begin{minipage}[t]{0.48\textwidth}
\centering
\includegraphics[width=\textwidth]{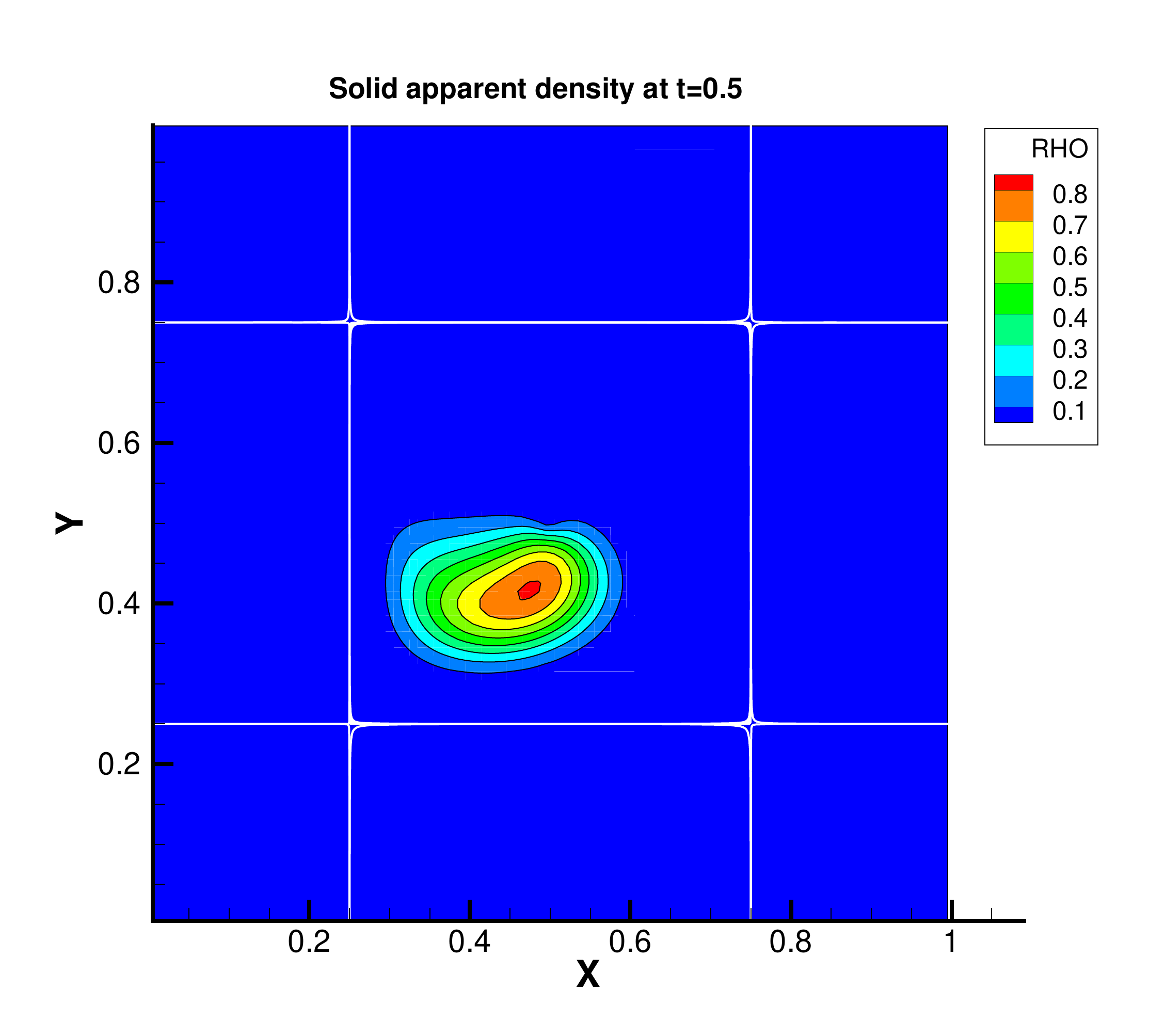}
\centerline{\footnotesize (a) t=0.6}
\end{minipage}
\hfill
\begin{minipage}[t]{0.48\textwidth}
\centering
\includegraphics[width=\textwidth]{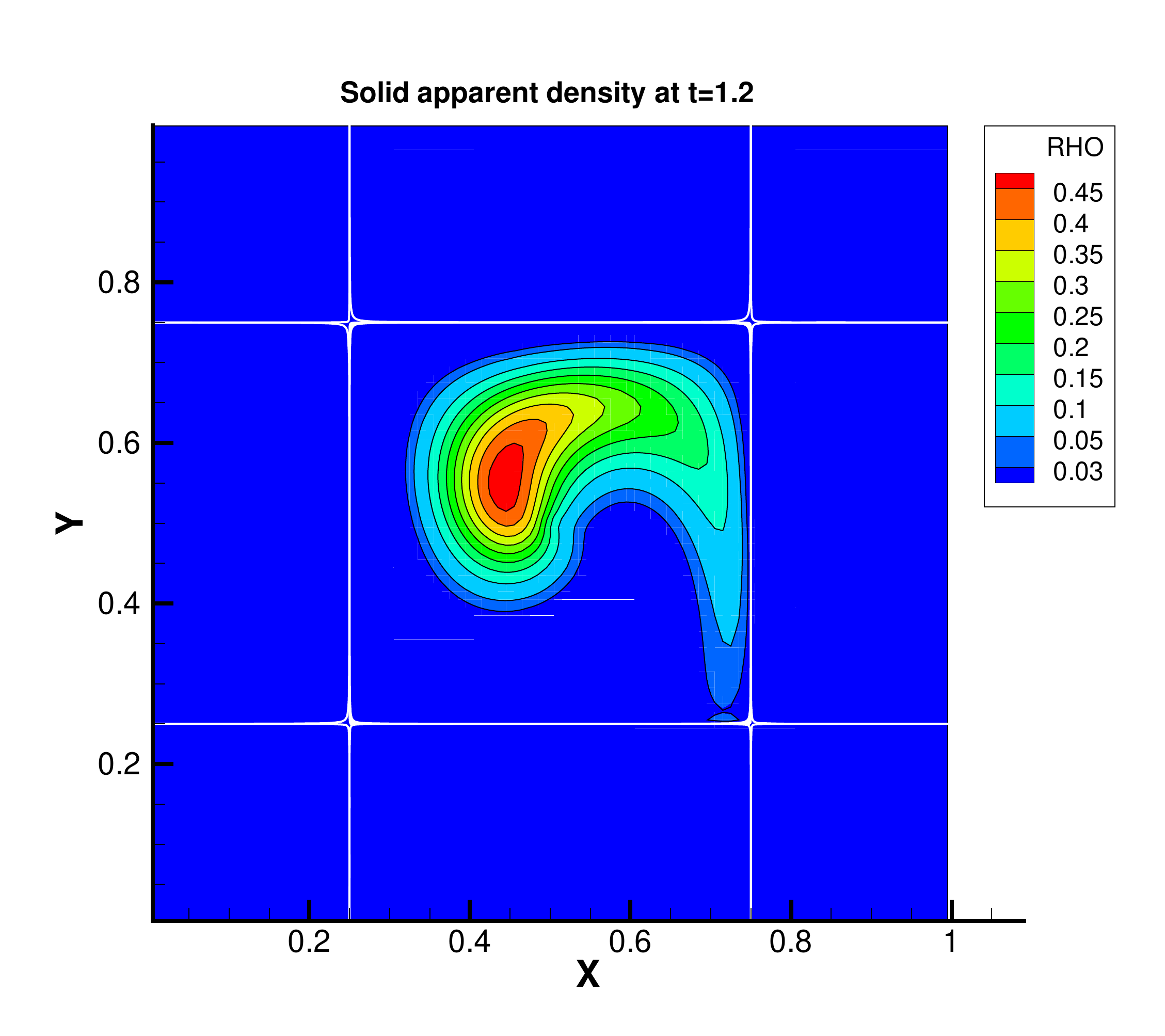}
\centerline{\footnotesize (b) t=1.2}
\end{minipage}
\caption{The UGKS-M (contour) and pressureless Euler (lines) results of particle apparent density distributions in {Taylor-Green} flow at $t=0.6$ and $t=1.2$, with parameter $Kn_s=10^{-4}$ and $St_s=10^{-3}$.}
\label{2taylor5}
\end{figure}

{
Preferential concentration describes the tendency of particles to cluster in regions of high strain or low vorticity due to their inertia.
The mechanisms which drive preferential concentration are centrifuging of particles away from vortex cores and accumulation of particles in convergence zones.}
In this example, we use UGKS-M to study the particle segregation in Taylor-Green flow, which is a 2D extension of the one dimensional particle concentration under a harmonic oscillatory flow.
{
Two initial conditions are considered as shown in Fig. \ref{2taylor0}.}
The gas field is assumed to be a two-dimensional Taylor-Green vortex with periodic boundary condition.
For the first initial condition, particles are uniformly distributed in space.
And for the second initial condition, the particles are set to be uniformly distributed in a circle centered in $(0.5,1-5/(4\pi))$ with radius $1/(4\pi)$ \cite{desjardins2008}.
For both test case, the initial particle velocity is the same as the initial gas flow velocity.
{
The reference length is the length of the computational domain, the reference velocity is the largest velocity magnitude of the initial gas field,
the reference density is the apparent density of the initial solid field, and the reference temperature is the temperature of the gas field.}
The initial gas density is $\rho_g=1$, initial solid phase apparent density is $\epsilon_s \rho_s=1.0$, the initial velocity field for both gas and solid phase is $U=\sin(2\pi x)\cos(2\pi y)$, $V=-\cos(2\pi x)\sin(2\pi y)$, the initial gas pressure is $p=1+(\sin(4\pi x)+\cos(4\pi y))/4$, initial granular temperature of solid phase is $T_s=10^{-8}$, and the restitution coefficient is $r=0$.
{
The physical domain is $[0,1]\times[0,1]$ divided equally into $200\times 200$ cells, and the velocity space is $[-1.2,1.2]\times[-1.2,1.2]$ with $42\times 42$ velocity cells.}
This problem is characterised by two important parameters, i.e., the Knudsen number and the Stokes number.
{
According to the analysis in \cite{chaisemartin2007}, the critical Stokes number is $St_c=1/8\pi$, below which the kinetic number density function will keep mono-kinetic, and above which the particle trajectory crossing can occur.
For the first initial condition, we first take $\tau_{st}=0.3>St_c$ and $Kn_s=10^{4}$. The solution of UGKS-M is compared with PIC solution at time $t=0.6$ and $t=2.0$ as shown in Fig. \ref{2taylor1}. The physical consistent particle trajectory crossing is captured, and UGKS-M gives satisfactory result comparing to PIC up to $t=0.6$, however, due to the numerical dissipation of finite volume scheme, the numerical resolution decreases for a long time calculation at $t=2$.
Next, we decrease the Stokes number to 0.03, which is less than the critical Stokes number, the solution of the UGKS-M and PIC results are shown in Fig. \ref{2taylor2}. Under this Stokes number, the velocity distribution will remain mono-kinetic and particles will concentrate on the edge of vortexes.
The UKGS-M solution agrees well with the PIC solution.
Then we reduce the Knudsen number to $Kn_s=10^{-4}$. In such regime, the intense inelastic collision will dissipate the kinetic energy of particles and even for large Stokes number $\tau_{st}=0.3$, an efficient preferential concentration occurs. The density distribution at $t=0.6$ of UGKS-M solution is shown in Fig. \ref{2taylor3} comparing with the pressureless Euler solution.
For the second initial condition, if the Stokes number smaller than $St_c$, the particles will remain inside of a {Taylor-Green} vortex forever
and no particle trajectory crossing will appear, and eventually the particles will accumulate at four corners of the vortex where the flow velocity is small. For the Stokes number larger than the critical stokes number, some particles will escape from the original vortex and enter into neighboring cells, and the particle trajectory crossing will appear \cite{desjardins2008,chaisemartin2007}. We first set the parameter as $Kn_s=10^{-4}$ and $\tau_{st}=0.1$. The solutions of UGKS-M at $t=0.6$ and $t=1.2$ are shown in Fig .\ref{2taylor4}, comparing with the pressureless Euler equation. Then we reduce the Stokes number to $\tau_{st}=10^{-3}$, and the UGKS-M solution and pressureless Euler solution are shown in Fig. \ref{2taylor5}. For both Stokes numbers, the solutions of UGKS-M are consistent with the theoretical analysis, and agrees well with Euler solution.}

\subsection{Shock-induced fluidization of a particles bed}
\begin{figure}
\centering
\includegraphics[width=0.9\textwidth]{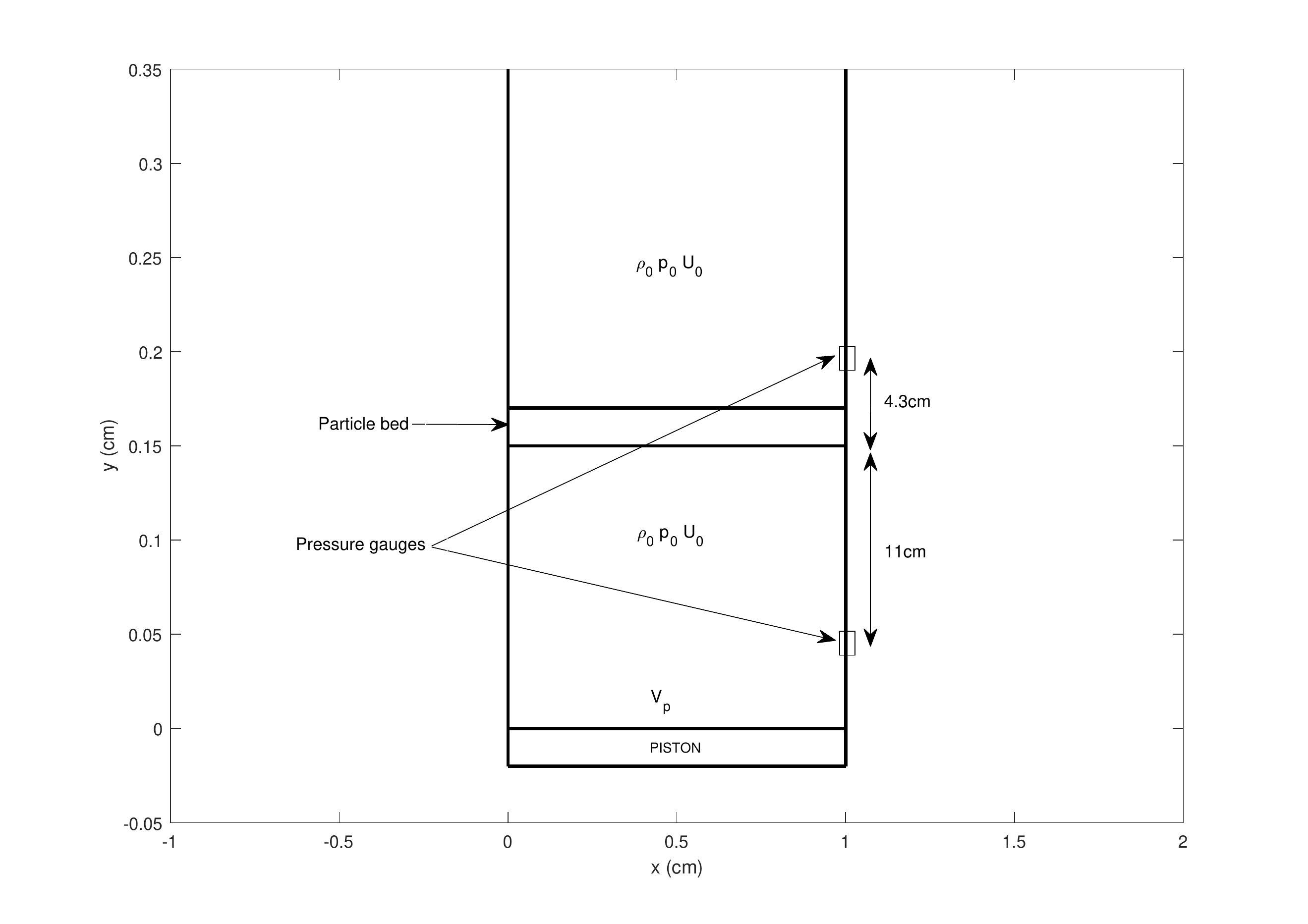}
\caption{Experiment set-up for fluidization shock tube test. A shock at Mach number 1.3 is created by the expansion of the high pressure gas through a moving piston at a speed of $151$m/s.}
\label{bed-initial}
\end{figure}

\begin{figure}
\centering
\includegraphics[width=0.6\textwidth]{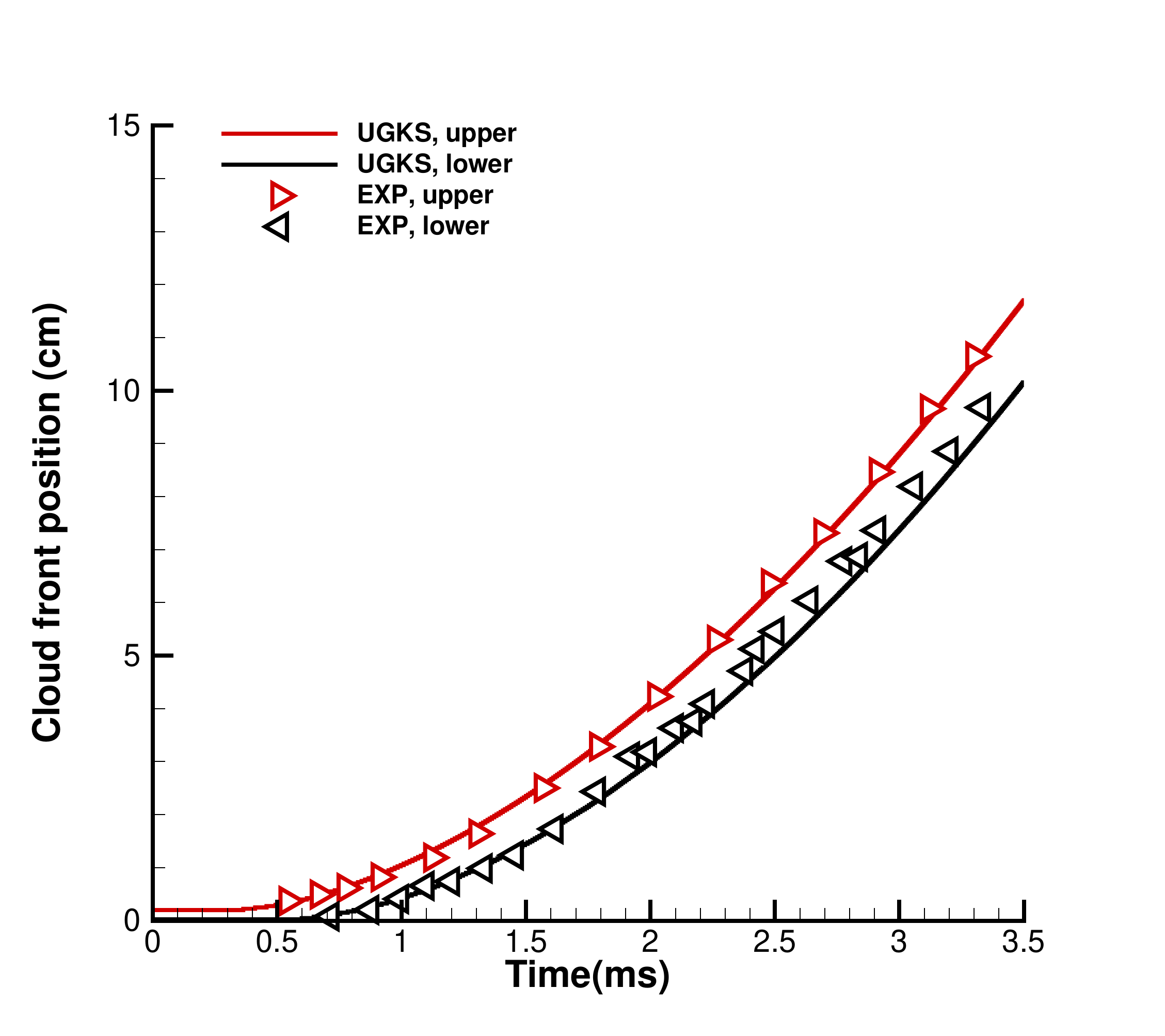}
\caption{Upper and lower front trajectories of the $2$mm bed impinged by a Mach 1.3 shock.}
\label{bed1}
\end{figure}

\begin{figure}
\centering
\includegraphics[width=0.6\textwidth]{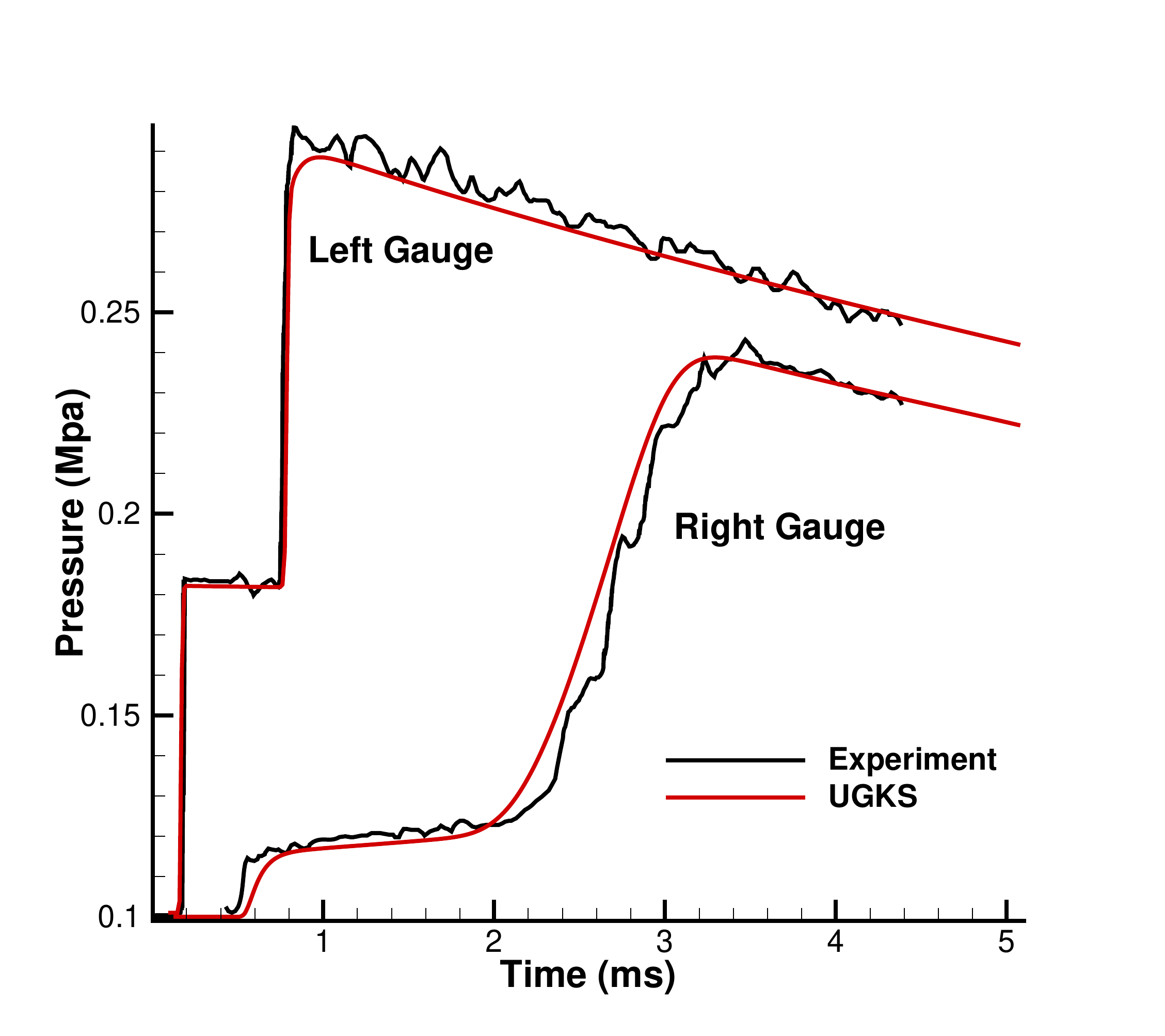}
\caption{Pressure signals from the transducers at the upstream and downstream of the bed shown in the experimental setup.}
\label{bed2}
\end{figure}

\begin{figure}
\centering
\includegraphics[width=0.6\textwidth]{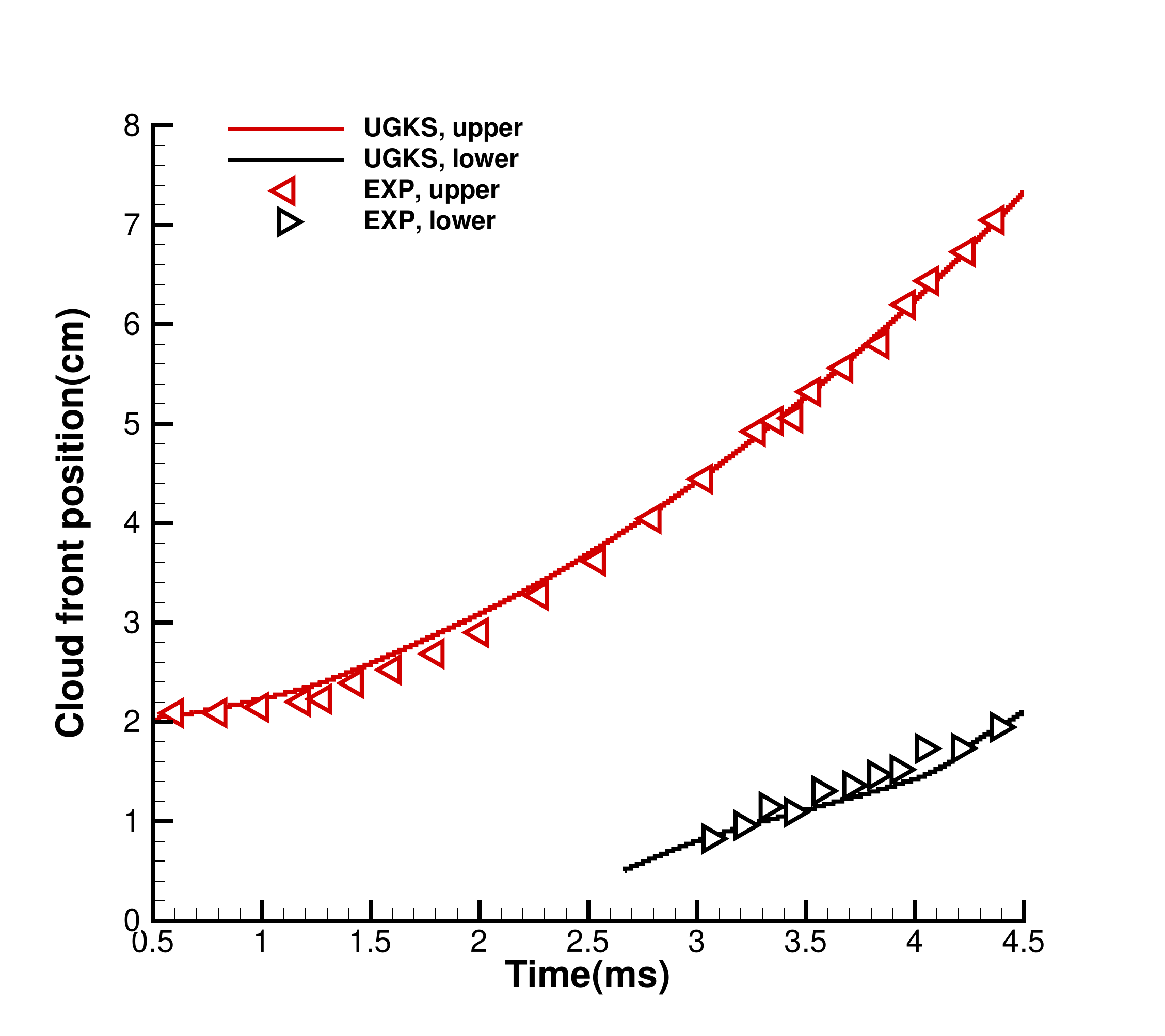}
\caption{Upper and lower bed front trajectories for the $2$cm bed with $1.5$mm diameter glass spheres impinged by a Mach 1.3 shock in air.}
\label{bed3}
\end{figure}

\begin{figure}
\centering
\includegraphics[width=0.6\textwidth]{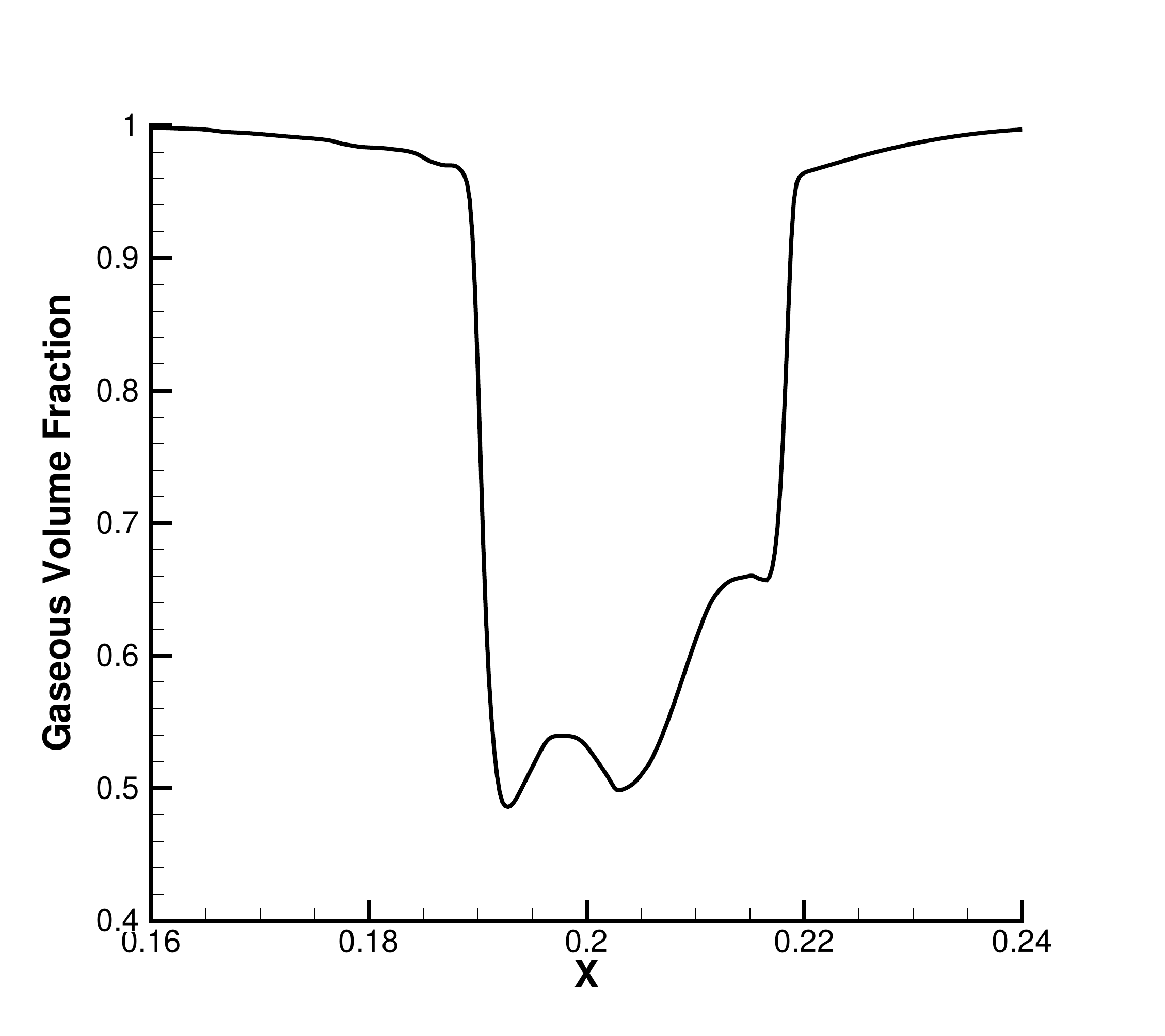}
\caption{Gaseous volume fraction distribution for the 2cm bed case with 1.5mm diameter glass spheres impinged by a Mach 1.3 shock in air at time $t=4.5ms$.}
\label{bed4}
\end{figure}

In this section, we study the experiment of shock induced fluidization of a particles bed \cite{rogue1998,saurel2017}.
The experiment set up is shown in Fig. \ref{bed-initial}, where initially a bed of particles locates at $x=15\text{cm}$, and two pressure gauges locate $11\text{cm}$ left and $4.3\text{cm}$ right to the particle bed.
A shock wave with Mach number $1.3$ is generated by a right moving piston with velocity $151\text{m/s}$.
In order to determine the Stokes number, we first calculate a single layer of particles and compare the simulation result with experiment data.
The parameters for the experiment and simulations are shown in Table \ref{1bed}.
From our numerical experiments, we find that the cloud front trajectories between UGKS-M and experiment matches well with $St=0.62$,
and the cloud front comparison is shown in Fig. \ref{bed1}.
Then, we calculate a dense $2\text{cm}$ bed composed of $1.5\text{mm}$ diameter glass particles. The initial volume fraction in the bed is $0.65$, and the initial pressure is $10^{5}$Pa.
When passing the dense particle bed, turbulence will be generated in gas phase.
The turbulent energy is treated as the internal energy
and in this calculation the internal degree of freedom of gas phase is modeled by $k(t)=k_0+0.15(t/t_{ref})^{1.5}$.
The Stokes number is set to be $0.62$, and Knudsen number is $1.0\times 10^{-3}$.
The computational domain is divided into $1000$ cells in physical space and $128$ cells in velocity space from $[-284\text{m/s},284\text{m/s}]$. The time dependent pressure signal is shown in Fig. \ref{bed2}, and the cloud front trajectories are shown in Fig. \ref{bed3}.
The UGKS-M gives satisfactory results in comparison with the experimental measurement. {Fig. \ref{bed4} presents the gas phase volume fraction at $t=4.5\text{ms}$.}

\begin{table}
  \centering
  \caption{Parameters of a single layer of particles}
\begin{tabular}{|l | c|}
  \hline
  Air pre-shock density  &$1.2\text{kg/m}^3$\\ \hline
  Incident shock Mach number &$1.3$\\ \hline
  Particle density &$2500\text{kg/m}^3$ \\ \hline
  Particle diameter &$2\text{mm}$ \\
  \hline
\end{tabular}
  \label{1bed}
\end{table}

\section{Conclusion}
In this paper, we propose a UGKS-M scheme for dilute disperse gas-particle multiphase flow.
The scheme is built in a finite volume framework. For the solid particle phase, the numerical flux is constructed by the UGKS
for preserving multiscale property.
For the gas phase, the GKS flux is used for the gas flow in the continuum regime.
The interaction between the solid and gas phase is calculated by a velocity space mapping method.
The UGKS-M calculates the flow in regimes from collisionless to two-fluid NS regime with different Knudsen number,
and from granular flow to dusty gas dynamics with different Stokes number.
The stability condition of UGKS-M is the CFL condition, and no requirement is imposed by the Knudsen and Stokes numbers.
By taking into account the material temperature, once the total energy loss in inelastic collision transfers into particle material thermal energy,
the whole system conserves the total mass, momentum, and energy.
The numerical experiments show that UGKS-M can capture the physical solution in different regimes,
such as the particle trajectory crossing, particle wall reflection, and particle scattering through elastic collision.
The simulation of Shock-induced fluidization test recovers the experiment measurement well.
In conclusion, the UGKS-M is an accurate multiscale numerical method for the gas-particle multiphase system, which
can be used confidently in many engineering applications.
The methodology of direct modeling in UGKS is a powerful tool for the construction of numerical method for simulating
multiscale transports.

\section*{Acknowledgements}
The authors would like to thank Mr. Zhu Yaju and Mr. Xiao Tianbai at HKUST for fruitful discussion.
The current research is supported by Hong Kong research grant council (16206617,16207715),
grants from NSFC {(Grant No. 11772281,91530319)} and science challenge project (No. TZ2016001).

{\printnomenclature[6em]}

\section*{Reference}
\bibliographystyle{acm}
\bibliography{multiphase}

\newpage

\clearpage

\end{document}